\documentclass[12pt]{iopart}%IOP
\pdfoutput=1
\usepackage{iopams}%IOP
\usepackage{amssymb,bm,graphicx}

\usepackage{multirow}

\newcommand{\sakujo}[1]{}

\newcommand{\ABS}[1]{#1}
\newcommand{\EIG}{\lambda}
\newcommand{\HALL}{\alpha}
\newcommand{\HEL}{\sigma}
\newcommand{\POL}{s}

\newcommand{\U}{\underline}

\newcommand{\Elsasser}{Els\"asser}

\newcommand{\LieBracket}[3]{{#1[}#2,#3{#1]}}
\newcommand{\LieBraket}{\LieBracket}

\newcommand{\Braket}[3]{{#1\langle}#2{#1|}#3{#1\rangle}}
\newcommand{\BraKet}{\Braket}

\newcommand{\ADJ}[3]{{\rm{ad}}^{#1}_{#2}{#3}}

\newcommand{\GEV}[2]{\vec{\bm{#1}}#2}
\newcommand{\GEC}[2]{\widehat{#1}#2}
\newcommand{\GEM}[2]{\tilde{#1}#2}

\newcommand{\CHM}[1]{\hat{#1}}
\newcommand{\CHW}[2]{{\bm{#1}}#2}
\newcommand{\CHC}[2]{\widehat{#1}#2}

\newcommand{\EIGEN}[2]{\lambda^{{#2}}(\widetilde{#1})}

\newcommand{\tripleHMHD}[4]{{#1(}\!{#1(}#2{#1|}\!{#1|}#3{#1|}\!{#1|}#4{#1)}\!{#1)}}
\newcommand{\tripleEuler}[4]{{#1(}#2{#1|}#3{#1|}#4{#1)}}
\newcommand{\LCC}[2]{\widetilde{\nabla}_{#1}{#2}}
\newcommand{\CC}[1]{\overline{#1}}

\newcommand{\dd}[2]{\frac{\partial #2}{\partial #1}}
\newcommand{\sdd}[2]{\frac{{\rm{d}} #2}{{\rm{d}} #1}}
\newcommand{\sDD}[2]{\frac{{\rm{D}} #2}{{\rm{D}} #1}}

\newcommand{\DivFree}[2]{{#1(}#2{#1)}_{S}}

\newcommand{\CalcNote}[1]{}%
\renewcommand{\CalcNote}[1]{*>#1<*}%

\newcommand{\q}{x}
\newcommand{\Q}{}

\newcommand{\EnsAve}[2]{{#1\langle}#2{#1\rangle}}

\newcommand{\curl}{\nabla\times}

\newcommand{\Vi}[1]{\bm{U}_{#1}}
\newcommand{\Ve}[1]{\bm{J}_{#1}}
\renewcommand{\Vi}[1]{\bm{V}_{i#1}}
\renewcommand{\Ve}[1]{\bm{V}_{e#1}}

\newcommand{\T}{}

\newcommand{\g}{}% dummy for Riemannian metric

\newcommand{\Exp}{}

\newcommand{\uncurl}{(\curl)^{-1}}

\newcommand{\p}{}

\newcommand{\COR}{}
\newcommand{\vExp}{}
\newcommand{\MSC}{}
\newcommand{\SAC}{}
\newcommand{\gMSC}{}
\newcommand{\gSAC}{}

\newcommand{\polarity}{linear-mode-branch}%
\newcommand{\myLineBreak}[1]{\nonumber\\\fl\hspace{#1em}}
\renewcommand{\gMSC}[2]{\widehat{R}_{#1}^{#2}}%
\renewcommand{\gSAC}[2]{\widehat{K}_{#1}^{#2}}%
\renewcommand{\p}{\underline{p}}%{p_*}%
\renewcommand{\q}{\underline{q}}%{q_*}%
\newcommand{\vs}{\vspace{1em}}
\newcommand{\extf}{jpg}

\begin{document}%IOP

\title[Differential-geometrical approach to HMHD dynamics II]{%
Differential-geometrical approach to the dynamics of dissipationless 
incompressible Hall magnetohydrodynamics: 
II.
Geodesic formulation 
and 
Riemannian curvature analysis 
of hydrodynamic and magnetohydrodynamic stabilities
}
\author{ARAKI Keisuke}

\address{%IOP
 Faculty of Engineering, Okayama University of Science,
 1-1 Ridai-cho, Kita-ku, Okayama 700-0005 JAPAN
}
\ead{araki@are.ous.ac.jp}%IOP
\vspace{10pt}%IOP
\begin{indented}%IOP
\item[]\today%IOP
\end{indented}%IOP

%\begin{document}%NON-IOP

%\maketitle%NON-IOP

\begin{abstract}
In this study, the dynamics of a dissipationless incompressible Hall magnetohydrodynamic (HMHD) medium are formulated as geodesics on a direct product of two volume-preserving diffeomorphism groups.
Formulations are given for the geodesic and Jacobi equations based on a linear connection with physically desirable properties, 
which agrees with the Levi--Civita connection.
Derivations of the explicit normal-mode expressions for the Riemannian metric, Levi--Civita connection, and related formulae and equations are also provided using the generalized {\Elsasser} variables (GEVs).
Examinations of the stabilities of the hydrodynamic (HD, $\HALL=0$) and
magnetohydrodynamic (MHD, $\HALL\to0$) motions and
the $O(\HALL)$ Hall-term effect 
in terms of the Jacobi equation and the Riemannian sectional curvature tensor are presented,
where $\HALL$ represents the Hall-term strength parameter.
It is very interesting that the sectional curvatures of the MHD and HMHD systems between two %($\HALL\to0$ limit of) 
GEV modes were found to take both the positive (stable) and negative (unstable) values, while that of the HD system between two complex helical waves was observed to be negative definite.
Moreover, for the MHD case, negative sectional curvatures were found to occur only when mode interaction was ``local,''
i.e., the wavenumber moduli of the main flow (say $p$) and perturbation (say $k$) were relatively close to each other.
However, in the nonlocal limit ($k\ll p$ or $k\gg p$),
the sectional curvatures were always positive.
This result leads to the conjecture that the MHD interactions mainly excite wavy or non-growing motions;
however, some local interactions cause dynamical instability 
that leads to chaotic or turbulent plasma motions.
Additionally, it was found that the tendencies of the $O(\HALL)$ effects are opposite between the ion cyclotron and whistler modes.
Comparison with the energy-Casimir method is also discussed using a remarkable constant of motion which relates the Riemannian curvature to the second variation of the Hamiltonian.

\end{abstract}

%\newpage

\section{Introduction\label{Introduction}}

In the present study, we addressed the geodesic formulation and related stability problems of a dissipationless, incompressible Hall magnetohydrodynamic (HMHD) system.
The purposes of the study are twofold: to revisit the geodesic formulation framework from a physical viewpoint, and to apply it to the linear stability of the hydrodynamic (HD), magnetohydrodynamic (MHD), and HMHD systems, with a consideration of its applicability to the analysis of fully-developed homogeneous and isotropic turbulence.
\vs

Since Arnold cast the study of dynamical systems on Lie groups and related hydrodynamic topics in a unified form \cite{arnold1966sur}, multiple fluid dynamical systems have been recognized to exist on appropriate Lie groups \cite{arnold1998topological}.
The key to the Lagrangian formalism of Lie groups is the appropriate choice of the inner product and the Lie bracket, which are denoted hereafter by $\Braket{\big}{*}{*}$ and $\LieBraket{\big}{*}{*}$, respectively.
Once these two mathematical structures are defined, the variational formulation is formally established, and the evolution equation known as the Euler--Poincar\'e equation is derived \cite{marsden2013introduction}.

However, the geodesic formulation, which was also discussed in \cite{arnold1966sur}, still seems to hold a somewhat bizarre position among the analytical mechanical theories of continuum mechanics because the Lagrangian or Hamiltonian mechanical descriptions of such systems work well without Riemannian geometrical notions such as the Levi--Civita connection and the Riemannian curvature tensor (e.g. \cite{marsden2013introduction}).

Despite this situation,
the geodesic formulation perspective provides a powerful tool.
The second variation of action, which yields the Jacobi equation, is known to provide information about the dynamical stabilities of the solutions (\cite{arnold1989mathematical}; app. I).
As an advantage, this approach provides a differential geometrical framework for the predictability problem of HD systems (\cite{arnold1966sur}; \S 11).
From this viewpoint, the attraction/repulsion (instability) or winding around (stability) of infinitesimally close paths is determined by the sign of the Riemannian curvature associated with the appropriate linear connection.
The value of the sectional curvature is determined by the snapshot pair of a (possibly non-stationary) reference solution and a perturbation field imposed upon it without solving the evolution equation or eigenvalue problem.

Thus, geodesic formulation and the associated sectional curvature analysis have attracted much research interest.
For example, Arnold \cite{arnold1966sur} performed curvature analysis of the stability of the flow of a neutral incompressible fluid on a two-dimensional torus.
These results were extended to three dimensions by Nakamura et al. \cite{nakamura1992geodesics}
and to the general $n$-dimensional case by Lukatskii \cite{lukatskii1981curvature}.
These researchers considered the sectional curvature between two Fourier modes and obtained negative values (i.e., instability of the basic flows).
Ohkitani \cite{ohkitani2010numerical} approached the problem numerically, adopting the formulation by Rouchon \cite{rouchon1992jacobi} and solving the initial-value problem of the Jacobi equation.
Zeitlin and Kambe \cite{ZeitlinKambe1993} applied the method to the standard (or reduced single-fluid) system of MHD in two dimensions, and Hattori \cite{hattori2011differential} later extended it to the $n$-dimensional case.

Moreover, as was reviewed by Vizman \cite{Vizman2008}, a wide variety of dynamical systems, including Euler, ideal MHD, and Korteweg-de Vries (KdV) systems, have been formulated as geodesic equations on appropriate Lie groups.
Despite the diversity of these dynamical systems, they did not include the geodesic equations on semidirect products of two groups, also due to Vizman \cite{vizman2001geodesics}.
Recently, dissipationless, incompressible Hall MHD (HMHD) systems were found to be dynamical systems on the semidirect product of two volume-preserving diffeomorphisms, $S$Diff($M$)$\ltimes S$Diff($M$) \cite{araki2015differential}.
Another interesting finding was that HMHD systems could also be formulated as dynamical systems on a direct product group, $S$Diff($M$)$\times S$Diff($M$), by changing the basic variables \cite{araki2015helicity}.

Many dynamical systems has been treated in this framework, however, non-uniqueness problem of the definition of parallel translation seems scarcely discussed.
A geodesic on a curved space (say $\gamma(t)$) is a curve such that its tangent vector $\dot{\gamma}$ is given by the parallel translation of itself along $\gamma(t)$.
However, in a curved space, the ``parallel translation'' of a vector between two close points is determined by the \textit{linear connection} around them and thus is not unique.
In fact, it is known that when a linear connection (say $\Gamma_{ij}^{k}$) 
and an induced geodesic are given, the same geodesic is given by the connection generated by the linear combination
$\Gamma_{ij}^{*k}=t\Gamma_{ij}^{k}+(1-t)\Gamma_{ji}^{k}$,
where the parameter $t$ is related to the \textit{torsion} associated with $\Gamma^*$
(\cite{KobayashiNomizu1963}; ch. III, \S 7).
This non-uniqueness of the connections implies that the ensemble of the curves determined by the Euler--Lagrange equation associated with a certain variational problem does not define the corresponding linear connection uniquely.
Thus, it is necessary to consider the physically appropriate linear connection, despite the torsion-free condition ($\Gamma_{ij}^{k}=\Gamma_{ji}^{k}$) often being imposed explicitly or assumed implicitly.

\vs

HMHD was firstly derived by Lighthill \cite{Lighthill1960studies}
and is based on the following approximations:
the MHD approximation of the Lorentz force by the electron component of the plasma ($ \bm{E} + \Ve{} \times \bm{B} = \bm{0} $);
and the approximation of the current density field generated by the difference between the ion and electron fluid motions ($ \Vi{} - \Ve{} = \HALL \bm{J}/en_{e}$); 
and an approximate form of Ampere's law ($ \nabla\times\bm{B}=\mu_{0}\bm{J}$).
The basic equations are as follows:
\begin{eqnarray}
&&
  \nabla\cdot\bm{u}=\nabla\cdot\bm{b}=0,
\nonumber
\\&&
  \partial_{t}{\bm{u}}
  =
  \bm{u} \times
  \bm{\omega}
  +
  \bm{j} \times \bm{b}
  -
  \nabla P
,
\label{eq of mot u and b}
\\&&
  \partial_{t}{\bm{b}}
  =
  \nabla \times ( \bm{u} \times \bm{b} )
  -
  \HALL
  \nabla \times ( \bm{j} \times \bm{b} )
,
\nonumber
\end{eqnarray}
where $\bm{u}$, $\bm{b}$, $\bm{\omega}$, $\bm{j}$, $P$, and $\HALL$
are the appropriately nondimensionalized variables corresponding to the ion velocity, magnetic field, vorticity ($\bm{\omega}=\curl\bm{u}$), current density 
($\bm{j}=\curl\bm{b}$),
generalized pressure, and Hall-term strength parameter, respectively.

As for the stability problems of the HMHD system, since Holm's pioneering work in this field \cite{Holm1987}, the analytical mechanical approaches have been mainly carried out from the Hamiltonian mechanical viewpoint (e.g. \cite{SahraouiETAL2003,HirotaETAL2006,YoshidaHameiri2013}).
The energy-Casimir method was one of the principal tools for the stability analysis (e.g. \cite{marsden2013introduction}; \S1.7).
The method treats only the stability of the \textit{stationary solutions}, though the analysis result states the Lyapunov stability, i.e., a priori estimation of perturbation amplitude for a sufficiently long time interval.
In the present study, since both the energy-Casimir method and the Riemannian curvature analysis are based on the second variation of some appropriate functional, we will discuss their conceptual correspondence.

As for the fully-developed turbulence of the HMHD systems, the effects of the Hall term on the properties of MHD turbulence have been attracted interests of many researchers from various viewpoints including the closure approach in the weak/wave turbulence framework \cite{galtier2006wave}, evaluation of turbulent energy transfer and dynamo action using direct numerical simulation (DNS) data \cite{MininniETAL2007}, and coherent structure formation by DNS \cite{MiuraAraki2014}.

In the present study, in order to evaluate characteristic time of turbulent flows, we will attempt to apply the Riemannian curvature analysis to the turbulence problem under the the statistical homogeneity and isotropy assumption.
Estimation of characteristic time scale (for example, Lyapunov exponent) has been recognized as one of the important issues of chaos and turbulence theories \cite{YamadaSaiki2007}.
As will be seen in the deriving process and interpretation of its implication in section \ref{linear stability analysis}, the sectional curvature operator will be shown to indicates the characteristic time of perturbation growth, since it has dimension of the reciprocal of the square of time $[{\rm{T}}^{-2}]$.
We utilize here the important advantage of geodesic approach that the Riemannian curvature is computable for non-stationary solutions.

As for the normal-mode analysis, it should be remarked that application of the geometrical method to the dissipationless, incompressible HD, MHD, and HMHD systems has its foundation on the following fact:
these three systems were found to have common dynamical system features \cite{2016arXiv160105477A};
that is, each system has its own action-preserving integro-differential operator,
and the corresponding eigenfunctions yield formally the same spectral representation as that of the related formulae and equations.
In particular, we found that the product of the Riemannian metric $g_{ij}$ with the structure constants $C^{i}_{jk}$ of the Lie group was given by the product of the eigenvalue $\Lambda(i)$ of the operator with a certain totally antisymmetric tensor $T_{ijk}$: 
$g_{i\alpha}C^{\alpha}_{jk}=\Lambda(i)T_{ijk}$.

The eigenfunctions in the HMHD system are generally given by double Beltrami flows (DBFs), namely, force-free stationary solutions for a two-fluid plasma, as determined by Mahajan and Yoshida \cite{mahajan1998double}.
Among the DBFs, considering the influence of a uniform background magnetic field and the Hall-term effect vanishing limit, the generalized {\Elsasser} variables (GEVs \cite{galtier2006wave}) have been found to be the most suitable for avoiding problems with singularities in the standard MHD limit \cite{araki2015helicity}.
\vs

This paper is organized as follows.
In section 2, the mathematical preliminaries for the geodesic formulation of the HMHD system are reviewed.
In section 3, the formulation of the linear stability problem as the Jacobi equation 
and the derivation of the Riemannian metric tensor are presented.
Stability analyses of the HD, MHD, and lowest-order of the Hall-term effect are provided in section 4.
A discussion is given in section 5.

\section{Geodesic formulation of a dissipationless, incompressible HMHD system}

In this section, we review some mathematical preliminaries of the geodesic formulation of
a dissipationless, incompressible HMHD system.
First,
the derivation of the equation of motion of an incompressible HMHD fluid from
Hamilton's principle is presented
as a review of the Lagrangian mechanical foundation.
Next,
the connection that causes 
the geodesic equation to agree with the equation of motion is discussed.

\subsection{Lagrangian mechanical foundations of a dissipationless, incompressible HMHD system}

The details of the mathematical backgrounds were described in \cite{araki2015differential,araki2015helicity,2016arXiv160105477A}.
In this section, we introduce some mathematical notions and notation which was not explained in the previous studies but will use in the following sections.
\vs

As is mentioned in the section \ref{Introduction}, the key mathematical structures are the Riemannian metric and the Lie bracket.
For the HMHD systems, they are defined as follows \cite{araki2015helicity}:
\begin{eqnarray}
\fl
  \Braket{\big}{\GEV{V}{_{1}}}{\GEV{V}{_{2}}}
  &:=&
  \int {\rm{d}}^3\vec{x} \Big[
    \Vi{1} \cdot \Vi{2}
    +
    \HALL^{-2}
    \uncurl(\Vi{1}-\Ve{1}) \cdot \uncurl(\Vi{2}-\Ve{2})
  \Big]
,
\label{def riemannian metric}
\\
\fl
 \LieBracket{\big}{\GEV{V}{_{1}}}{\GEV{V}{_{2}}}
 &:=&
 \Big(
  \nabla \times ( \Vi{1} \times \Vi{2} )
 ,
  \nabla \times ( \Ve{1} \times \Ve{2} )
 \Big)
,
\label{basic Lie bracket}
\end{eqnarray}
where $\curl$ and $\uncurl$ are the curl operator and its inverse%
\footnote{%
Here we use the relation
$
  \curl(\bm{a}\times\bm{b})=
  ( b^{k}\partial_{k}a^{i} - a^{k}\partial_{k}b^{i} ) \partial_{i}
$
that holds when $\bm{a}$ and $\bm{b}$ are divergence-free.
The sign of the Lie bracket is chosen to satisfy the Hausdorff formula
$
\renewcommand{\Exp}[1]{{\rm{e}}^{#1}}
  \Exp{\bm\xi}\Exp{\bm\eta}
  =
\renewcommand{\Exp}[1]{\exp(#1)}
  \Exp{\bm\xi+\bm\eta+\frac12[\bm\xi,\bm\eta]+\cdots}
,
$
where e and exp denote the exponential map of a vector field (i.e. $\exp(t\bm{V})$ is a solution of the ODE $\partial_t\vec{X}=\bm{V}$ for a fixed $\bm{V}$).
}%
, and the \textit{generalized velocity}, ${\GEV{V}{}}=({\Vi{}},{\Ve{}})$%
\footnote{%
In this paper, an arrow above a symbol denotes its multifunctional character. For example, a position vector is expressed by $\vec{x}=(x^1,x^2,x^3)$, and a pair of vector fields by $\GEV{V}{}=(\Vi{},\Ve{})$. Boldface letters are used to denote vector fields on $M$.
}%
, is the pair of the ion and electron velocity fields, which are related to the variables used in Eq. (\ref{eq of mot u and b}) by
\begin{eqnarray}
%\fl
  \bm{u}=\Vi{}, \ 
  \bm{\omega}=\curl\Vi{}, \ 
  \bm{j}=\HALL^{-1}(\Vi{}-\Ve{}), \ 
  \bm{b}=\HALL^{-1}\uncurl(\Vi{}-\Ve{})
.
\label{relation vi ve to u w j b}
\end{eqnarray}
Using these variables, the combination of these mathematical structures reads as
\begin{eqnarray}
\fl
  \Braket{\big}{\GEV{V}{_{3}}}{\LieBraket{\big}{\GEV{V}{_{1}}}{\GEV{V}{_{2}}}}
  =
  \int {\rm{d}}^3\vec{x} \bigg[
    {\bm{\omega}_{3}} \cdot ({\bm{u}_1}\times{\bm{u}_{2}})
    +
    {\bm{b}_{3}} \cdot \big(
      {\bm{u}_1}\times{\bm{j}_{2}}+{\bm{j}_1}\times{\bm{u}_{2}}
      -\alpha{\bm{j}_1}\times{\bm{j}_{2}}
    \big)
  \bigg]
.
\label{<a|[b,c]> in uwjb}
\end{eqnarray}
The term $\GEV{V}{}$-variable indicates the elements of the function space of the ion and electron velocity-field pairs.%, which we denote by $\mathfrak{g}$: $\mathfrak{g}=\mathfrak{X}_{\Sigma}(M)\times\mathfrak{X}_{\Sigma}(M)$, where $\mathfrak{X}_{\Sigma}(M)$ is the function space of the solenoidal vector fields on $M$.

Note that, when $\HALL=0$, the second term of the inner product Eq. (\ref{def riemannian metric}), which gives the magnetic energy in physical context, vanishes, because the difference between the ion and electron fluids is exactly zero ($\Vi{}-\Ve{}=0$).
Thus, the system describes the HD one (see \ref{CHW mode representation of the Euler sectional curvature}).
In the $\HALL\to0$ limit, however, the term remains at $O(1)$ in amplitude and the system is reduced to the MHD one.
\vs

The \textit{Lie derivatives}
are defined as the ``advections'' of functions, vector fields, and tensor fields.
They provide tools for proving certain conservation laws associated with the Lagrangian invariants \cite{araki2009comprehensive,TurYanovsky1993}.
The Lie derivative of the vector fields on the configuration space (say $G$) is defined by an extension of the basic definition (e.g., \cite{KobayashiNomizu1963}; ch. I, \S 3) to the direct product group:
$$
  \vec{L}_{\GEV{V}{}}{\GEV{\xi}{}}
  =
  \lim_{t\to0}\frac{\GEV{\xi}{}-\phi^{*}(t)\GEV{\xi}{}}{t}
  =
  \lim_{s\to0}\lim_{t\to0}\frac{\psi(s)-\phi(t)\circ\psi(s)\circ(\phi(t))^{-1}}{st}
  =
  \LieBracket{\big}{\GEV{\xi}{}}{\GEV{V}{}}
,
$$
where $\psi(s)$ and $\phi(t)$ are the one-parameter subgroups of $G$ 
that satisfy $\psi(0)=\phi(0)=e$, 
$\GEV{\xi}{}:=\partial_{s}\psi(s)|_{s=0}$, and
$\GEV{V}{}=\partial_{t}\phi(t)|_{t=0}$.

The action $S$ is defined along a path $\gamma(t;\epsilon)\subset G$,
where $t$ and $\epsilon$ are a line parameter and
a small parameter for variation, respectively.
Using the generalized velocities $\GEV{V}{}=(\Vi{}(t),\Ve{}(t))$ 
and the fluid particle displacement fields 
$\GEV{\xi}{}=(\bm{\xi}_{i}(t),\bm{\xi}_{e}(t))$,
the path $\gamma$ can be approximated locally by 
\renewcommand{\Exp}[1]{{\rm{e}}^{#1}}%
\renewcommand{\vExp}[1]{{\vec{\rm{e}}}^{\,#1}}%
\begin{eqnarray}
%\fl
  \gamma(t+\tau;0)
  \approx
  \vExp{\tau\GEV{V}{(t)}}
  \circ
  \gamma(t;0)
,
\hspace{1em}
  \gamma(t;\epsilon)
  \approx
  \vExp{\epsilon\GEV{\xi}{(t)}}
  \circ
  \gamma(t;0)
,
\nonumber
\end{eqnarray}
where
$\vExp{\tau\GEV{V}{(t)}}=\big(\Exp{\tau\Vi{}(t)},\Exp{\tau\Ve{}(t)}\big)$, 
$
  \vExp{\epsilon\GEV{\xi}{(t)}}=
  \big(\Exp{\epsilon\bm{\xi}_{i}(t)},\Exp{\epsilon\bm{\xi}_{e}(t)}\big)
$.
Let 
${\GEV{V}{_{\epsilon}}}=\GEV{V}{}+\epsilon\,\GEV{v}{}$ 
be the tangent vector of $\gamma(t;\epsilon)$.
The perturbation, $\GEV{v}{}$, 
is related to the variation of path 
$\GEV{\xi}{}=(\bm{\xi}_{i},\bm{\xi}_{e})$
by \textit{Lin constraints} \cite{marsden2013introduction}
\begin{eqnarray}
\renewcommand{\dot}{\partial_t}%
  \GEV{v}{}
  =
  \dot{\GEV{\xi}{}}+\LieBracket{\big}{\GEV{\xi}{}}{\GEV{V}{}}
  =
 \renewcommand{\LieBracket}[3]{\vec{L}_{#3}{#2}}%
  (\dot{}+\LieBracket{\big}{}{\GEV{V}{}}){\GEV{\xi}{}}
,
\label{lin constraints}
\end{eqnarray}
which corresponds to the $O(\epsilon\tau)$ 
terms of the asymptotic relation
$$
  \vExp{\tau\GEV{V}{_{\epsilon}(t)}} \approx
  \vExp{\epsilon\GEV{\xi}{(t+\tau)}} \circ
  \vExp{\tau\GEV{V}{(t)}} \circ
  \vExp{-\epsilon\GEV{\xi}{(t)}}
,
$$
(e.g., \cite{araki2015helicity}; app. A).
The value of the action on the path $\{\gamma(t;\epsilon)\}$ is
$
 S_{\epsilon}=
 \frac{1}{2}\int_{0}^{1}
  \Braket{\big}{\GEV{V}{_{\epsilon}}}{\GEV{V}{_{\epsilon}}}
 dt 
.
$
\renewcommand{\LieBracket}[3]{\vec{L}_{#3}{#2}}
Its first variation becomes
\renewcommand{\ADJ}[3]{\vec{L}{}^{#1}_{#2}{#3}}
\begin{eqnarray}
  \left.\dd{\epsilon}{S_{\epsilon}}\right|_{\epsilon=0}
  & = & 
  \int_{0}^{1}{\rm{d}}t
  \Braket{\big}{\GEV{V}{}
  }{
    \dot{\GEV{\xi}{}}+\LieBraket{\big}{\GEV{\xi}{}}{\GEV{V}{}}
  }
\nonumber\\
  & = & 
  \Braket{\big}{{\GEV{V}{}}}{\GEV{\xi}{}}\Big|^{t=1}_{t=0}
  %-
  %\Braket{\big}{{\GEV{V}{}}}{\GEV{\xi}{}}_{t=0}
  -
  \int_{0}^{1}{\rm{d}}t
  \Braket{\big}{\dot{\GEV{V}{}}}{\GEV{\xi}{}}
  +
  \int_{0}^{1}{\rm{d}}t
  \Braket{\big}{\ADJ{\dag}{\GEV{V}{}}{\GEV{V}{}}}{\GEV{\xi}{}}
,
\end{eqnarray}
where the adjoint operator $\ADJ{\dag}{}{}$ is defined by%
\footnote{%
Note that the adjoint operator $\ADJ{\dag}{}{}$ is denoted by $B(*,*)$ in Arnold's work
\cite{arnold1966sur,arnold1998topological,arnold1989mathematical}.
}%
\begin{eqnarray}
\fl
  \Braket{\Big}{\ADJ{\dag}{\GEV{V}{_{1}}}{\GEV{V}{_{2}}}}{\GEV{V}{_{3}}}
  & := &
 \renewcommand{\LieBracket}[3]{\vec{L}_{#3}{#2}}
  \Braket{\Big}{\GEV{V}{_{2}}}{\LieBraket{\Big}{\GEV{V}{_{3}}}{\GEV{V}{_{1}}}}
  =
 \renewcommand{\LieBracket}[3]{{#1[}#2,#3{#1]}}
  \Braket{\Big}{\GEV{V}{_{2}}}{\LieBraket{\big}{\GEV{V}{_{3}}}{\GEV{V}{_{1}}}}
\\ \fl
  & = &
  \int {\rm{d}}^3\vec{x} \bigg\{
    {\bm{u}_{3}} \cdot (
      {\bm{u}_{1}}\times{\bm{\omega}_{2}}
      +
      {\bm{j}_{1}}\times{\bm{b}_{2}}
    )
    +
    {\bm{b}_{3}} \cdot \curl \big[
      ({\bm{u}_{1}}-\alpha{\bm{j}_{1}})\times{\bm{b}_{2}}
    \big]
  \bigg\}
.
\label{adj in uwjb}
\end{eqnarray}
Equation (\ref{adj in uwjb}) yields the following
explicit expression of the operator $\ADJ{\dag}{}{}$:
\begin{eqnarray}
\fl
  \big(
    \ADJ{\dag}{\GEV{V}{_{1}}}{\GEV{V}{_{2}}}
  \big)_{i}
  =
  \DivFree{\bigg}{
    \Vi{1} \times ( \nabla \times \Vi{2} )
    +
    \HALL^{-2}
    (\Vi{1}-\Ve{1})
    \times
    (\curl)^{-1}(\Vi{2}-\Ve{2})
  }
,
\label{adji}
\\
\fl
  \big(
    \ADJ{\dag}{\GEV{V}{_{1}}}{\GEV{V}{_{2}}}
  \big)_{i}
  -
  \big(
    \ADJ{\dag}{\GEV{V}{_{1}}}{\GEV{V}{_{2}}}
  \big)_{e}
  =
  (\nabla\times)^2 \Big(
    \Ve{1} \times
    (\curl)^{-1}(\Vi{2}-\Ve{2})
  \Big)
.
\label{adjiadje}
\end{eqnarray}
Hereafter $\DivFree{}{*}$ denotes the solenoidal component of vector field obtained by Hodge decomposition (e.g., \cite{goldberg1962curvature}; \S 2.10).
Thus, Hamilton's principle yields the following Euler--Lagrange equation:
\begin{eqnarray}
  \dot{\GEV{V}{}}=\ADJ{\dag}{\GEV{V}{}}{\GEV{V}{}}
,
\label{euler-lagrange eq in ad-operator}
\end{eqnarray}
which is the well-known \textit{Euler--Poincar\'e equation}
(\cite{marsden2013introduction}; ch. 13).
Application of Equation (\ref{adj in uwjb}) to this equation derives Eq. (\ref{eq of mot u and b}).

\subsection{Linear connection with physically desirable properties
\label{Linear connection with physically desirable properties}}

As was discussed in the section \ref{Introduction}, the Euler--Lagrange equation derived in the previous section does not determine the parallel translation of a vector uniquely.
Thus, we present here the derivation of a linear connection on $\GEV{V}{}$-variable space 
$\LCC{}{}$
that satisfies physically desirable features.

To begin with,
we postulate that the connection is defined by a combination of the Riemannian metric and Lie bracket:
\renewcommand{\LieBracket}[3]{{#1[}#2,#3{#1]}}%
\begin{eqnarray}
&&\hspace{-3em}
 \Braket{\big}{\GEV{V}{_{3}}}{\LCC{\GEV{V}{_{1}}}{\GEV{V}{_{2}}}}
 :=
   C_{1} \Braket{\big}{\GEV{V}{_{3}}}{ \LieBraket{\big}{\GEV{V}{_{1}}}{\GEV{V}{_{2}}} }
 + C_{2} \Braket{\big}{\GEV{V}{_{1}}}{ \LieBraket{\big}{\GEV{V}{_{2}}}{\GEV{V}{_{3}}} }
 + C_{3} \Braket{\big}{\GEV{V}{_{2}}}{ \LieBraket{\big}{\GEV{V}{_{3}}}{\GEV{V}{_{1}}} }
,
\label{postulate linear connection}
\end{eqnarray}
where $C_{i}$ are constants.
This postulate leads to the expression of ${\LCC{}{}}$ 
in terms of the Lie derivative $L$ and its adjoint $\ADJ{\dag}{}{}$
as
\renewcommand{\LieBracket}[3]{\vec{L}_{#3}{#2}}%
\begin{eqnarray}
 {\LCC{\GEV{V}{_{1}}}{\GEV{V}{_{2}}}}
 =
   C_{1}\, { \LieBraket{\big}{\GEV{V}{_{1}}}{\GEV{V}{_{2}}} }
 - C_{2}\, { \ADJ{\dag}{\GEV{V}{_{2}}}{\GEV{V}{_{1}}} }
 + C_{3}\, { \ADJ{\dag}{\GEV{V}{_{1}}}{\GEV{V}{_{2}}} }
.
\end{eqnarray}
$C_{i}$ are determined 
by the following three physical conditions.
\begin{enumerate}
\item\label{energy conserving}
\textit{The connection $\LCC{}{}$ is metric-preserving:}
\begin{eqnarray}
\label{metric preserving}
  \LCC{\GEV{V}{_{3}}}{ \Braket{\big}{\GEV{V}{_{1}}}{\GEV{V}{_{2}}} }
  =
  \Braket{\big}{\LCC{\GEV{V}{_{3}}}{\GEV{V}{_{1}}}}{\GEV{V}{_{2}}}
  +
  \Braket{\big}{\GEV{V}{_{1}}}{\LCC{\GEV{V}{_{3}}}{\GEV{V}{_{2}}}}
  =
  0
.
\end{eqnarray}
Mathematically, 
this relation simply reflects the right-invariance of the Riemannian metric.
In a physical context, however,
this condition guarantees 
not only energy conservation as a whole,
but also the \textit{detailed energy balance} between the interacting modes.
This condition leads to the relation $C_{1} = C_{3}$.
\item\label{EL is geodesic}
\textit{%
The Euler--Lagrange Equation (\ref{euler-lagrange eq in ad-operator})
agrees with
the geodesic equation under the connection $\LCC{}{}$:}
$
  \LCC{\GEV{V}{}}{\GEV{V}{}}
  =
  -\ADJ{\dag}{\GEV{V}{}}{\GEV{V}{}}
.
$
The relation $C_{2} - C_{3}=1$ is derived here.
\item\label{galilean invariant}
\textit{The substantial derivative
$
 \partial_t+\widetilde{\nabla}_{\GEV{V}{}}
$
is covariant against the Galilean boost
for arbitrary frozen-in vector fields}.
The calculation is given in 
\ref{Galilean invariance of substantial derivative},
and results in $C_{1}+C_{3}=-1$.
\end{enumerate}
These three conditions uniquely determine
the coefficients to be $C_{1}=-C_{2}=C_{3}=-1/2$.
The connection that satisfies these conditions is given by
\begin{eqnarray}
\label{levi-civita connection in general}
  \LCC{\GEV{V}{_{1}}}{\GEV{V}{_{2}}}
  =
  -\frac{1}{2}
  \LieBraket{\big}{\GEV{V}{_{1}}}{\GEV{V}{_{2}}}
  -\frac{1}{2}
  \ADJ{\dag}{\GEV{V}{_{1}}}{\GEV{V}{_{2}}}
  -\frac{1}{2}
  \ADJ{\dag}{\GEV{V}{_{2}}}{\GEV{V}{_{1}}}
.
\end{eqnarray}
Substituting Equations (\ref{<a|[b,c]> in uwjb}) and (\ref{adj in uwjb}),
we can obtain the combination of the Riemannian metric and 
the following physically desirable connection:
\begin{eqnarray}
\fl
  \Braket{\big}{\GEV{V}{_{3}}}{\LCC{\GEV{V}{_{1}}}{\GEV{V}{_{2}}}}
  =\frac12\int {\rm{d}}^3\vec{x} \bigg\{
    {\bm{v}_{3}}\cdot\Big[
      {\bm{\omega}_{1}}\times{\bm{v}_{2}}
      -{\bm{v}_{1}}\times{\bm{\omega}_{2}}
      -\nabla\times({\bm{v}_{1}}\times{\bm{v}_{2}})
      +{\bm{b}_{1}}\times{\bm{j}_{2}}
      -{\bm{j}_{1}}\times{\bm{b}_{2}}
    \Big]
   \myLineBreak{0.5}
    +
    {\bm{b}_{3}}
      \cdot\Big[
      \nabla\times[
        {\bm{b}_{1}}\times({\bm{v}_{2}}-\HALL{\bm{j}_{2}})
        -({\bm{v}_{1}}-\HALL{\bm{j}_{1}})\times{\bm{b}_{2}}
      ]
      -
      {\bm{v}_{1}}\times{\bm{j}_{2}}-{\bm{j}_{1}}\times{\bm{v}_{2}}
      +\HALL{\bm{j}_{1}}\times{\bm{j}_{2}}
    \Big]
  \bigg\}
.
%\nonumber
\end{eqnarray}
Note that this connection satisfies the torsion-free condition
\renewcommand{\LieBracket}[3]{{#1[}#2,#3{#1]}}%
\begin{eqnarray}
  \LieBraket{\big}{\GEV{V}{_{1}}}{\GEV{V}{_{2}}}
  =
  \LCC{\GEV{V}{_{2}}}{\GEV{V}{_{1}}}
  -
  \LCC{\GEV{V}{_{1}}}{\GEV{V}{_{2}}}
,
\label{torsion free}
\end{eqnarray}
i.e., 
it is the Levi--Civita connection on $G$ 
associated with the Riemannian metric
(\ref{def riemannian metric}).
Conversely,
the torsion-free condition itself requires $2C_{1}=-1$ and $C_{2}+C_{3}=0$
and it can surrogate the physical conditions 
(\ref{EL is geodesic}) and (\ref{galilean invariant}).

%\newpage
%%%%%%%%%%%%%%%%%%%%%%%%%%%%%%%%%%%%%%%%%%%%%%%%%%%%%%%%%%%%%%%%%%%%%%%%
\section{Linear stability analysis in the geodesic formulation
\label{linear stability analysis}}
%%%%%%%%%%%%%%%%%%%%%%%%%%%%%%%%%%%%%%%%%%%%%%%%%%%%%%%%%%%%%%%%%%%%%%%%

%-----------------------------------------------------------------------
\subsection{Formulation of the linear stability analysis and a remark on its applicability
\label{Jacobi equation formulation}}
%-----------------------------------------------------------------------

In this section, we discuss the linear stability problem in the geodesic-formulation framework.
As is well known, the second variation of the curve length yields the Jacobi equation.

Let
$\GEV{V}{}$
be a ``reference'' solution that satisfies the equation of motion
(\ref{euler-lagrange eq in ad-operator}):
\begin{eqnarray}
  \sDD{t}{}{\GEV{V}{}}=0
,
\label{eq of mot}
\end{eqnarray}
where 
$
 \sDD{t}{} := \dd{t}{} + \LCC{\GEV{V}{}}{}
$
is the ``substantial derivative''
with respect to the Levi--Civita connection 
(\ref{levi-civita connection in general}).
The linear stability of the reference solution $\GEV{V}{}$
is determined by the evolution equation
\begin{eqnarray}
  \sDD{t}{}{\GEV{v}{}}+
  \LCC{\mbox{$\GEV{v}{}$}}{\GEV{V}{}}
  =0,
\label{perturbation equation}
\end{eqnarray}
where
$\GEV{v}{}$ is a small perturbation of $\GEV{V}{}$.
Remember that
the Jacobi field $\GEV{\xi}{}$
and the perturbation $\GEV{v}{}$
are closely related by Lin constraints (\ref{lin constraints}):
\begin{eqnarray}
&&
 {\GEV{v}{}}
 =
 \sDD{t}{}{\GEV{\xi}{}}
 -
 \LCC{\GEV{\xi}{}}{\GEV{V}{}}
,
\label{lin constraint}
\end{eqnarray}
where the torsion-free condition (\ref{torsion free}) 
has been used% to derive the second equation
.
Substituting the Lin constraints (\ref{lin constraint}) 
into the perturbation Equation (\ref{perturbation equation})
and
using Equations (\ref{eq of mot}) and (\ref{lin constraint}),
we can 
eliminate ${\GEV{v}{}}$
and
obtain the following evolution equation for $\GEV{\xi}{}$:
\begin{eqnarray}
  \left(\sDD{t}{}\right)^2{\GEV{\xi}{}}
  +
  R(\GEV{\xi}{},\GEV{V}{})\,\GEV{V}{}
  = 0
\label{jacobi equation}
\end{eqnarray}
where $R$ is the Riemannian curvature tensor given by
\begin{eqnarray}
  R(\GEV{\xi}{},\GEV{V}{})
  =
  \LCC{
    \GEV{\xi}{}
  }{
    \LCC{ \GEV{V}{} }{}% \GEV{V}{} }
  }
  -
  \LCC{
    \GEV{V}{}
  }{
    \LCC{ \GEV{\xi}{} }{}% \GEV{V}{} }
  }
  -
  \LCC{
    \LCC{ \GEV{\xi}{} }{ \GEV{V}{} }
  }{
    %\GEV{V}{}
  }
  +
  \LCC{
    \LCC{ \GEV{V}{} }{ \GEV{\xi}{} }
  }{
    %\GEV{V}{}
  }
\end{eqnarray}
\cite{KobayashiNomizu1963}.
This is the Jacobi equation associated with the reference solution
${\GEV{V}{}}$.
It is convenient to rewrite this equation 
as a first order simultaneous differential equations as follows:
\begin{eqnarray}
  \sDD{t}{}
  \left(\begin{array}{c}
    {\GEV{\xi}{}} \\ {\GEV{\eta}{}}
  \end{array}\right)
  =
  \left(\begin{array}{cc}
    O & I \\ -R^{\prime}({\GEV{V}{}},{\GEV{V}{}}) & O
  \end{array}\right)
  \left(\begin{array}{c}
    {\GEV{\xi}{}} \\ {\GEV{\eta}{}}
  \end{array}\right)
,
\label{jacobi equation matrix}
\end{eqnarray}
where 
$
 \GEV{\eta}{}:=\sDD{t}{}\GEV{\xi}{}
$
and the multilinear operator $R^{\prime}$ is defined by
$
  R^{\prime}({\GEV{V}{}},{\GEV{V}{}}){\GEV{\xi}{}}
  :=
  R({\GEV{\xi}{}},{\GEV{V}{}}){\GEV{V}{}}
.
$

Note that, since the PDE (\ref{jacobi equation matrix}) requires a pair of $\GEV{V}{}$-variables, ${\GEV{\xi}{}}$ and ${\GEV{\eta}{}}$, as an initial condition, by setting them appropriately, one can treat somewhat wider classes of stability problems than those commented in Arnold's textbook \cite{arnold1989mathematical} (see \ref{Remark on applicable stability problems}).

\subsubsection*{Physical implication of the Riemannian curvature:}
\begin{figure}
\begin{center}
\includegraphics[width=0.49\textwidth]{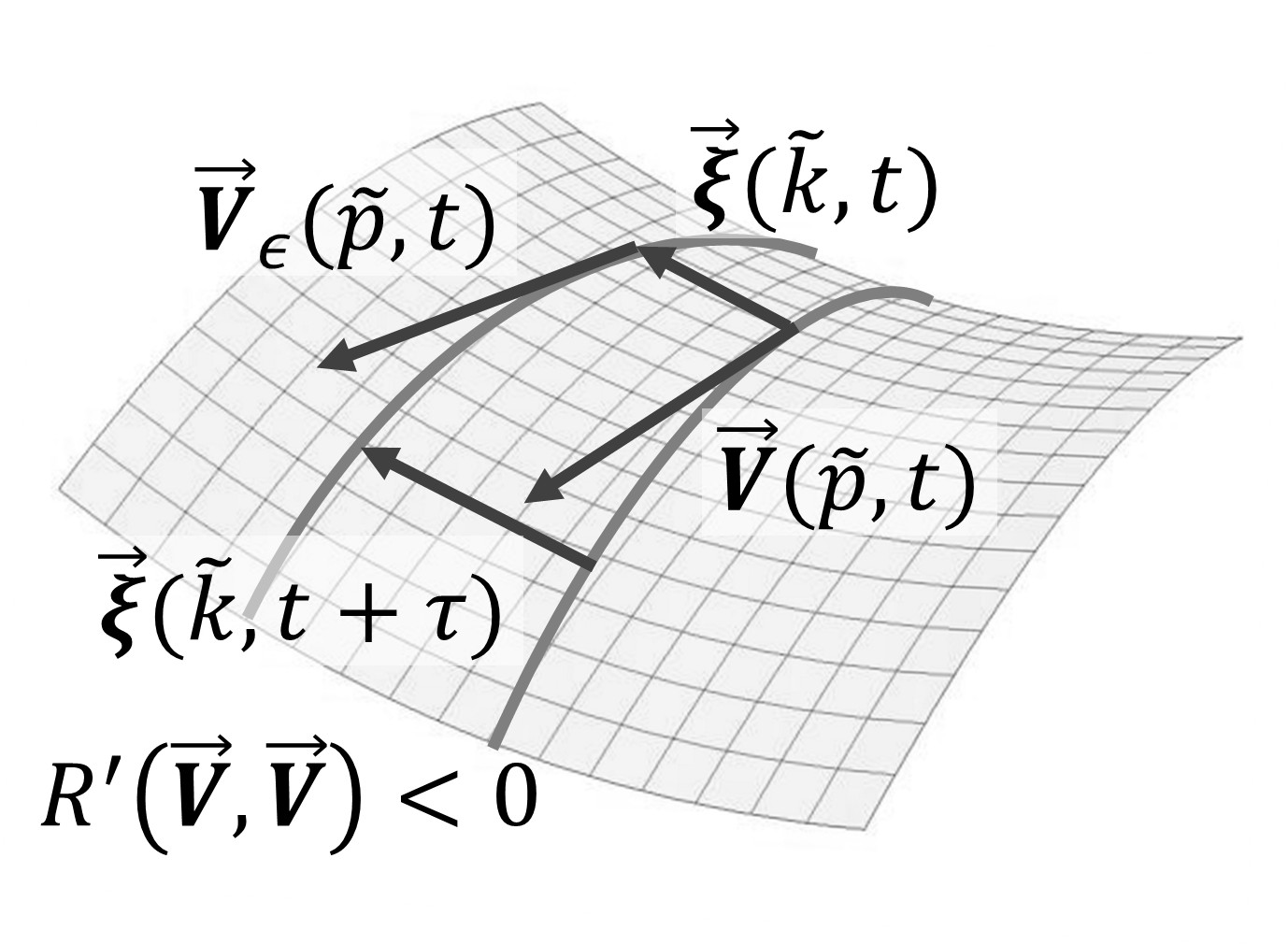}
\includegraphics[width=0.49\textwidth]{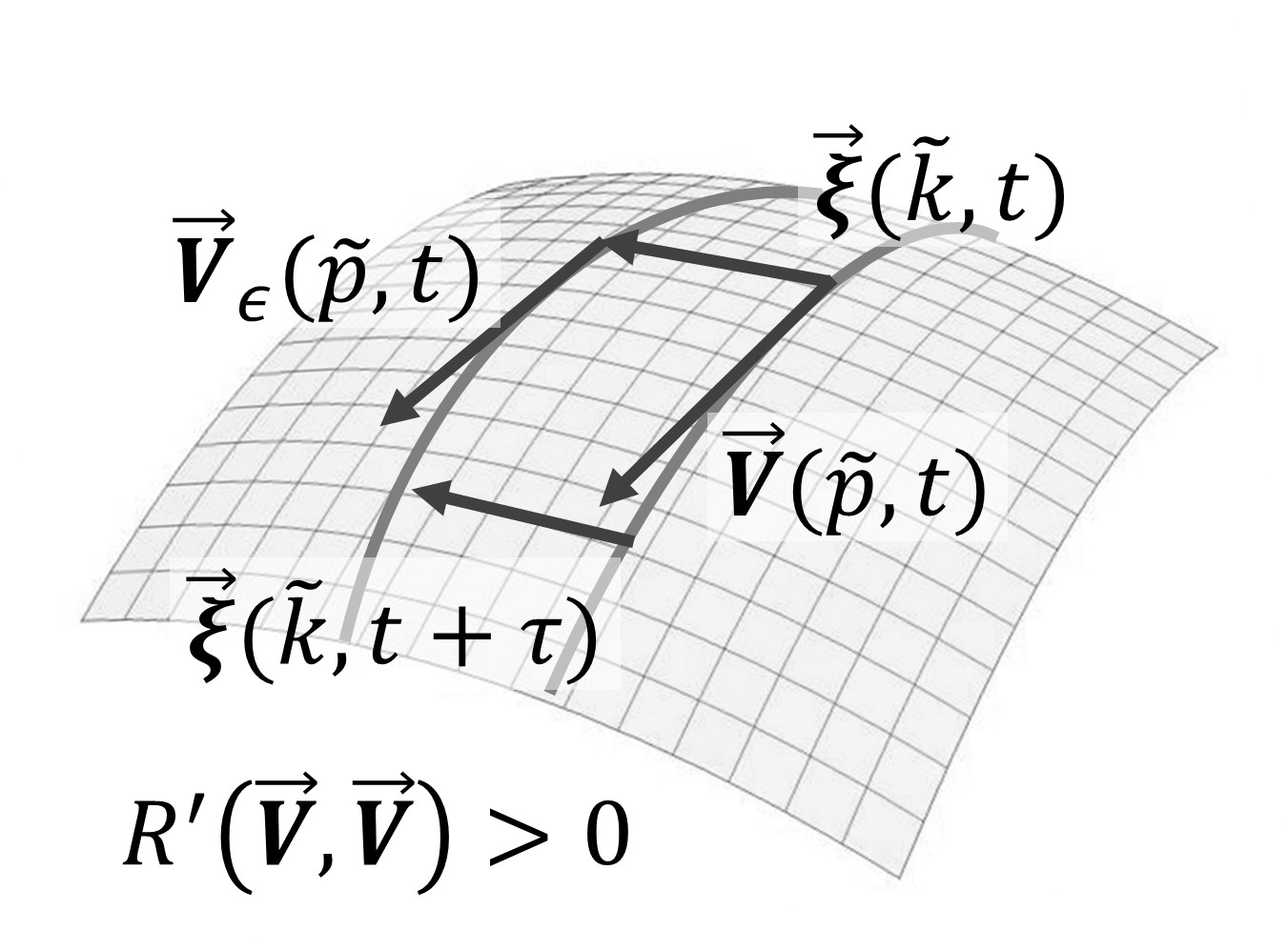}
\end{center}
\caption{\label{how to interpret}
Illustration of the implication of the Riemannian curvature analysis.
The pictures represent two dimensional surfaces 
locally spanned by the family of the solution paths
$
  \gamma(t;\epsilon)
,
$
where $t$ and $\epsilon$ are the time and perturbation parameters.
Differentiations with respect to $t$ and $\epsilon$ give the generalized velocity
$
  \GEV{V}{_{\epsilon}(t)}=\dot{\gamma}(t;\epsilon)
$
and the perturbation field
$
  \GEV{\xi}{(t)}=\partial_{\epsilon}{\gamma}(t;\epsilon)
,
$
respectively.
When the curvature is negative (left), 
there is a perturbation field that grows exponentially,
and thus,
the solution paths separate each other gradually as time goes by.
When the curvature is positive (right), 
the norm of perturbation field does not grow,
and thus,
the solution paths remain close each other.
}%
\end{figure}

Temporal behaviors of $\GEV{\xi}{}$ and $\GEV{\eta}{}$
are determined by the eigenvalues (say $\Lambda$) 
of the eigenequation of the matrix operator in Equation (\ref{jacobi equation matrix}),
$$
  \Lambda^2 I + R^{\prime}({\GEV{V}{}},{\GEV{V}{}}) =0
.
$$
If the curvature term is \textit{negative}, 
the eigenvalues are real and the norm of the solution grows exponentially.
However, when the curvature term is \textit{positive}, 
the eigenvalues are purely imaginary, i.e., the solution is expected to be oscillatory 
(cf. \cite{arnold1989mathematical}; app. 1, \S I).
In Figure \ref{how to interpret}, we present
the schematic picture of the implications of sectional curvature.
%%%%%%%%%%%%%%%%%%%%%%%%%%%%%%%%%%%%%%%%%%%%%%%%%%%%%%%%%%%%%%%%%%%%%%%%
%%%%%%%%%%%%%%%%%%%%%%%%%%%%%%%%%%%%%%%%%%%%%%%%%%%%%%%%%%%%%%%%%%%%%%%%

Since dimension of the eigenvalue $\Lambda$ 
is reciprocal of time, square root of the moduli of the Riemannian curvature indicate characteristic time scale of the disturbance fields.
Thus, the large moduli of the curvature implies the rapid temporal variations irrespective of oscillatory or growing.

Note that, since the Riemannian curvature is defined by the covariant derivatives, the result describes the behavior of solution paths for a very short time interval in general.
Thus, one should be very careful to interpret the results, though applicability to predictability problems has been claimed.
Despite this shortcoming, the Riemannian curvature analysis has a powerful advantage that it can be calculated for arbitrary snapshot pair of (possibly non-stationary) reference solution and perturbation without solving the evolution equation or eigenvalue problem.

%-----------------------------------------------------------------------
%\newpage
\subsection{Normal mode representation of the Jacobi equation and the Riemannian curvature tensor}
%-----------------------------------------------------------------------

We describe here the time evolution of the norm of solutions to 
Equation (\ref{jacobi equation matrix}) 
that is defined by
$
  ||{\GEV{\xi}{}}||
  :=
  \Braket{\big}{\GEV{\xi}{}}{\GEV{\xi}{}}
  +
  \Braket{\big}{\GEV{\eta}{}}{\GEV{\eta}{}}
.
$
Taking the inner product of Equation (\ref{jacobi equation matrix}) 
with $({\GEV{\xi}{}}\ {\GEV{\eta}{}})$
and 
using
the metric preserving property of the Levi--Civita connection
(\ref{metric preserving}),
$
  \Braket{\big}{\GEV{\xi}{}}{\LCC{\GEV{V}{}}{\GEV{\xi}{}}}
  =
  \Braket{\big}{\GEV{\eta}{}}{\LCC{\GEV{V}{}}{\GEV{\eta}{}}}
  =
  0
,
$
we can obtain the following time evolution of the norm of the Jacobi field:
\begin{equation}
 \frac12\sdd{t}{}\left(
  \Braket{\big}{\GEV{\xi}{}}{\GEV{\xi}{}}
  +
  \Braket{\big}{\GEV{\eta}{}}{\GEV{\eta}{}}
 \right)
 =
 \Braket{\big}{\GEV{\xi}{}}{\GEV{\eta}{}}
 -
 \Braket{\big}{\GEV{\eta}{}}
  {R({\GEV{\xi}{}},{\GEV{V}{}}){\GEV{V}{}}}
.
\label{norm equation}
\end{equation}
Substituting the normal mode expansions of $\GEV{V}{}$-variables, e.g.
$
  {\GEV{V}{}}
  =\sum_{\GEM{k}{}}
  {\GEC{V}{(\GEM{k}{};t)}} {\GEV{\Phi}{(\GEM{k}{};\vec{x})}}
,
$
where
$
  {\GEV{\Phi}{}}
$
is the GEV given by (\ref{eigenfunction of L}),
we obtain the GEV representation of the equation as
\renewcommand{\GEM}[2]{\widetilde{#1}}%
\renewcommand{\g}[1]{g(\GEM{#1}{})}%
\begin{eqnarray}
\fl
  \frac{1}{2}
  \dd{t}{}
  \left(
    |\GEC{\xi}{(\GEM{k}{};t)}|^2
    +
    |\GEC{\eta}{(\GEM{k}{};t)}|^2
  \right)
  =
  \CC{\GEC{\xi}{(\GEM{k}{};t)}}\,
  {\GEC{\eta}{(\GEM{k}{};t)}}
 \myLineBreak{2}
  -
  \g{k}^{-1}
  \mathop{
    \sum_{\GEM{p}{}}
    \sum_{\GEM{q}{}}
    \sum_{\GEM{r}{}}
  }^{\vec{k}=\vec{p}+\vec{q}+\vec{r}}
  \Braket{\Big}{
    \CC{\GEV{\Phi}{(\GEM{k}{})}}
  }{
    R({\GEV{\Phi}{(\GEM{p}{})}},{\GEV{\Phi}{(\GEM{q}{})}})
    {\GEV{\Phi}{(\GEM{r}{})}}
  }
  {\GEC{V}{(\GEM{q}{};t)}}\,
  {\GEC{V}{(\GEM{r}{};t)}}\,
  {\GEC{\xi}{(\GEM{p}{};t)}}\,
  \CC{\GEC{\eta}{(\GEM{k}{};t)}}
\label{jacobi equation for a single GEV mode}
\end{eqnarray}
for each mode.
Hereafter, 
the tilde notation
$
 \GEM{k}{}
$
stands for the set of wavenumber, helicity, and {\polarity} index,
$
 \renewcommand{\GEM}[2]{\vec{#1},\HEL_{#1},{#2}\POL_{#1}}%
 (\GEM{k}{})
,
$
overbar denotes complex conjugate,
and 
$
  \g{k}:=
  \Braket{\big}{\CC{\GEV{\Phi}{(\GEM{k}{})}}}{{\GEV{\Phi}{(\GEM{k}{})}}}
$
is the value of the Riemannian metric for mode $\GEM{k}{}$.
%(see (\ref{Riemannian metric in Z-coefficient})):
\renewcommand{\EIGEN}[2]{\EIG^{{#2}}(\GEM{#1}{})}%
The component of the Riemannian curvature tensor for the GEV modes is given by
\begin{eqnarray}
\fl
  \g{k}^{-1}
  \Braket{\Big}{
    \CC{\GEV{\Phi}{(\GEM{k}{})}}
  }{
    R({\GEV{\Phi}{(\GEM{p}{})}},{\GEV{\Phi}{(\GEM{q}{})}})
    {\GEV{\Phi}{(\GEM{r}{})}}
  }
\myLineBreak{1}
  =
  \g{k}^{-1}
  \Braket{\bigg}{
    \GEV{\Phi}{(-\GEM{k}{})}
  }{
    \LCC{
      \GEV{\Phi}{(\GEM{p}{})}
    }{
      \LCC{ \GEV{\Phi}{(\GEM{q}{})} }{ \GEV{\Phi}{(\GEM{r}{})} }
    }
    -
    \LCC{
      \GEV{\Phi}{(\GEM{q}{})}
    }{
      \LCC{ \GEV{\Phi}{(\GEM{p}{})} }{ \GEV{\Phi}{(\GEM{r}{})} }
    }
    +
    \LCC{
      \LieBraket{\big}{
        \GEV{\Phi}{(\GEM{p}{})}
      }{
        \GEV{\Phi}{(\GEM{q}{})}
      }
    }{
      \GEV{\Phi}{(\GEM{r}{})}
    }
  }
\myLineBreak{1}
  =
  \frac{1}{4}
  \sum_{\HEL_{l},\POL_{l}}^{\vec{l}=\vec{k}-\vec{p}}
  \frac{
    \tripleHMHD{\big}{-\GEM{k}{}}{\GEM{p}{}}{\GEM{l}{}}
  }{
    \g{k}
  }
  \frac{
    \tripleHMHD{\big}{-\GEM{l}{}}{\GEM{q}{}}{\GEM{r}{}}
  }{
    \g{l}
  }
  (\EIGEN{p}{} - \EIGEN{l}{} - \EIGEN{k}{})
  (\EIGEN{q}{} - \EIGEN{r}{} - \EIGEN{l}{})
 \myLineBreak{2}
  -
  \frac{1}{4}
  \sum_{\HEL_{m},\POL_{m}}^{\vec{m}=\vec{k}-\vec{q}}
  \frac{
    \tripleHMHD{\big}{-\GEM{k}{}}{\GEM{q}{}}{\GEM{m}{}}
  }{
    \g{k}
  }
  \frac{
    \tripleHMHD{\big}{-\GEM{m}{}}{\GEM{p}{}}{\GEM{r}{}}
  }{
    \g{m}
  }
  (\EIGEN{q}{} - \EIGEN{m}{} - \EIGEN{k}{})
  (\EIGEN{p}{} - \EIGEN{r}{} - \EIGEN{m}{})
 \myLineBreak{2}
  +
  \frac{1}{2}
  \sum_{\HEL_{n},\POL_{n}}^{\vec{n}=\vec{k}-\vec{r}}
  \frac{
    \tripleHMHD{\big}{-\GEM{k}{}}{\GEM{n}{}}{\GEM{r}{}}
  }{
    \g{k}
  }
  \frac{
    \tripleHMHD{\big}{-\GEM{n}{}}{\GEM{p}{}}{\GEM{q}{}}
  }{
    \g{n}
  }
  (\EIGEN{n}{} - \EIGEN{r}{} - \EIGEN{k}{})
  \EIGEN{n}{}
,
\label{normal-mode expansion of the Riemannian curvature}
\end{eqnarray}
where 
$\EIG$
is the eigenvalue of the helicity-based, particle-relabeling operator $\widehat{W}$
(see (\ref{particle relabeling operator}) and (\ref{eigenvalue of W})),
the tilde notation with a minus sign ($-\GEM{k}{}$) represents
$
 \renewcommand{\GEM}[2]{\vec{#1},\HEL_{#1},{#2}\POL_{#1}}%
 (-\GEM{k}{})
.
$
Note that the relation
$
  \EIG(-\GEM{k}{})=\EIGEN{k}{}
$
holds.
The explicit expression of the three-mode parenthesis symbol 
$
  \tripleHMHD{\big}{*}{*}{*}
$
is given by (\ref{fac str cont HMHD}).
The details of the expressions, formulae, and equations related to the GEV representation are summarized in \ref{Generalized Elsasser variable representation of differential geometrical quantities}.
As can be seen from Equation (\ref{jacobi equation for a single GEV mode}), the Riemannian curvature term contains four-wave resonances, while the other terms are decoupled for each GEV mode.

%-----------------------------------------------------------------------
%\newpage
\subsection{Statistical homogeneity and isotropy assumption and normal mode 
representation of sectional curvature\label{statistical homogeneity}}
%-----------------------------------------------------------------------

Since our principal interest is 
the statistical features of turbulent solutions of HMHD systems,
we address the ensemble average of reference solutions $\GEV{V}{(t)}$.
\renewcommand{\COR}[2]{Q(\ABS{#1},\HEL_{#1},\POL_{#1}#2)}
Assuming that the reference solutions are random vector fields that are statistically homogeneous and isotropic and that main flow and perturbation are independent each other, we introduce here the correlation function 
$\COR{k}{;t}$
that satisfies
\begin{eqnarray}
\renewcommand{\ABS}[1]{|\vec{#1}|}
  \EnsAve{\Big}{\GEC{V}{(\GEM{k}{};t)}\GEC{V}{(\GEM{p}{};t)}}
  =
  \COR{k}{;t}\,
  \delta_{\HEL_{k}\HEL_{p}}\, \delta_{\POL_{k}\POL_{p}}\,
  \delta^3_{-\vec{k},\vec{p}}
,
\end{eqnarray}
where the angled brackets without vertical bars $\big<*\big>$
denote the ensemble average
and
$\delta$ and $\delta^3$ denote the Kronecker delta and its triple product, 
respectively.
This assumption imposes the wavenumber relation 
$\vec{r}=-\vec{q}$, and $\vec{p}=\vec{k}$
and
simplifies the curvature term considerably.
%
%
%\newpage
Thus,
the Jacobi Equation (\ref{jacobi equation for a single GEV mode}) 
is reduced to
\renewcommand{\MSC}[2]{R_{#1}^{#2}}%{R_{H}}% GEV-mode sectional curvature, _H means HMHD
\begin{eqnarray}
\fl
  \frac{1}{2}
  \dd{t}{}
  \left(
    |\GEC{\xi}{(\GEM{k}{};t)}|^2
    +
    |\GEC{\eta}{(\GEM{k}{};t)}|^2
  \right)
  =
  \CC{\GEC{\xi}{(\GEM{k}{};t)}}\,
    {\GEC{\eta}{(\GEM{k}{};t)}}
 \nonumber\\\fl\hspace{2em}
  -
  \sum_{\vec{p},\HEL_{p},\POL_{p}}
    \MSC{H}{}(\ABS{k},\HEL_{k},\POL_{k},\ABS{p},\HEL_{p},\POL_{p},q,\HALL)\,
    \COR{p}{;t}\,
    {\GEC{\xi}{(\GEM{k}{};t)}}\,
    \CC{\GEC{\eta}{(\GEM{k}{};t)}}
\label{jacobi equation matrix norm in GEC}
\end{eqnarray}
for each GEV mode $\GEM{k}{}$, where ${q}:=|\vec{k}-\vec{p}|$, and $\MSC{H}{}$
is defined by
\begin{eqnarray}
\fl
  \MSC{H}{}(\ABS{k},\HEL_{k},\POL_{k},\ABS{p},\HEL_{p},\POL_{p},{q},\HALL)
  :=
  \g{k}^{-1}
  \Braket{\Big}{\CC{\GEV{\Phi}{(\GEM{k}{})}}}
    { R({\GEV{\Phi}{(\GEM{k}{})}},{\GEV{\Phi}{(\GEM{p}{})}})
      {\GEV{\Phi}{(-\GEM{p}{})}} }
,
\nonumber\\
\fl\hspace{3em}
  =
  \MSC{1}{}(\ABS{k},\HEL_{k},\POL_{k},\ABS{p},\HEL_{p},\POL_{p},q,\HALL)
  +
  \MSC{2}{}(\ABS{k},\HEL_{k},\POL_{k},\ABS{p},\HEL_{p},\POL_{p},f(q),\HALL)
,
\label{def R_H}
\\
\fl
  \MSC{1}{}(\ABS{k},\HEL_{k},\POL_{k},\ABS{p},\HEL_{p},\POL_{p},
        \ABS{m},\HALL)
  =
  \sum_{\HEL_{m},\POL_{m}}^{\vec{m}=\vec{k}-\vec{p}}
  \frac{
    \left|\tripleHMHD{\big}{-\GEM{k}{}}{\GEM{p}{}}{\GEM{m}{}}\right|^2
  }{
    4\, \g{k}\, \g{m}
  }
  \Big[
    \Big( \EIGEN{p}{} - \EIGEN{k}{} \Big)^2 - \EIGEN{m}{}^2
  \Big]
,
\label{def R_1}
\\
\fl
  \MSC{2}{}(\ABS{k},\HEL_{k},\POL_{k},\ABS{p},\HEL_{p},\POL_{p},
        \ABS{n},\HALL)
  =
  \sum_{\HEL_{n},\POL_{n}}^{\vec{n}=\vec{k}+\vec{p}}
  \frac{
    \left|\tripleHMHD{\big}{-\GEM{n}{}}{\GEM{k}{}}{\GEM{p}{}}\right|^2
  }{
    2\, \g{k}\, \g{n}
  }
  \EIGEN{n}{} \Big( \EIGEN{p}{} + \EIGEN{k}{} - \EIGEN{n}{} \Big)
,
\label{def R_2}
\end{eqnarray}
$f(q):=|\vec{k}+\vec{p}|=\sqrt{2k^2+2p^2-q^2}$.
The reduced curvature tensor $\MSC{H}{}$
is the sectional curvature between two GEV modes%
\footnote{
Mathematically, the sectional curvature is defined by 
$
  K=
  \Braket{\big}{\GEV{\xi}{}}
    { R\big({\GEV{\xi}{}},{\GEV{V}{}}\big){\GEV{V}{}} }
  \Big/\Big(
  \Braket{\big}{\GEV{\xi}{}}{\GEV{\xi}{}}
  \Braket{\big}{\GEV{V}{}}{\GEV{V}{}}
  -
  \Braket{\big}{\GEV{\xi}{}}{\GEV{V}{}}^2
  \Big)
$,
and does not agree with $\MSC{H}{}$ 
unless each of ${\GEV{\xi}{}}$ and ${\GEV{V}{}}$ is a single GEV-mode.
}.
We call $\MSC{H}{}$ the GEV-mode sectional curvature in the following.

Equation (\ref{jacobi equation matrix norm in GEC}) 
can be rewritten as simultaneous equations 
for each GEV mode as follows:
\begin{eqnarray}
\fl
  \sdd{t}{}
  \left(\begin{array}{c}
    {\GEC{\xi}{(\GEM{k}{};t)}} \\ {\GEC{\eta}{(\GEM{k}{};t)}}
  \end{array}\right)
  =
  \left(\begin{array}{cc}
    0 & 1 \\
    -\sum_{\GEM{p}{}} \MSC{H}{}(\GEM{k}{},\GEM{p}{},q,\HALL)\,
    \COR{p}{;t}\,
    & 0
  \end{array}\right)
  \left(\begin{array}{c}
    {\GEC{\xi}{(\GEM{k}{};t)}} \\ {\GEC{\eta}{(\GEM{k}{};t)}}
  \end{array}\right)
.
\label{jacobi equation matrix in gev}
\end{eqnarray}
Note that
the GEV-mode sectional curvatures
do not depend directly on their wavenumber vectors,
but rather on their moduli ($\ABS{k}$ and $\ABS{p}$) and the angle between them 
($q=\ABS{k}^2+\ABS{p}^2-2\ABS{k}\ABS{p}\cos\theta$).
The explicit expression of the the GEV-mode sectional curvature 
is given by Equation (\ref{GEV curvature general}) 
in \ref{GEV mode representation of the HMHD sectional curvature}.

Assuming isotropy and 
approximating the summation with respect to $\vec{p}$
by the three-dimensional integral, 
($%\displaystyle
  \sum_{\vec{p}}
  \approx
  \int\!\!\int\!\!\int{\rm{d}}^{3}\vec{p}
$),
we can obtain the following formula:
\renewcommand{\SAC}[2]{K_{#1}^{#2}}%{K_{H}}% shell-averaged curvature, _H means HMHD
\begin{eqnarray}
\fl
  \sum_{\vec{p},\HEL_{p},\POL_{p}}
    \MSC{H}{}(\ABS{k},\HEL_{k},\POL_{k},\ABS{p},\HEL_{p},\POL_{p},{q},\HALL)\,
    \COR{p}{;t}\,
 \myLineBreak{1}
  \approx
  \sum_{\HEL_{p},\POL_{p}}
  \int_{0}^{\infty} {\rm{d}}p\, 
    \SAC{H}{}(\ABS{k},\HEL_{k},\POL_{k},\ABS{p},\HEL_{p},\POL_{p},\HALL)\,
    \COR{p}{;t}
,
\label{integrated curvature}
\end{eqnarray}
where the integration kernel function $\SAC{H}{}$ is given by
\begin{eqnarray}
\fl
  \SAC{H}{}(\ABS{k},\HEL_{k},\POL_{k},\ABS{p},\HEL_{p},\POL_{p},\HALL)\,
  =
  \frac{2 \pi p}{k}
    \int_{|\vec{k}-\vec{p}|}^{k+p} {\rm{d}}q \bigg[
      q\,
      \MSC{H}{}(\ABS{k},\HEL_{k},\POL_{k},\ABS{p},\HEL_{p},\POL_{p},{q},\HALL)
    \bigg]
.
\label{curvature kernel}
\end{eqnarray}
Hereafter we call $\SAC{H}{}$ the shell-averaged curvature kernel.
The explicit expression of the shell-averaged curvature kernel is given by Equation (\ref{GEV curvature kernel general})
in \ref{GEV mode representation of the HMHD sectional curvature}.
Note that the GEV-mode sectional curvature and the shell-averaged curvature kernel are, by definition (\ref{def R_H}), 
symmetric with exchange of $k$ and $p$.

Since the Fourier transform of the correlation function, $\COR{p}{;t}$,
is positive definite, the sign of the curvature is determined solely by the sign of the shell-averaged curvature kernel $\SAC{H}{}$.
Moreover, the contribution of the interaction between the perturbation ($\GEM{k}{}$) 
and the main flow ($\GEM{p}{}$) to $\SAC{H}{}$
is measured by the ($p$,$q$)-plane distribution of the GEV-mode sectional curvature 
$\MSC{H}{}$ for assigned $\GEM{k}{}$.

%\newpage
%%%%%%%%%%%%%%%%%%%%%%%%%%%%%%%%%%%%%%%%%%%%%%%%%%%%%%%%%%%%%%%%%%%%%%
\section{Sectional curvature analysis of hydrodynamic and 
magnetohydrodynamic stabilities}
%%%%%%%%%%%%%%%%%%%%%%%%%%%%%%%%%%%%%%%%%%%%%%%%%%%%%%%%%%%%%%%%%%%%%%
\renewcommand{\EIGEN}[2]{\EIG(\tilde{#1}_{#2})}%

In this section, we observe the functional features of the sectional curvatures and discuss their implications for the stabilities of the basic flows.
Before embarking on a detailed analysis, some remarks should be made.
\vs

%%%%%%%%%%%%%%%%%%%%%%%%%%%%%%%%%%%%%%%%%%%%%%%%%%%%%%%%%%%%%%%%%%%%%%%%
\begin{figure}
\begin{center}
\includegraphics[width=0.49\textwidth]{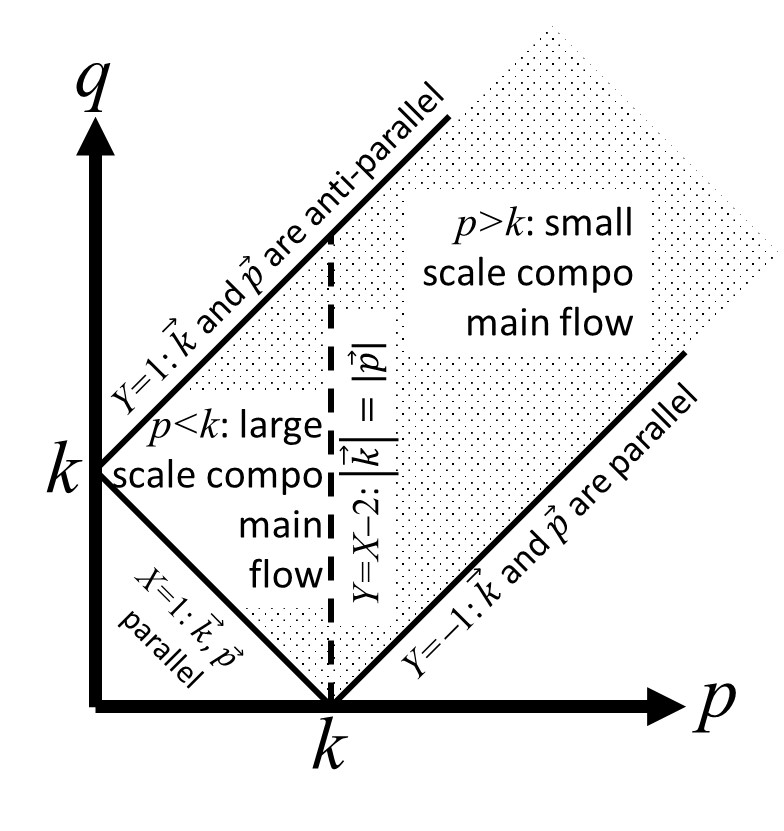}
\includegraphics[width=0.49\textwidth]{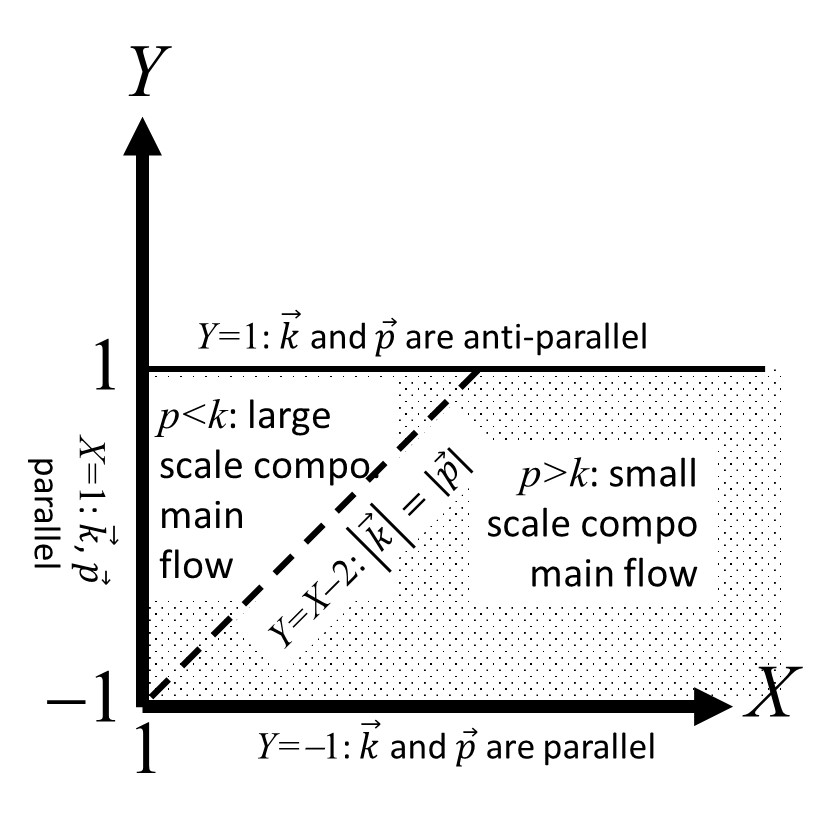}
\end{center}
\caption{\label{how to read}
How to read the following figures of geometric factors of 
normal-mode sectional curvature $\gMSC{}{}$.
Left: Region wherein moduli ($k$, $p$, $q$) of resonant three waves
constitute a triangle in ($p$, $q$)-space for an assigned $k$.
Right: Region is inclined clockwise by 45 degrees,
and the coordinates are normalized by $k$;
abscissa: $X={\p}+{\q}>1$, ordinate: $-1<Y={\q}-{\p}<1$.
Top left corner region close to $(X,Y)\approx(1,1)$ corresponds to 
$0\approx p\ll k\approx q$, i.e, the influence of large-scale components (or
``nonlocal'' interaction) of main flow on perturbation.
In contrast, region close to line $Y=X-2$ reads 
``local'' ($k\approx p$) interaction.
Region to right of $Y=X-2$ describes influence
of small-scale components of main flow.
}%
\end{figure}
%%%%%%%%%%%%%%%%%%%%%%%%%%%%%%%%%%%%%%%%%%%%%%%%%%%%%%%%%%%%%%%%%%%%%%%%

The normal-mode sectional curvatures for the HMHD and HD systems are decomposed into a product of the square of the wavenumber of the perturbation $\pi^2k^2$ and a dimensionless function $\gMSC{}{}$:
$\MSC{}{}=\pi^2k^2\gMSC{}{}$.
Furthermore,
the GEV-mode sectional curvature for the HMHD system 
(\ref{GEV curvature general}) 
is given by a series of functions of the wavenumber moduli ratios
${\p}:=\ABS{p}/\ABS{k}$ and ${\q}:=\ABS{q}/\ABS{k}$ as
$
  \MSC{H}{}(k,p,q)
  =\pi^2k^2\sum_{n}(\pi\HALL\ABS{k})^n
  \gMSC{H}{(n)}({\p},{\q})
,
$
where the dimensionless expansion parameter $\pi\HALL\ABS{k}$
measures the scale disparity ratio of the ion skin depth to the observed perturbation scale.
As for the HD case, the CHW-mode sectional curvature 
(\ref{Euler curvature general})
has the following form:
$
  \MSC{E}{}(k,p,q)
  =\pi^2k^2
  \gMSC{E}{}({\p},{\q})
.
$

This leads to two important consequences:
first,
the stability features of the systems, i.e., the sign of $\MSC{}{}$s 
are determined by the dimensionless functions, $\gMSC{H}{(n)}$ or $\gMSC{E}{}$.
Second,
these functions are scale-independent because they are the function of the wavenumber moduli ratios
${\p}:=\ABS{p}/\ABS{k}$ and ${\q}:=\ABS{q}/\ABS{k}$.
In other words, the stability features do not depend directly on the wavenumber vectors of the main flow $\vec{p}$ and the perturbation $\vec{k}$, 
but on their relative scale disparity and angle between them, i.e., the geometrical shape of the triangle formed by $\{\vec{k},\vec{p},\vec{k}-\vec{p}\}$ (irrespective of its orientation).

Thus, we will present in the following sections the dimensionless functions $\gMSC{}{}$s, which are called ``geometric factors'' hereafter, instead of $\MSC{}{}$s.
An explanation of how to read the functional profiles of the geometric factors is given in Fig. \ref{how to read}.
\vs

In the following we use the term ``nonlocal'' when the considered spatial scales of main flow and perturbation differ significantly each other.
In terms of the disparity ratio of wavenumbers ${\p}=p/k$, 
the term ``nonlocal'' implies the characteristic features around 
${\p}\approx0$ (or ${\p}\gg1$), 
or the tendencies of some properties for ${\p}\to0$ (or ${\p}\to+\infty$).
The term ``local'' is used when the features or tendencies are not ``nonlocal.''
The use of these terms are not so strict.

%Secondly, the profiles of the geometric factors of the shell-averaged curvature kernels $\gSAC{}{}$ are plotted in the range $0<{\p}<1$. Because of the symmetry of the mode sectional curvature $\MSC{}{}$ with respect to the exchange of $k$ and $p$, the curvature kernels $\SAC{}{}$ can be written in the form of $\SAC{}{}({\p})=\pi^3k^4{\p}^3S({\p})$, where the function $S({\p})$ satisfies $S({\p})=S(1/{\p})$. Thus, the plot of $({\p},{\p}^{\gamma}\SAC{}{}({\p}))$ for ${\p}\in(0,1)$ agrees with that of $\left(\frac{1}{\p},{\p}^{\gamma+6}\SAC{}{}\left(\frac{1}{\p}\right)\right)$ for $\frac{1}{\p}\in(0,1)$, and the plot on the interval $(0,1)$ is sufficient to determine the behavior on $(1,\infty)$. The multiplication factor ${\p}^{\gamma}$ is introduced to obtain the regularity of the curvature kernel at sufficiently small ${\p}$.

%%%%%%%%%%%%%%%%%%%%%%%%%%%%%%%%%%%%%%%%%%%%%%%%%%%%%%%%%%%%%%%%%%%%%%%%
\subsection{Stability of Euler dynamics}

For comparison with the HMHD case, 
we examine the HD case, i.e., the case wherein the magnetic field is absent.

As was discussed in \cite{2016arXiv160105477A},
the dissipationless, incompressible HD, MHD, and HMHD systems 
have common mathematical structures.
Thus, 
the formal derivations of differential-geometrical properties and equations 
such as given in the previous two sections 
are also applicable to the Euler equation case.
The explicit expressions of the Riemannian metric, Lie bracket, 
Levi-Civita connection, and sectional curvature for the representation 
using the complex helical wave (CHW), 
which is known as the basis function of three-dimensional 
solenoidal fields \cite{waleffe1992nature}, 
are summarized in
\ref{CHW mode representation of the Euler sectional curvature}.

First of all,
the sectional curvature 
$
  \MSC{E}{}(k,\HEL_k,p,\HEL_p,q)
$ 
is negative definite for arbitrary two CHWs
(see Eq. (\ref{Euler curvature general})).
In the left panel of Figure \ref{contour Euler},
we presented 
the functional form of the geometric factor of the sectional curvature 
$
  \gMSC{E}{}({\p},{\q})
$.
This plausibly implies that the Euler dynamics is always unstable for 
arbitrary perturbation.
The negative sectional curvature between two monochromatic waves 
was already reported by Nakamura et al. \cite{nakamura1992geodesics}.
Our novel findings are that 
this feature is independent of the helicities $\HEL$ 
of the main flow and perturbation,
and
that the (${\p}$,${\q}$)-plane 
distribution of sectional curvature amplitude is presented.
%
%%%%%%%%%%%%%%%%%%%%%%%%%%%%%%%%%%%%%%%%%%%%%%%%%%%%%%%%%%%%%%%%%%%%%%%%
\newcommand{\gHeight}{0.22\textwidth}%{0.16\textwidth}
\begin{figure}
\begin{center}
\includegraphics[height=\gHeight]{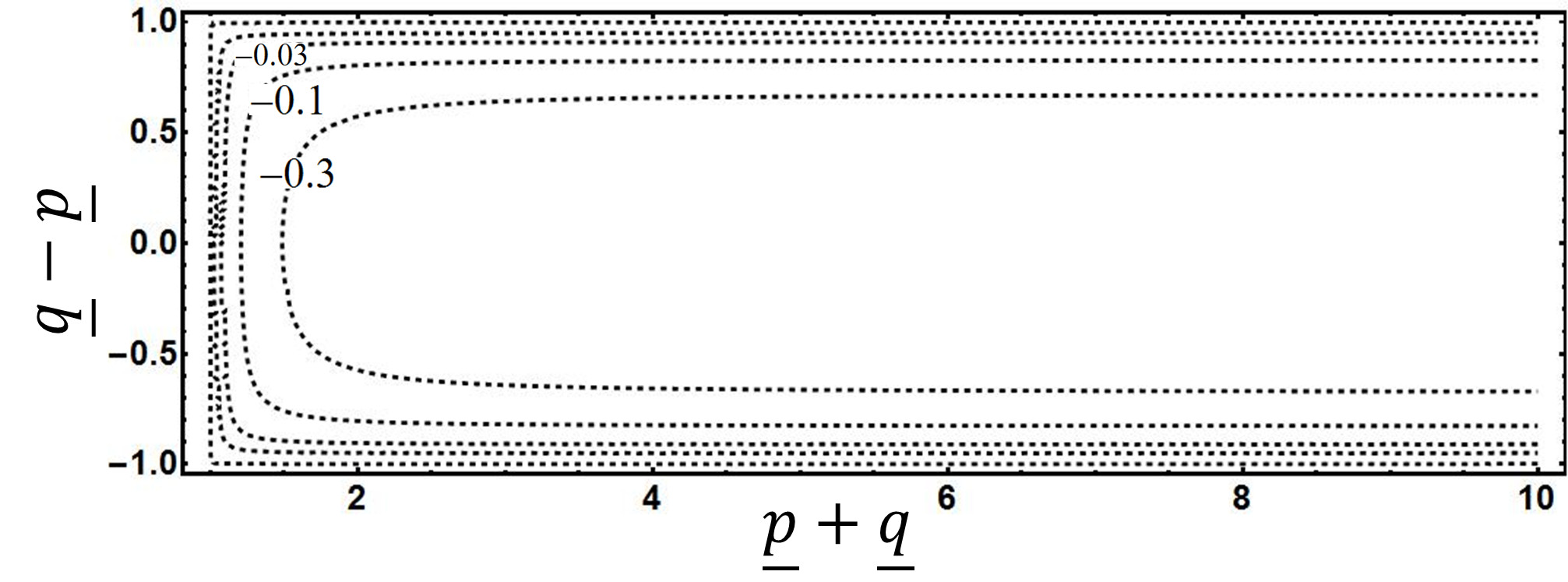}
\includegraphics[height=\gHeight]{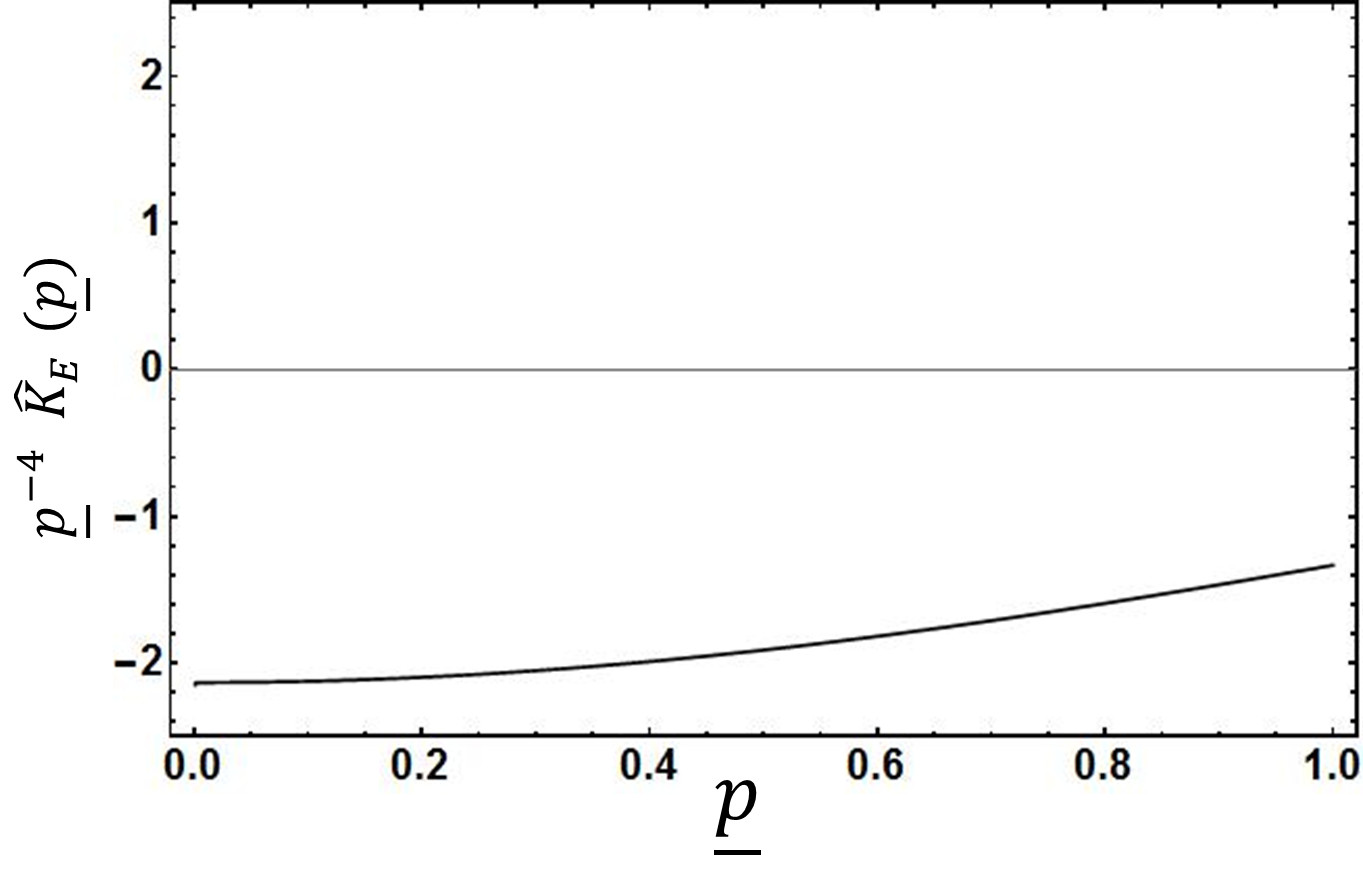}
\end{center}
\caption{\label{contour Euler}
Left: Geometric factor of CHW-mode sectional curvature
of Euler dynamics, $\gMSC{E}{}({\p},{\q})$,
drawn in range $1<{\p}+{\q}<10$, $-1<{\q}-{\p}<1$.
Contour levels are set to $-10^{N/2}$ ($N\geq-7$).
Right: Geometric factor of shell-averaged curvature kernel 
(divided by ${\p}^{4}$), 
${\p}^{-4}\gSAC{E}{}({\p})$,
drawn in range $0<{\p}<1$.
All of the values are independent of helicity parameters $\HEL_{k}$ and $\HEL_{p}$.
}
\end{figure}
%%%%%%%%%%%%%%%%%%%%%%%%%%%%%%%%%%%%%%%%%%%%%%%%%%%%%%%%%%%%%%%%%%%%%%%%
\vs

As for the application to homogeneous and isotropic turbulet flows,
on the other hand,
the sectional curvature should be accumulated by the spherical shells 
in wavenumber space (see Equation (\ref{integrated curvature})).
In the right panel of Figure \ref{contour Euler}, 
the geometric factor of the shell-averaged curvature kernel for the Euler dynamics 
(\ref{Euler curvature kernel}) 
is presented.

Note that in the right panel of the Fig. {contour Euler} the infrared side ($0<{\p}<1$) of 
$\big({\p},{\p}^{-4}\gSAC{E}{}({\p})\big)$
is plotted where the factor ${\p}^{-4}$
is multiplied to grasp the ${\p}\to0$ tendency of $\gSAC{E}{}$.
As for the ultraviolet side (${\p}>1$), 
it is checked that the plot of
$\big({\p}^{-1},{\p}^{2}\gSAC{E}{}({\p}^{-1})\big)$
for $0<{\p}^{-1}<1$
gives the same functional profile as this graph.

Since the regularity of the curvature kernel is $O({\p}^4)$ for small ${\p}$,
the sectional curvature (\ref{integrated curvature})
for the correlation function with the scaling 
$Q(\ABS{p})\propto p^{\gamma}$
converges in the infrared side if $\gamma>-5$.
For the ultraviolet side, the sectional curvature integral converges for 
$\gamma<-3$
due to the $O({\p}^{-2})$ tendency for large ${\p}$.
It is very interesting that $\gamma=-11/3$,
which corresponds to Kolmogorov's 1941 inertial range scaling law,
is included in this convergent range.
Thus,
the sectional curvature of the Euler dynamics is negative definite,
and
the interaction between the main flow and the perturbation is ``local'' 
in the sense that the integration does not diverge on either the infrared side or the ultraviolet side.

%\newpage%{blank}\newpage
%%%%%%%%%%%%%%%%%%%%%%%%%%%%%%%%%%%%%%%%%%%%%%%%%%%%%%%%%%%%%%%%%%%%%%%%
\subsection{$O(\HALL^0)$ features of HMHD dynamics: MHD system stability
\label{O(a0) features of HMHD dynamics: MHD system stability}}
%%%%%%%%%%%%%%%%%%%%%%%%%%%%%%%%%%%%%%%%%%%%%%%%%%%%%%%%%%%%%%%%%%%%%%%%

\begin{figure}%[h]
\begin{center}
\includegraphics[height=\gHeight]{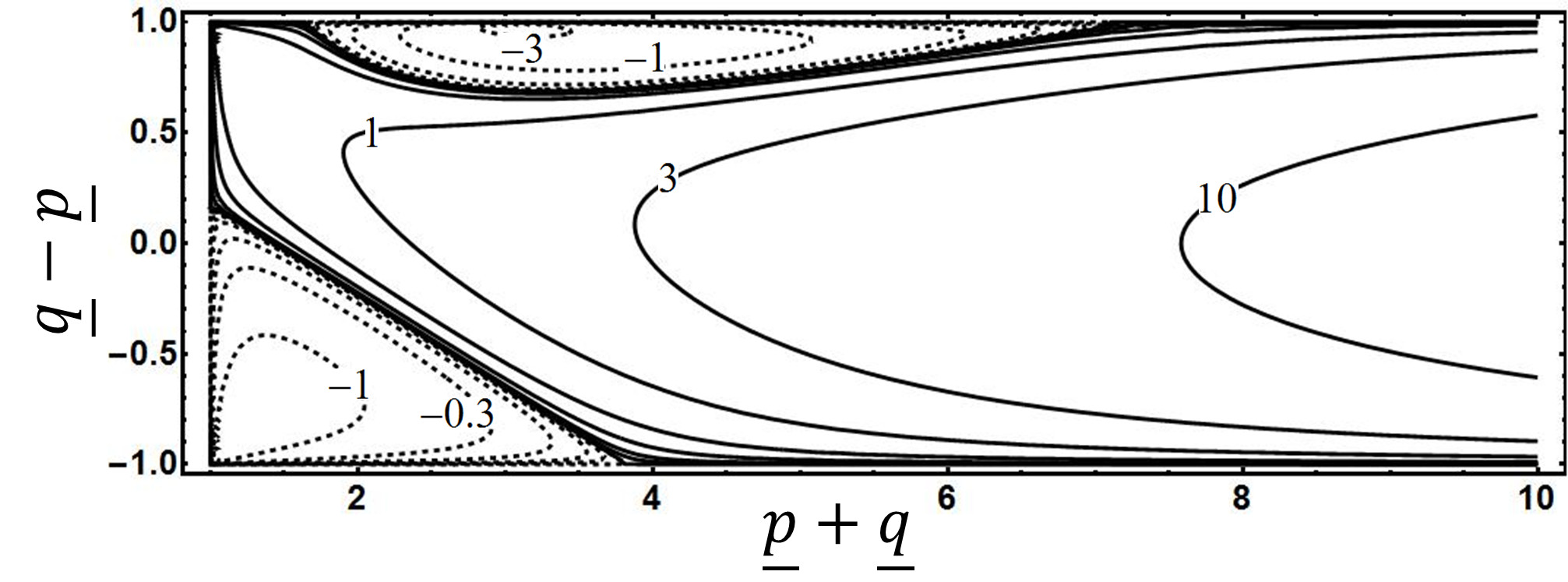}
\includegraphics[height=\gHeight]{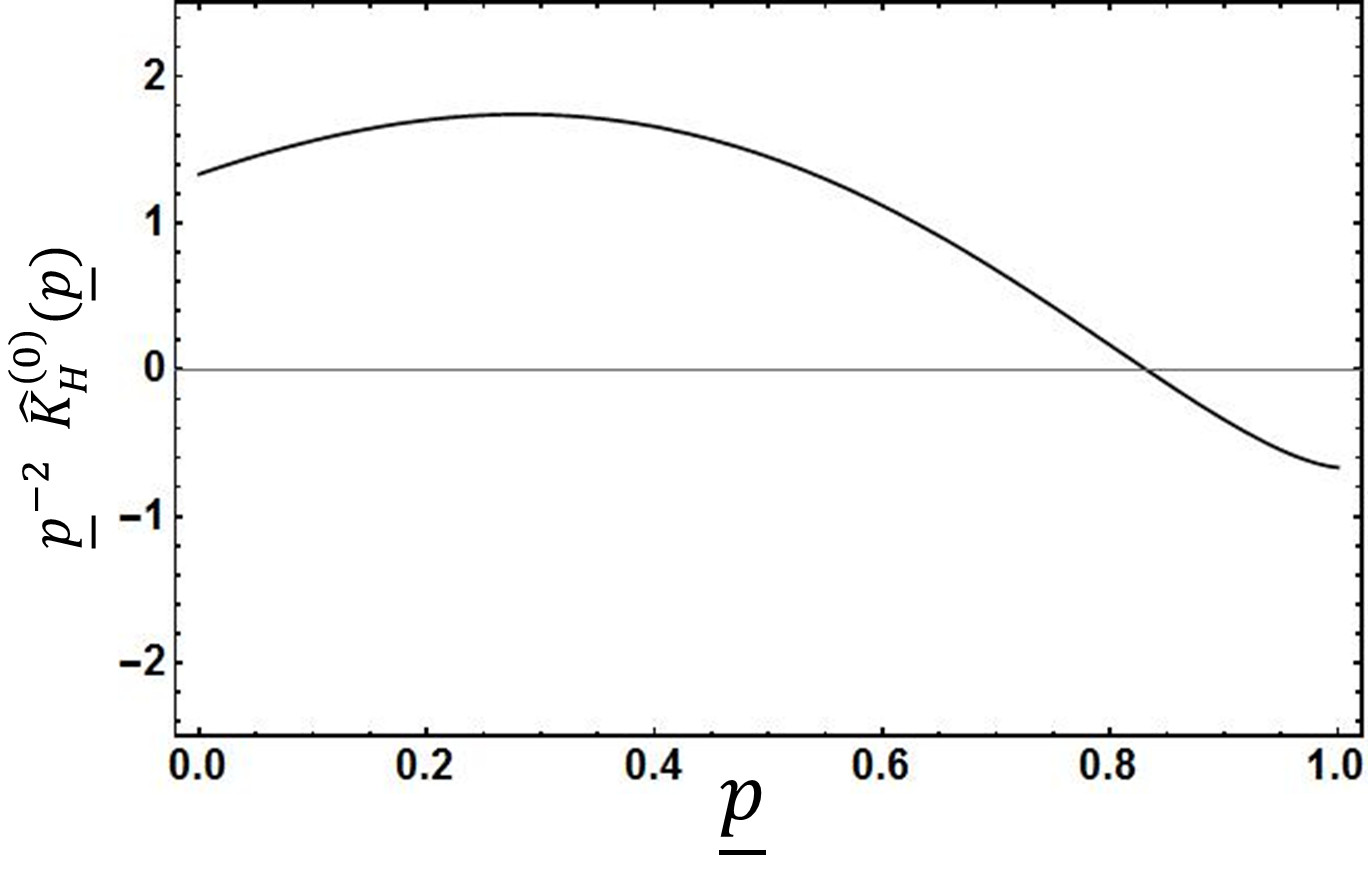}
\\
\includegraphics[height=\gHeight]{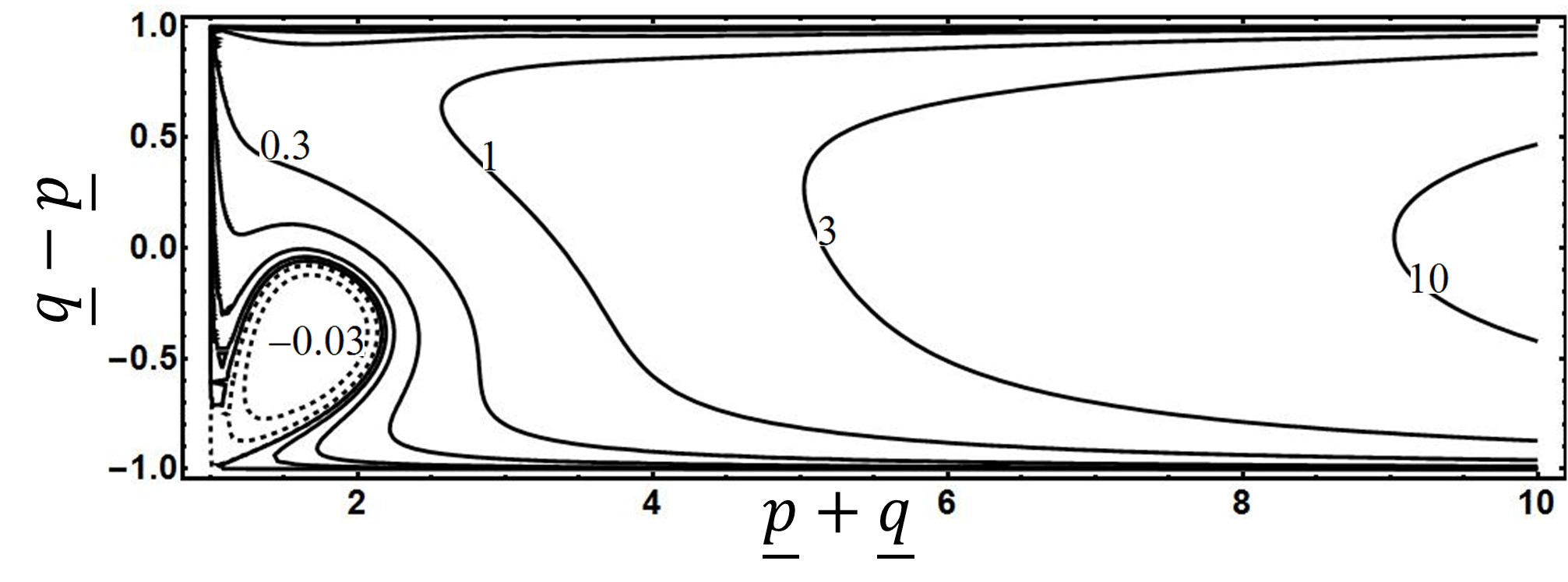}
\includegraphics[height=\gHeight]{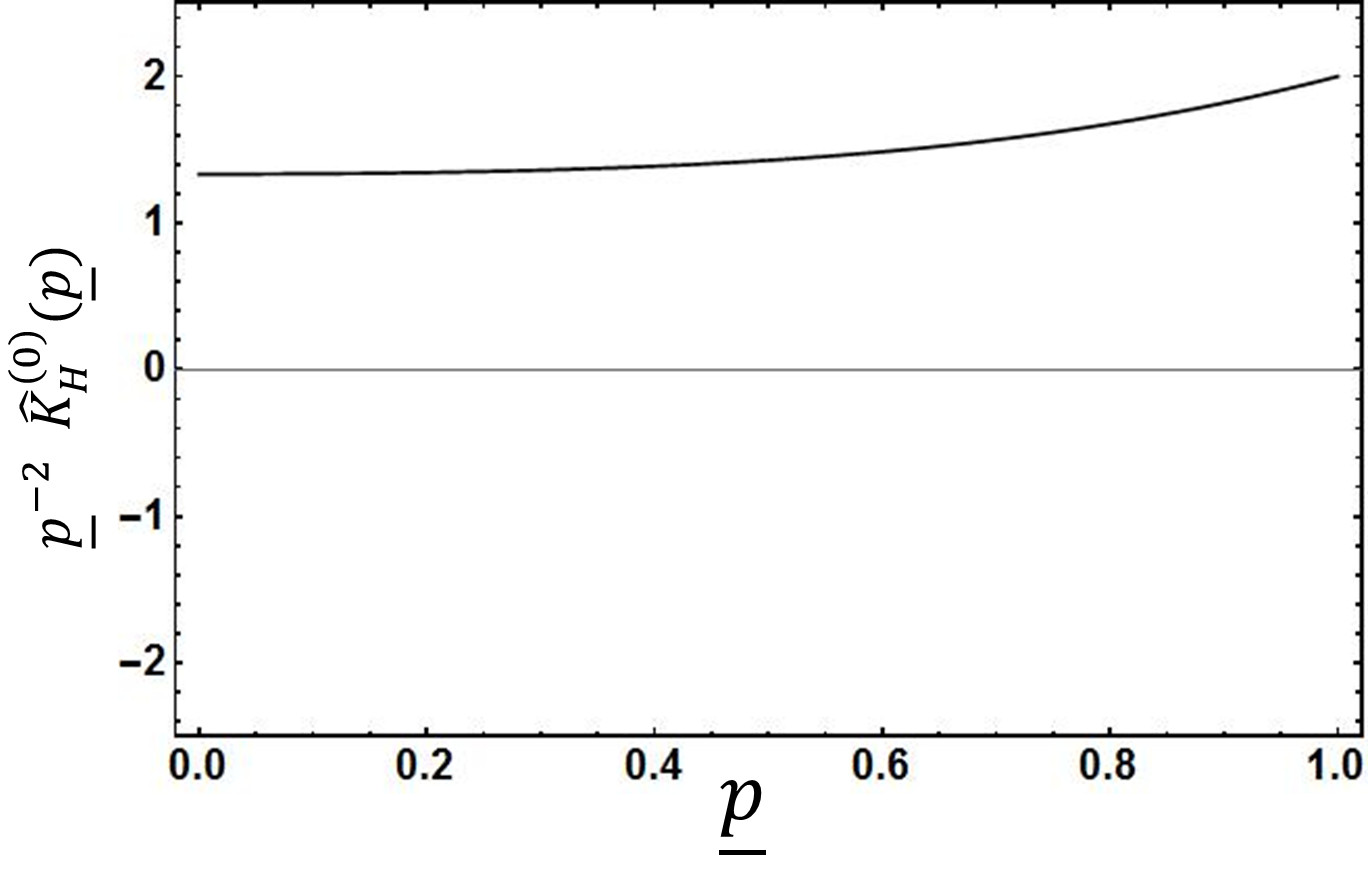}
\\
\includegraphics[height=\gHeight]{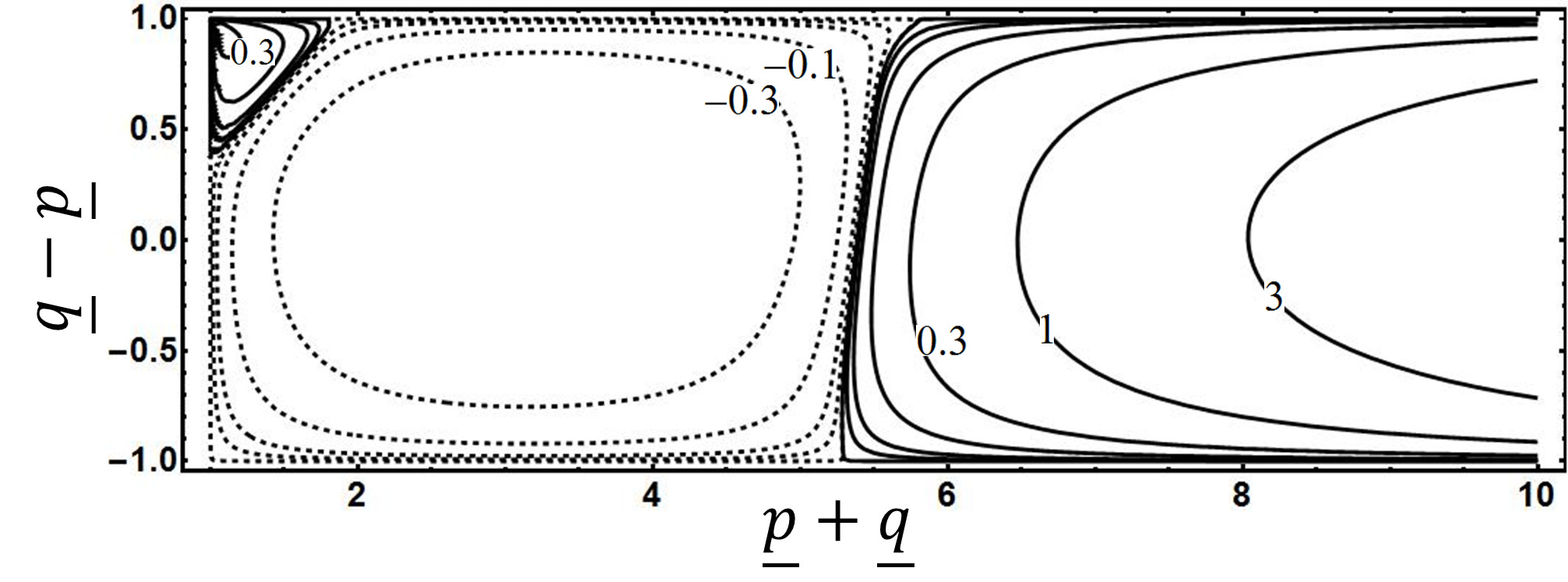}
\includegraphics[height=\gHeight]{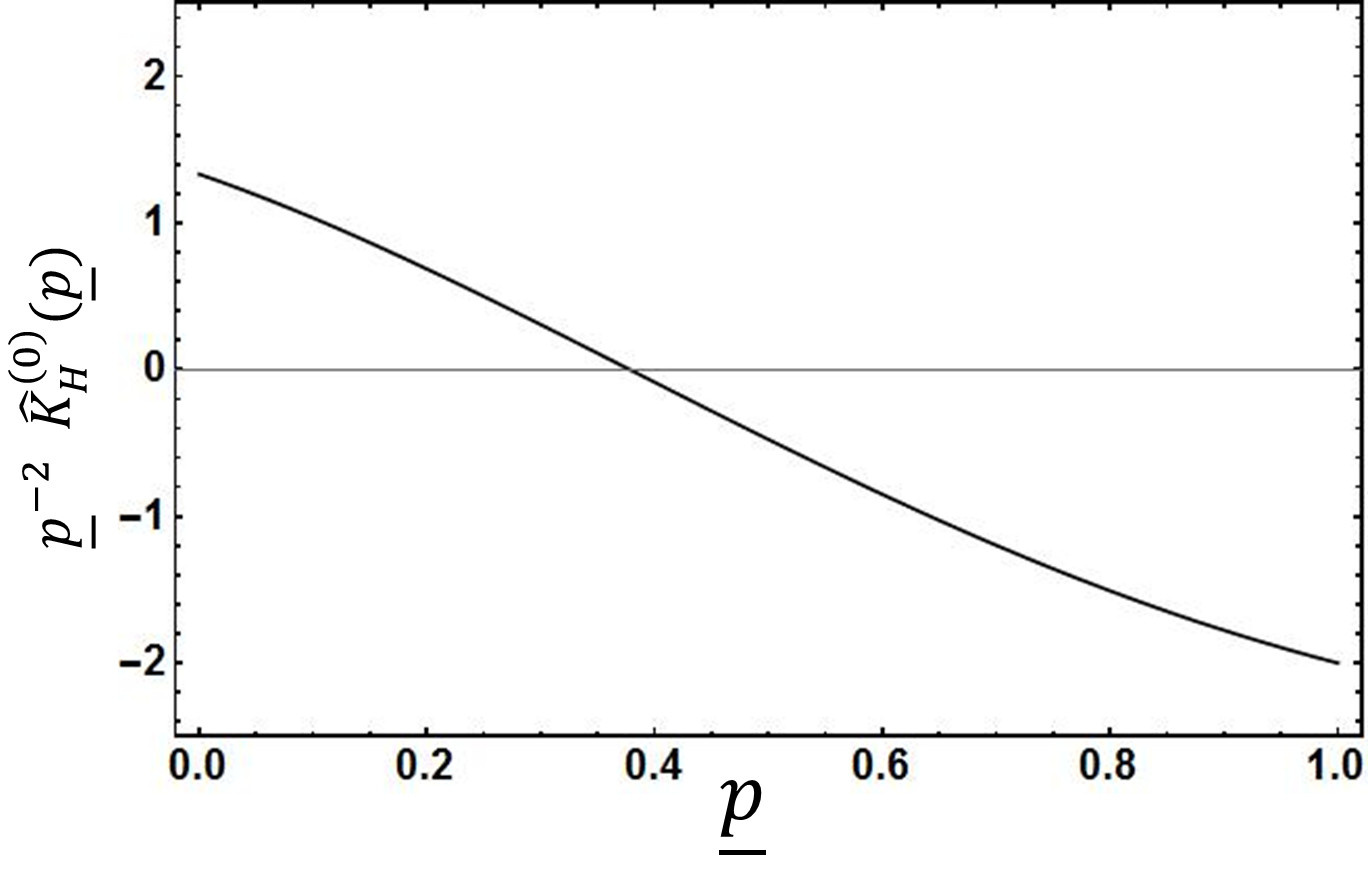}
\end{center}
\caption{\label{O(1) curvature}
Left: Geometric factor of GEV-mode sectional curvature 
of the MHD dynamics, 
$\gMSC{H}{(0)}({\p},{\q},\HEL_{k}\HEL_{p},\POL_{k}\POL_{p})$.
The range shown is same as that in Fig. \ref{contour Euler}.
Solid (dashed) contour lines denote positive (negative) 
values; their levels are set to $10^{N/2}$ ($-10^{N/2}$) for $N\geq-7$.
Right: Corresponding geometric factors of 
shell-averaged curvature kernel (divided by ${\p}^{2}$)
${\p}^{-2}\gSAC{H}{(0)}({\p},{\q},\HEL_{k}\HEL_{p},\POL_{k}\POL_{p})$
in range $0<{\p}<1$.
Mode combinations are 
$\POL_{k}\POL_{p}=\HEL_{k}\HEL_{p}=1$ (upper),
$\POL_{k}\POL_{p}=-\HEL_{k}\HEL_{p}=\pm1$ (middle),
and
$\POL_{k}\POL_{p}=\HEL_{k}\HEL_{p}=-1$ (lower).
}
\end{figure}

In Figure \ref{O(1) curvature},
the GEV-mode sectional curvature for the MHD case 
(i.e. the $\HALL\to0$ limit case) is presented.

In contrast to the Euler dynamics case,
two remarkable features are seen:
firstly,
the stability of the MHD dynamics 
depends on the helicity combinations of the main flow and perturbation,
as well as depending on their {\polarity} combinations.

Furthermore,
the mode sectional curvatures take both positive and negative values,
i.e.,
there are both stable and unstable wavenumber combinations.
In particular, all of the sectional curvatures are positive around the regions with 
${\p}\approx{\q}\approx0$ 
(upper left corner of each panel) and those with ${\p}$, ${\q}\gg1$
(far beyond the right side of each panel),
which implies that such the interactions that is ``nonlocal'' 
in wavenumber space do not lead to the growth of small perturbations.
In other words, 
this conjectures that the main flow destabilization is essentially caused by the 
``local'' mode interactions in the MHD case.

%%%%%%%%%%%%%%%%%%%%%%%%%%%%%%%%%%%%%%%%%%%%%%%%%%%%%%%%%%%%%%%%%%%%%%%%
%%%%%%%%%%%%%%%%%%%%%%%%%%%%%%%%%%%%%%%%%%%%%%%%%%%%%%%%%%%%%%%%%%%%%%%%

Let us see some of the details of the sectional curvature profiles.

The upper and middle panels of Fig. \ref{O(1) curvature}
depict the stability of waves that belong to the same {\polarity}
($\POL_{k}=\POL_{p}$).
Negative values of $\gMSC{H}{(0)}$ 
appear around $\ABS{k}\approx\ABS{p}\gg\ABS{q}$ for $\HEL_{k}=\pm\HEL_{p}$
and  $\ABS{k}\approx\ABS{p}\approx\ABS{q}/2$ for $\HEL_{k}=\HEL_{p}$.
These wavenumber modulus relations indicate that the interactions between the waves whose wavenumbers are nearly parallel ($\vec{k}\parallel\vec{p}$) destabilize the main flow.

However, the stable wavenumber combinations are roughly perpendicular, i.e., they satisfy
$\vec{k}\perp\vec{p}$ approximately.
It is well known that Alfv\'en waves (with wavenumber $\vec{k}$) 
propagate most efficiently in the direction of the ambient magnetic field
$\bm{B}$, i.e., $\bm{B}\parallel\vec{k}$.
Since $\bm{B}$ is solenoidal, $\bm{B}\perp\vec{p}$, and thus
it is expected that the stable region is dominated mainly by the propagation of 
Alfv\'en waves.

The middle and lower panels of Fig. \ref{O(1) curvature} 
give the stability of the waves in the opposite {\polarity}
($\POL_{k}=-\POL_{p}$).
It is remarkable that the interactions between the different helicities
($\HEL_{k}=-\HEL_{p}$)
are unstable irrespective of the angles between the wavenumbers of the main flow and the perturbation.
\vs

As for the application to homogeneous and isotropic turbulent flows,
on the other hand,
the sectional curvature should be accumulated by the spherical shells 
in wavenumber space (see Equation (\ref{integrated curvature})).
The right-hand panels of Fig. \ref{O(1) curvature} 
show the functional profiles of the geometric factor of the shell-averaged sectional curvature kernel
$\gSAC{H}{(0)}$ (multiplied by ${\p}^{-2}$) for $0<{\p}<1$.
Note that, the plot of
$\big({\p}^{-1},{\p}^{4}\gSAC{E}{}({\p}^{-1})\big)$
for $0<{\p}^{-1}<1$
gives the same functional profile.

Since
$\gMSC{H}{(0)}({\p},{\q})$ 
takes both the positive and negative values for some assigned ${\p}$s,
the kernel function $\gSAC{H}{(0)}({\p})$
tends to have positive (stable) value even for such wavenumbers ${\p}$
that have negative sectional curvature (unstable) combinations 
$({\p},{\q})$.
Due to this summation process, the negative values of the kernel function appear only for those cases with 
$\HEL_{k}\HEL_{p}=\POL_{k}\POL_{p}$.
The unstable ranges are 
$0.8318<{\p}<1/0.8318$ for $\HEL_{k}\HEL_{p}=\POL_{k}\POL_{p}=1$
and
$0.3784<P<1/0.3784$ for $\HEL_{k}\HEL_{p}=\POL_{k}\POL_{p}=-1$.
This result suggests that the contribution of the mode interactions to the plasma motion stability should be analyzed more carefully than by integrating the sectional curvature simply.

It is interesting that the shell-averaged curvature kernel behaves as 
%$\displaystyle\lim_{{\p}\to0}{\p}^{-2}\gSAC{H}{(0)}({\p})=4/3$, 
%$\lim_{{\p}\to0}\gSAC{H}{(0)}({\p})/{\p}^{2}=4/3$, 
%i.e.,
$\SAC{H}{(0)}(k,p)\approx\frac43\pi^3k^4{\p}^2+o({\p}^2)$ 
irrespective of the values of $\HEL$ and $\POL$
for sufficiently small $p$.
The value is analytically obtained by expanding Equation (\ref{gSACH(0)})
in power series of ${\p}$ around ${\p}=0$.
As will be discussed in the final section, this conjectures that the interaction with large scale components of plasma motions is dominated by the propagation of Alfv\'en waves.

%%%%%%%%%%%%%%%%%%%%%%%%%%%%%%%%%%%%%%%%%%%%%%%%%%%%%%%%%%%%%%%%%%%%%%%%
\subsection{$O(\HALL^1)$ features of the HMHD dynamics: 
lowest-order Hall-term effect}
%%%%%%%%%%%%%%%%%%%%%%%%%%%%%%%%%%%%%%%%%%%%%%%%%%%%%%%%%%%%%%%%%%%%%%%%
\begin{figure}%[h]
\begin{center}
\includegraphics[width=0.49\textwidth]{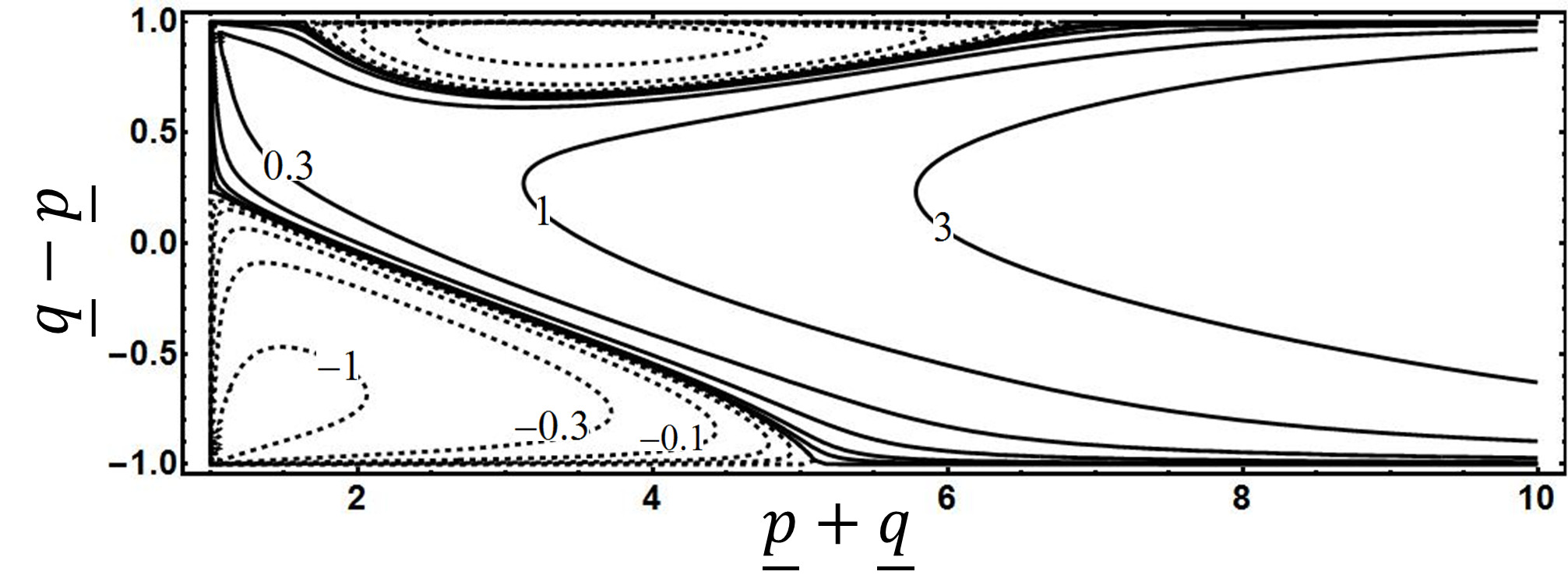}
\includegraphics[width=0.49\textwidth]{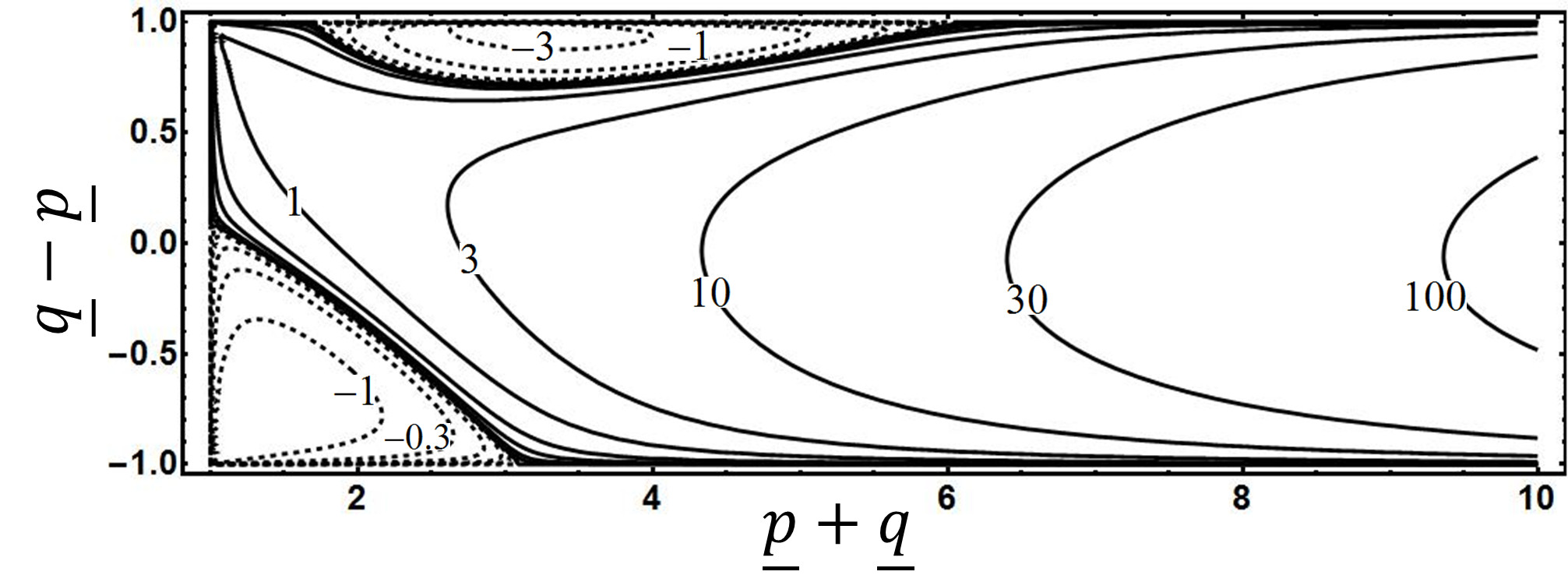}
\\
\includegraphics[width=0.49\textwidth]{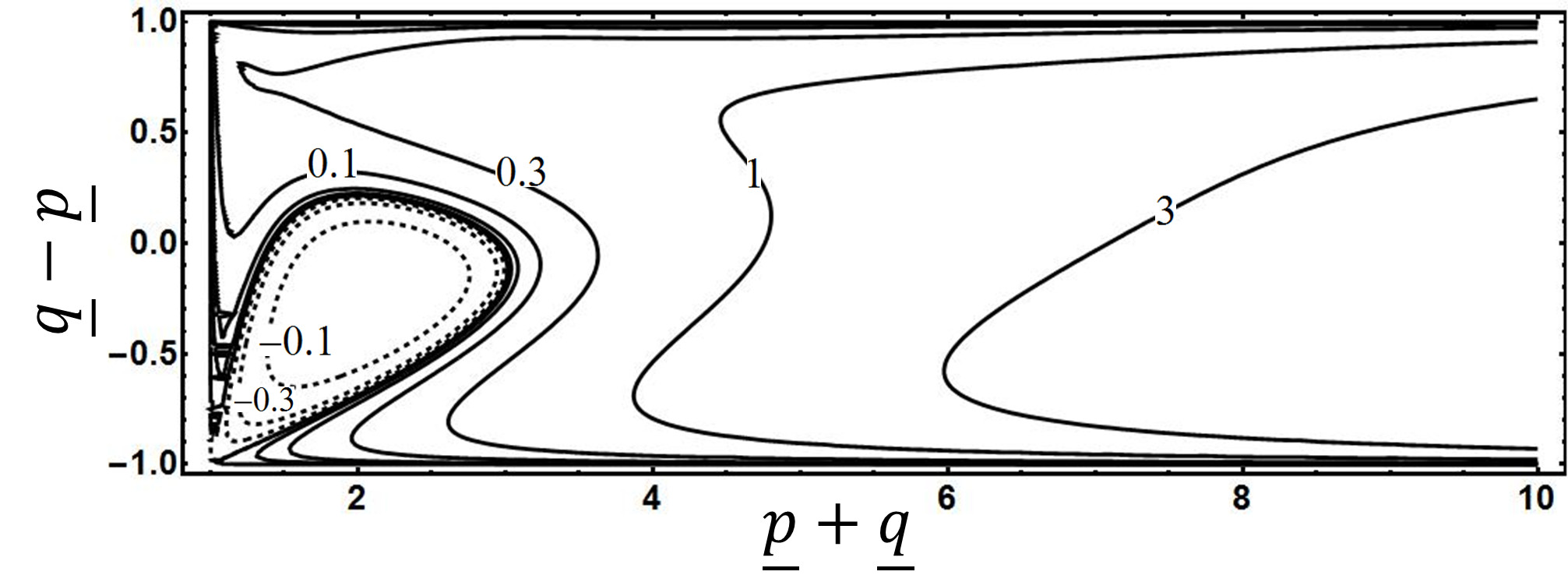}
\includegraphics[width=0.49\textwidth]{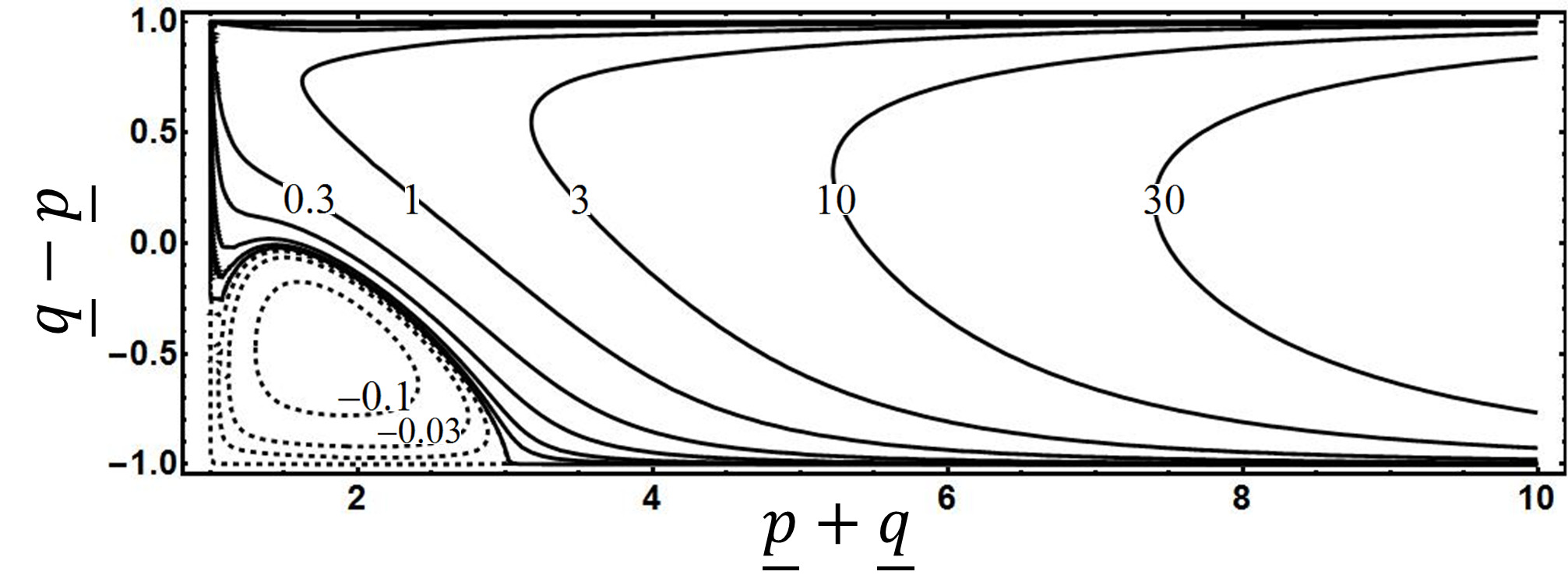}
\\
\includegraphics[width=0.49\textwidth]{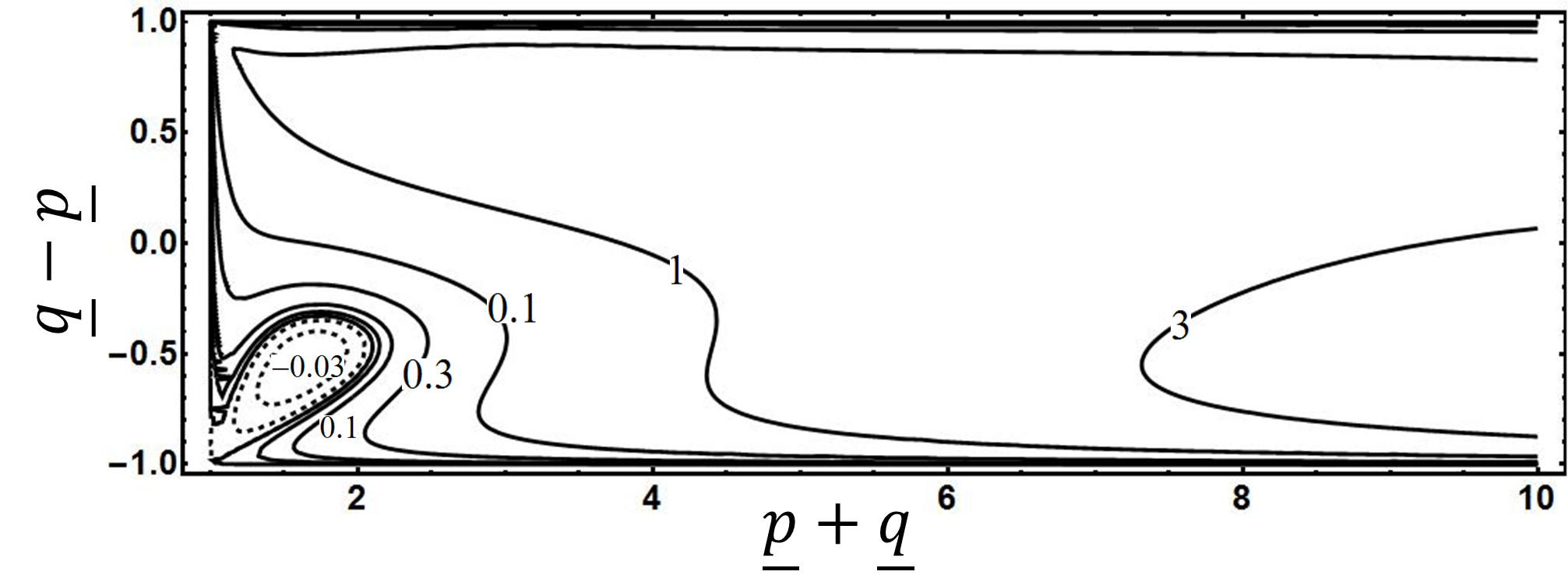}
\includegraphics[width=0.49\textwidth]{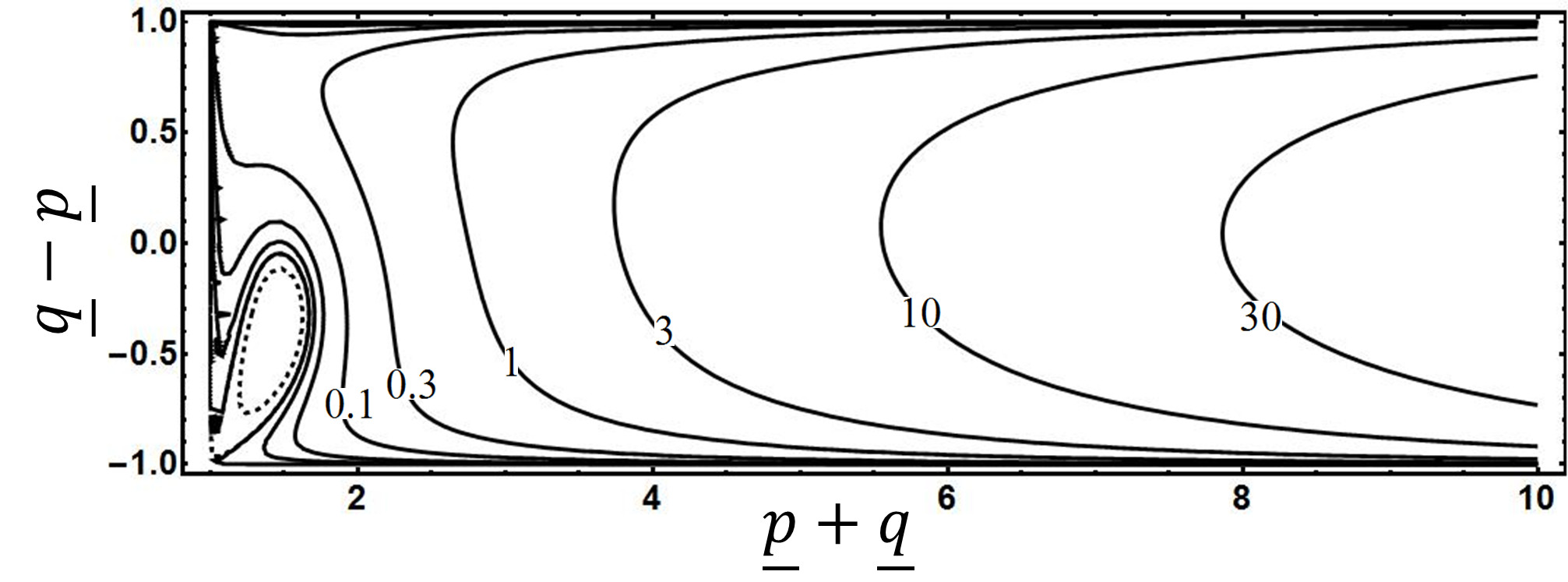}
\\
\includegraphics[width=0.49\textwidth]{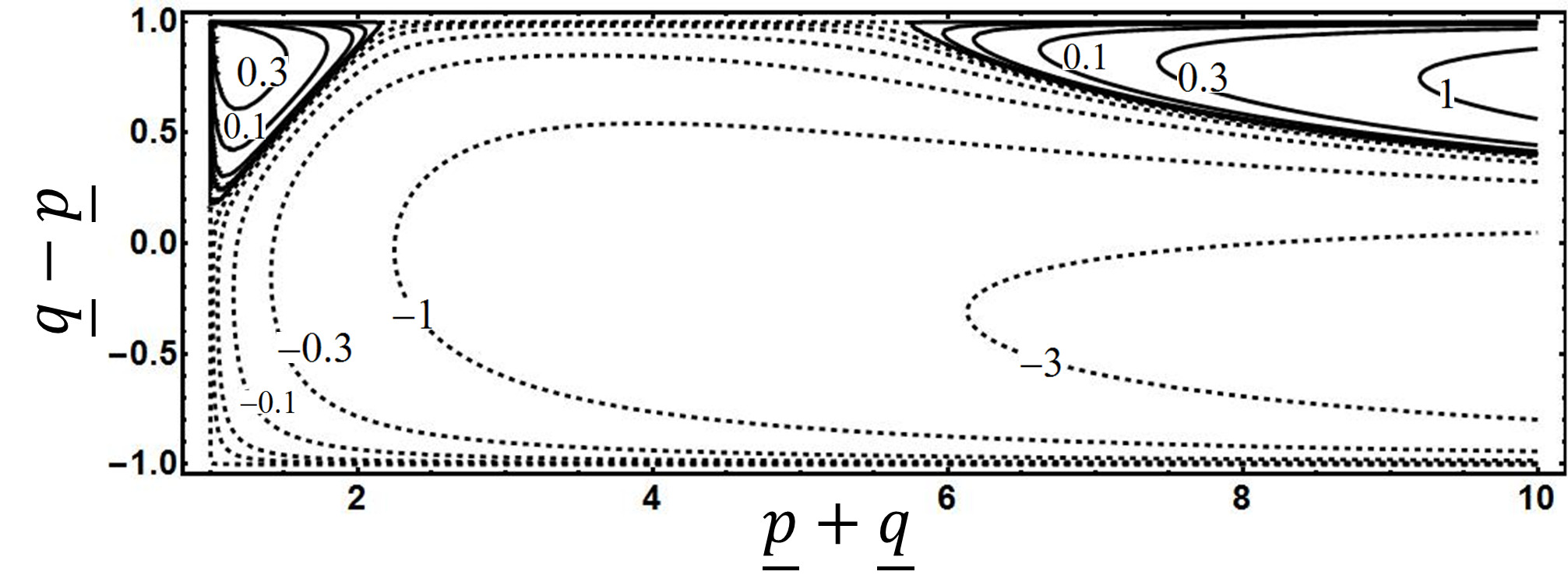}
\includegraphics[width=0.49\textwidth]{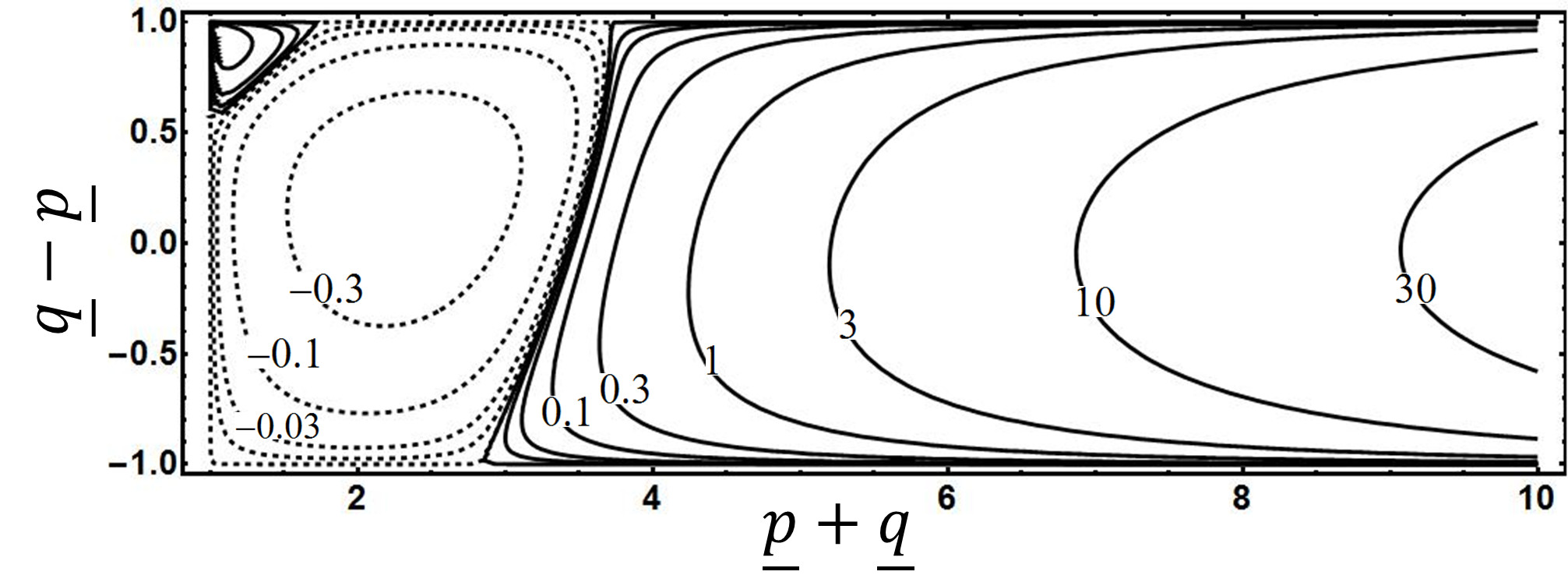}
\end{center}
\caption{\label{geometric factor for O(HALL)}
\\
\begin{minipage}{0.45\textwidth}
Geometric factor including the $O(\HALL)$ terms of GEV-mode sectional curvature, 
$\gMSC{H}{(0)}+0.1\pi\gMSC{H}{(1)}$.
Helicity and {\polarity} parameters are summarized in table on the right.
Range and contour levels plotted are same as those in Fig. 
\ref{O(1) curvature}.
\end{minipage}\ \ 
\begin{minipage}{0.34\textwidth}
\begin{tabular}{c||r|rr|rr}
   \multirow{2}{*}{row}
 & \multirow{2}{*}{$\HEL_{k}\HEL_{p}$}
 & \multicolumn{2}{c|}{left}
 & \multicolumn{2}{c}{right}
\\ \cline{3-6}
  & & $\POL_{k}$ & $\POL_{p}$ & $\POL_{k}$ & $\POL_{p}$
\\\hline\hline
  1 &  $1$ &   1  & 1 & $-1$ & $-1$
\\
  2 & $-1$ &   1  & 1 & $-1$ & $-1$
\\
  3 &  $1$ & $-1$ & 1 &   1  & $-1$
\\
  4 & $-1$ & $-1$ & 1 &   1  & $-1$
\end{tabular}
\end{minipage}
}
\end{figure}
%%%%%%%%%%%%%%%%%%%%%%%%%%%%%%%%%%%%%%%%%%%%%%%%%%%%%%%%%%%%%%%%%%%%%%%%

In this section, we examine the geometric factor including the $O(\HALL)$ 
terms of the GEV-mode sectional curvature, 
$\gMSC{H}{(0)}+\pi\HALL\ABS{k}\gMSC{H}{(1)}$, 
for the dimensionless parameter $\HALL\ABS{k}=0.1$.
The detailed expression of $\gMSC{H}{(1)}$ is given by Eq. (\ref{gMSCH(1)}).
We focus here how the Hall-term effect at the lowest order changes the stability diagram of the MHD system (see Fig. \ref{O(1) curvature}).
In Figure \ref{geometric factor for O(HALL)},
the (${\p},{\q}$)-space distributions for all possible combinations of 
$\HEL_{k}$, $\HEL_{p}$, $\POL_{k}$, and $\POL_{p}$
are presented.

The panels are arranged according to the displaying order of Fig. \ref{O(1) curvature} in the vertical direction.
The left (right) four panels show the contribution of ion cyclotron (whistler) modes to the stability.
As is seen from the figure, the contribution of the ion cyclotron mode ($\POL_{p}=1$) by the Hall-term effect is somewhat opposite to that of the whistler mode ($\POL_{p}=-1$).

In the three of four cases (except for $(\POL_{k},\POL_{p})=(-1,1)$),
the ion cyclotron mode of the reference flow enhances the unstable mode combination regions.
Especially, for the combination with whistler mode with opposite helicity (the 4-th row of Fig. \ref{geometric factor for O(HALL)}), the unstable region is significantly spread over and the amplitude of the negative sectional curvature is enhanced (compared with other seven panels).
This tendency physically implies that, due to the Hall-term effect, the ion cyclotron modes tend to excite the whistler modes more efficiently than other mode combinations.

On the other hand, in the three of four cases of (except for $(\POL_{k},\POL_{p})=(-1,-1)$),
the Hall-term effect suppresses the instability due to the whistler mode of the reference solution, i.e., reduces the unstable mode combination regions.
The amplitudes of the positive (stable) sectional curvature region, i.e., the characteristic time or frequency of the perturbation fields significantly increased compared with the MHD and the ion cyclotron mode of HMHD cases.
It is guessed that, since the whistler modes have large phase velocities, they disperse more quickly than the unstable modes grow.

As a whole, the result conjectures an interesting picture that the ion cyclotron modes excite the whistler mode, while the excited whistler modes disperse quickly.
Thus the energy is transferred from the ion cyclotron modes to the whistler ones  in an almost one-sided way.
This picture is partly supported by the direct numerical simulation result of the fully-developed HMHD turbulence \cite{arakiMiura2016ICTAM}.

%\newpage%{blank}\newpage
%%%%%%%%%%%%%%%%%%%%%%%%%%%%%%%%%%%%%%%%%%%%%%%%%%%%%%%%%%%%%%%%%%%%%%
\section{Summary and discussion}
%%%%%%%%%%%%%%%%%%%%%%%%%%%%%%%%%%%%%%%%%%%%%%%%%%%%%%%%%%%%%%%%%%%%%%

In the present study, we approached the linear stability problem of the HMHD system using geodesic formulation and considered its applicability to turbulence theory.

\subsubsection*{Geodesic formulation, normal-mode expansion, and application to turbulence:}
The evolution equation of a dissipationless, incompressible HMHD fluid was formulated as a geodesic equation from a direct product of two volume-preserving diffeomorphisms.
Instead of usual mathematical postulation, we considered the linear connection ($\LCC{}{}$) from the viewpoint of physically desirable properties that
1) reproduced the Euler--Lagrange equation as its geodesic, 
2) guaranteed the detailed energy balance, and 
3) made the associated substantial derivative $\partial_t+\LCC{\GEV{V}{}}{}$ 
covariant against the Galilean boost.
The obtained connection agreed with the Levi-Civita connection.
It was an interesting finding that the Levi-Civita connection had these physically desirable properties for the HMHD (finite $\HALL$), MHD ($\HALL\to0$), and HD ($\HALL=0$) cases.

We performed stability analysis in terms of the Jacobi equation.
Lin constraints
($
 \sDD{t}{}{\GEV{\xi}{}}
 -
 \LCC{\GEV{\xi}{}}{\GEV{V}{}}
 =
 {\GEV{v}{}}
$)
and the linearized equation of motion
($
  \sDD{t}{}{\GEV{v}{}}+
  \LCC{\mbox{$\GEV{v}{}$}}{\GEV{V}{}}
  =0
$)
gave the Jacobi equation
($
  \left(\sDD{t}{}\right)^2{\GEV{\xi}{}}
  +
  R(\GEV{\xi}{},\GEV{V}{})\,\GEV{V}{}
  = 0
$).
It was also discussed that the analysis method based on the Jacobi equation is applicable to the problems of both the linear stability of basic flow and the relative dispersion of passively advected particle pairs.
The sign of the Riemannian curvature tensor
$
  R(*,{\GEV{V}{}}){\GEV{V}{}}
$
was used to determine the stability of the reference solution $\GEV{V}{}$.

Normal-mode stability analysis was carried out by using the GEV-mode expansion.
The difficulty that the Riemannian curvature term contains four-wave resonance was reduced by taking ensemble average of random vector fields (fully-developed turbulence).
Since the reduced curvature term naturally includes the sectional curvature between two GEV-mode, the curvature analysis was also able to be interpreted as the stability of a single GEV-mode.

\subsubsection*{Remark on applicability of normal-mode expansion other than GEVs:}
the normal-mode expansion expression of the Levi--Civita connection Eq. (\ref{Levi--Civita in HMHD}) and the Riemannian curvature tensor Eq. (\ref{normal-mode expansion of the Riemannian curvature}) had its foundation on the formula that the combination of the Riemannian metric and the Lie bracket is equal to the product of the eigenvalue of the helicity-based, particle-relabeling operator and a certain totally antisymmetric tensor,
$
  \Braket{\big}{\GEV{\Phi}{(\GEM{k}{})}}
    {\LieBraket{\big}{\GEV{\Phi}{(\GEM{p}{})}}{\GEV{\Phi}{(\GEM{q}{})}}}
  =
  \EIGEN{k}{}
  \tripleHMHD{\big}{\GEM{k}{}}{\GEM{p}{}}{\GEM{q}{}}
.
$
As was described in Ref. \cite{2016arXiv160105477A}, this decomposition formula is based on the general eigenvalue problem that derives the DBFs under appropriate boundary condition.
Thus, the expression of the curvature tensor is applicable to the systems with arbitrary shape, for example, to cylindrical configuration.

\subsubsection*{Comparison with Hamiltonian mechanical stability analyses:}

The stability of the HMHD system had often been analyzed in the Hamiltonian mechanical framework.
One of the well-known method is the energy-Casimir method, in which the second variation of an appropriate functional, i.e., the Hamiltonian with some other constants of motion is calculated as a sufficient condition for the stability of stationary solution (e.g. \cite{marsden2013introduction}; \S1.7).

As for the HMHD system, the Hamiltonian formulation was developed by Holm and the stability conditions were investigated \cite{Holm1987}.
An important improvement of the variational procedure was made by Hirota et al., in which the kind of variations were categorized in a sophisticated manner \cite{HirotaETAL2006}.
They distinguished among the following three kinds of perturbation fields:
arbitrary perturbation, Lagrangian displacements (LD), and dynamically accessible variation (DAV).

Our stability analysis is based on the Lin constraints (\ref{lin constraint}) which correspond to their LD variation given by Eqs. (12) and (13) of \cite{HirotaETAL2006}; that is, the variation of the generalized velocity is given by
$
  \GEV{V}{} \mapsto
  \GEV{V}{}+\epsilon(\partial_{t}+L_{\GEV{V}{}}){\GEV{\xi}{}}+o(\epsilon)
.
$
They derived a conservation law of the Lagrangian displacement
$\GEV{\xi}{}$ and the stationary solution $\GEV{V}{}$
($\partial_t\GEV{V}{}=\ADJ{\dag}{\GEV{V}{}}{\GEV{V}{}}=0$)
as follows:
\begin{eqnarray}
\fl
  \BraKet{}{\partial_t\GEV{\xi}{}}{
    \partial_t\GEV{v}{}
    - \ADJ{\dag}{\GEV{V}{}}{\GEV{v}{}} - \ADJ{\dag}{\GEV{v}{}}{\GEV{V}{}}
  }
  &=&
  \frac12\dd{t}{}\left(
    \BraKet{}{\partial_t\GEV{\xi}{}}{\partial_t\GEV{\xi}{}}
    -
    \BraKet{}{\ADJ{}{\GEV{\xi}{}}{\GEV{V}{}}}{\ADJ{\dag}{\GEV{\xi}{}}{\GEV{V}{}}}
    -
    \BraKet{}{\ADJ{}{\GEV{\xi}{}}{\GEV{V}{}}}{\ADJ{}{\GEV{\xi}{}}{\GEV{V}{}}}
  \right)
  =0
.
\label{conservation LD}
\end{eqnarray}
The obtained constant of motion gives the second variation of the Hamiltonian (Eq. (35) of \cite{HirotaETAL2006}):
\begin{eqnarray}
\fl
  \delta^2 H_{LD}
  =
    \BraKet{}{\partial_t\GEV{\xi}{}}{\partial_t\GEV{\xi}{}}
    -
    \BraKet{}{\ADJ{}{\GEV{\xi}{}}{\GEV{V}{}}}{\ADJ{\dag}{\GEV{\xi}{}}{\GEV{V}{}}}
    -
    \BraKet{}{\ADJ{}{\GEV{\xi}{}}{\GEV{V}{}}}{\ADJ{}{\GEV{\xi}{}}{\GEV{V}{}}}
 \myLineBreak{3}
  =
  g(\GEM{q}{})|\dot{\GEC{\xi}{}}(\GEM{q}{})|^2
  +
  |\hat{V}(\GEM{k}{})|^2\,
  |\hat{\xi}(\GEM{q}{})|^2\,
  \mathop{\sum_{\HEL_{p}=\pm1}\sum_{\POL_{p}=\pm1}}^{\vec{p}=-\vec{k}-\vec{q}}
  \frac{ \EIG(\GEM{p}{}) (\EIG(\GEM{k}{}) - \EIG(\GEM{p}{})) }{ g(\GEM{p}{}) }
    |\tripleHMHD{}{\GEM{k}{}}{\GEM{p}{}}{\GEM{q}{}}|^2
,
\label{2nd var LD}
\end{eqnarray}
where the second equation is derived by substituting single-mode GEVs for $\GEV{V}{}$ and $\GEV{\xi}{}$.

However, consideration of the conservation laws of the Casimirs naturally leads to the DAV, whose formulae for the velocity and magnetic fields are given by Eqs. (52) and (55) of \cite{HirotaETAL2006} as follows:
\begin{eqnarray}
  \delta\bm{v}_{da}
  &=&
  \bm{\xi}\times(\curl\bm{v})
  +
  \varepsilon^{-1}(\bm{\xi}-\bm{\xi}_{e})\times\bm{B}
  -\nabla\sigma_{1}+\sigma_{2}\nabla s
,
\nonumber\\
  \delta\bm{B}_{da}
  &=&
  \curl(\bm{\xi}_{e}\times\bm{B})
,
\nonumber
\end{eqnarray}
where $\varepsilon$ is the Hall term strength parameter here.
Imposing the incompressibility condition and the constancy of thermodynamic quantities and using $\ADJ{\dagger}{}{}$ operator, the formulae are reduced to
\begin{eqnarray}
  \delta\bm{v}_{da}=\big(\ADJ{\dag}{\GEV{\xi}{}}{\GEV{V}{}}\big)_i
  &,\ \ &
  \delta\bm{B}_{da}=(\HALL\curl)^{-1}\bigg[
    \big(\ADJ{\dag}{\GEV{\xi}{}}{\GEV{V}{}}\big)_i
    -
    \big(\ADJ{\dag}{\GEV{\xi}{}}{\GEV{V}{}}\big)_e
  \bigg]
.
\nonumber
\end{eqnarray}
Thus, using our notation, the DAV of the generalized velocity is given by 
$
  \GEV{V}{} \mapsto
  \GEV{V}{}+\epsilon\ADJ{\dag}{\GEV{\xi}{}}{\GEV{V}{}}+o(\epsilon)
.
$
The principal difference between these two perturbation field classes is that the DAV does not have the time dependence, which implies that the DAV evaluates directly the Hamiltonian around an assigned fixed point (i.e. stationary solution) in the phase space without solving initial value problems.

%%%%%%%%%%%%%%%%%%%%%%%%%%%%%%%%%%%%%%%%%%%%%%%%%%%%%%%%%%%%%%%%%%%%%%%%
\begin{figure}
\begin{center}
\includegraphics[width=0.49\textwidth]{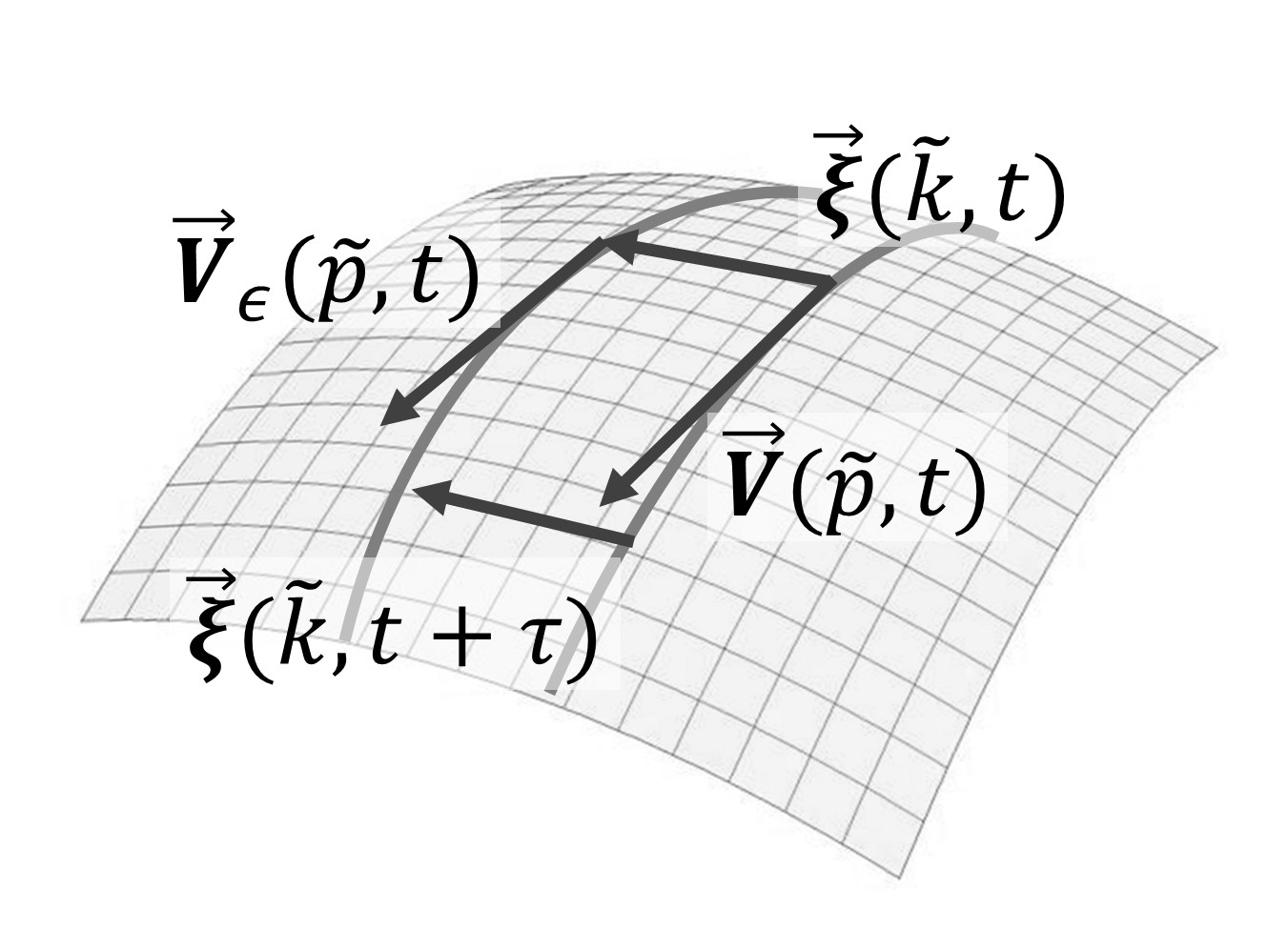}
\includegraphics[width=0.49\textwidth]{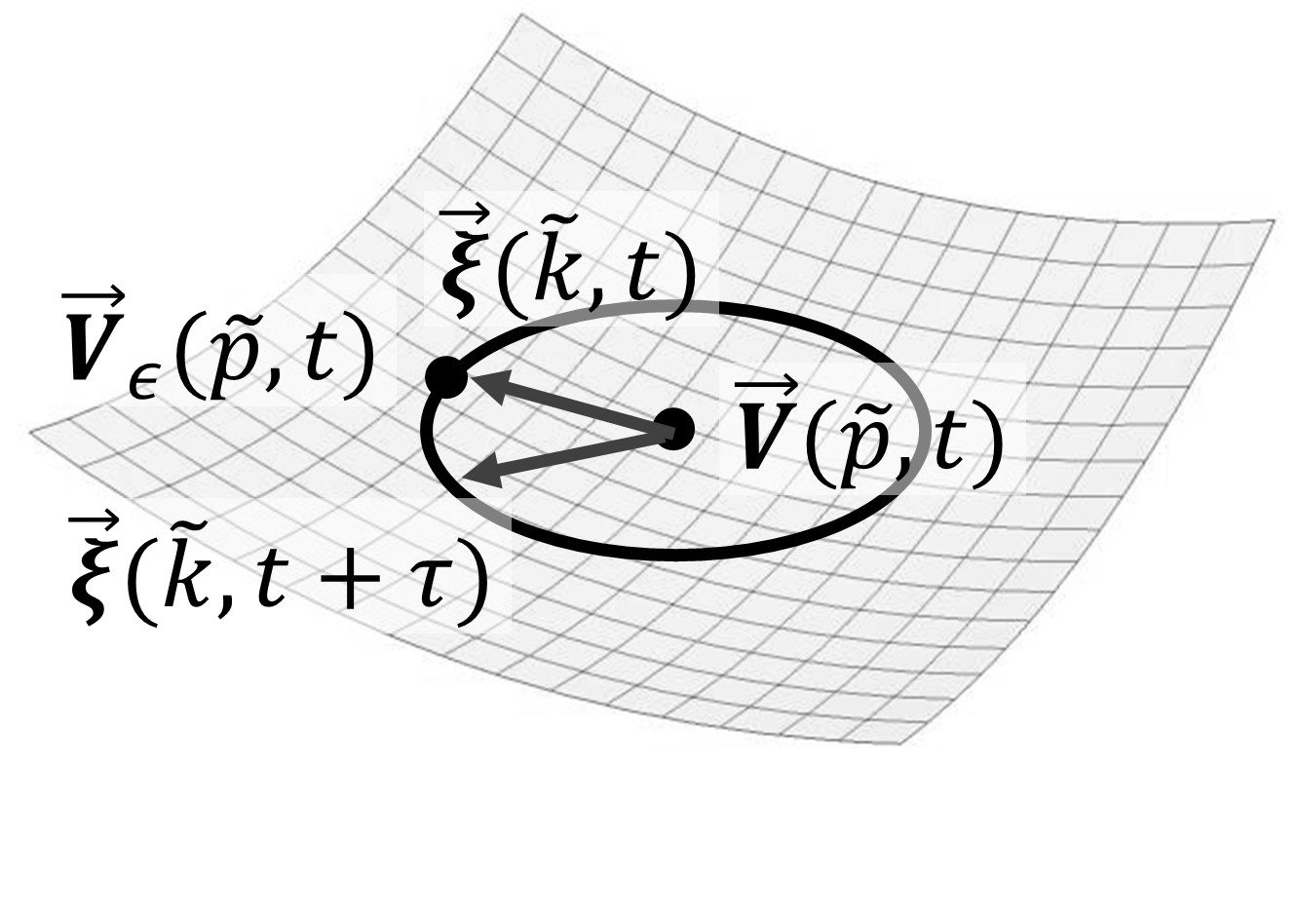}
\end{center}
\caption{\label{compare L and H}
In the Lagrangian mechanics, a stationary solution is expressed by a path
$
  e^{t\GEV{V}{}}
$
on a \textit{configuration} space.
In the Hamiltonian mechanics, on the other hand, a stationary solution is expressed by a fixed point on a \textit{phase} space.
In both cases, the stability condition is evaluated by the growth of some appropriate norm of perturbation fields $||\GEV{\xi}{(t)}||$.
If the perturbation norm is proved to be bounded, the solution satisfy the sufficient condition for the stability.
}%
\end{figure}
%%%%%%%%%%%%%%%%%%%%%%%%%%%%%%%%%%%%%%%%%%%%%%%%%%%%%%%%%%%%%%%%%%%%%%%%

It is an interesting attempt to express the second variation of the Hamiltonian 
$H=H(\U{\GEV{M}{}})=\frac12\BraKet{}{\GEV{V}{}}{\GEV{V}{}}$
in the framework of our formulation.
When both the stationary reference flow 
$\left(\GEV{V}{}=\frac{\delta H}{\delta\U{\GEV{M}{}}}\right)$ 
and the perturbation 
$\left(\GEV{\xi}{}=\frac{\delta K}{\delta\U{\GEV{M}{}}}\right)$ 
are single-mode GEVs, the second variation of the Hamiltonian is given by
\begin{eqnarray}
\fl
  \delta^2 H_{da}
  =
  \{\{H,K\},K\}(\U{\GEV{M}{}})
  =
  \EIG(\GEM{k}{})\,
    |\hat{V}(\GEM{k}{})|^2\,
    |\hat{\xi}(\GEM{q}{})|^2\,
  \mathop{\sum_{\HEL_{p}=\pm1}\sum_{\POL_{p}=\pm1}}^{\vec{p}=-\vec{k}-\vec{q}}
  \frac{ \EIG(\GEM{k}{}) - \EIG(\GEM{p}{}) }{ g(\GEM{p}{}) }
    |\tripleHMHD{}{\GEM{k}{}}{\GEM{p}{}}{\GEM{q}{}}|^2
,
\label{2nd var}
\end{eqnarray}
where $K$ is an arbitrary perturbation superimposed on $H$.
The calculation is carried out using the relation between the Poisson and Lie brackets \cite{marsden2013introduction}:
$$
  \{F,G\}(\U{\GEV{M}{}})
  =
  \left(\U{\GEV{M}{}}\middle|
    \LieBracket{\bigg}{
      \frac{\delta F}{\delta\U{\GEV{M}{}}}
    }{
      \frac{\delta G}{\delta\U{\GEV{M}{}}}
    }
  \right)
,
$$
where the parentheses symbol denotes the natural inner product between the generalized momentum and velocity.

The obtained equations (\ref{2nd var LD}) and (\ref{2nd var}) 
have significantly simpler expressions than that of the sectional curvature.
The signs of the variations are determined by their summation part,
which are verified to take both positive and negative values for assigned $\GEM{k}{}$.
This implies that, as for the GEVs, sufficient condition for stability is not guaranteed.

Besides the physical pictures of the Lagrangian and Hamiltonian mechanics are qualitatively different each other (see Fig. \ref{compare L and H}),
the GEV-mode analyses of the sectional curvature and Hamiltonian variational formulations give quantitatively different results.
What causes the differences?
In the DAV approach, initial value problem is not solved, but the upper bound of a certain appropriate solution norm is sought.
Thus, the result does not have direct information of temporal behavior of perturbed solution.
In the LD approach, the evolution equation of the displacement field ($\GEV{\xi}{}$)
is considered, and its inner product with the field $\partial_t\GEV{\xi}{}$
derives a conservation law (\ref{conservation LD}).
Using the Levi-Civita connection $\LCC{}{}$,
the constants of motion is rewritten as
\begin{eqnarray}
  \delta^2H_{LD}
  &=&
  \BraKet{}{\partial_t\GEV{\xi}{}}{\partial_t\GEV{\xi}{}}
  -
  \BraKet{}{\LCC{\GEV{V}{}}{\GEV{\xi}{}}}{\LCC{\GEV{V}{}}{\GEV{\xi}{}}}
  +
  \BraKet{}{\GEV{\xi}{}}{R({\GEV{\xi}{}},{\GEV{V}{}}){\GEV{V}{}}}
.
\end{eqnarray}
This remarkable expression enables us to discuss about the difference between the energy-Casimir and sectional curvature analyses.
The signs of the terms of $\delta^2H_{LD}$ are positive, negative, and indefinite (depends on the combination of the normal-mode parameters), respectively.
Although time development of perturbation field is taken into account and the sum is a constant of motion, it is difficult to read the temporal behavior.
On the other hand, the sectional curvature analysis evaluates the time development of the norm of the perturbation field
$
\left(
  \BraKet{}{\GEV{\xi}{}}{\GEV{\xi}{}}
  +
  \BraKet{}{\sDD{t}{}\GEV{\xi}{}}{\sDD{t}{}\GEV{\xi}{}}
\right)
,
$
and thus, provides us with the information about its temporal behavior.
The principal difference between these approaches is that the energy-Casimir method investigates the long term behavior of solutions irrespective of the time development details, while the sectional curvature one aims at the details of instantaneous mode interaction behaviors, i.e., the onset of instabilities.

%\newpage
\subsubsection*{Results of normal-mode analysis and implication for turbulence:}
We examined the HD ($\HALL=0$) and MHD ($\HALL\to0$) systems and the $O(\HALL)$ contribution of the Hall-term effect.

The most important and unexpected finding was that the sectional curvature analysis of the MHD system between two GEV modes derives both the stable (positive curvature) and unstable (negative curvature) mode interactions.
This characteristic was in sharp contrast to the behavior of the HD system in which the value of the sectional curvature between two arbitrary CHW modes was negative, i.e., the hydrodynamic flow was always unstable.

Furthermore, the GEV-mode solution was always stable if the combination of the wavenumbers of the reference solution ($\vec{p}$) and perturbation ($\vec{k}$) was nonlocal ($|\vec{k}|/|\vec{p}|\approx0$ or $|\vec{k}|/|\vec{p}|\gg1$), i.e., their characteristic spatial scales differ significantly.
Conversely, the instability occurred only when their spatial scales were relatively close, i.e., the interaction was local.

One possible explanation for the positive sectional curvatures in the MHD case is the propagation of waves.
For example, the positive curvatures at small ${\p}$ can be explained by the Alfv\'en waves (see \ref{physical picture of the nonlocal interaction in the MHD case}).
It is expected that the qualitative physical feature is retained when the wavenumbers gradually change without change the sign of the sectional curvature.

As for the negative curvature, besides wavy plasma motion, the evolution equation 
(\ref{eq of mot u and b}) can be rewritten as
\begin{eqnarray}
  \partial_t(\HALL\bm{\omega}+\bm{b})
  =
  \curl[\bm{v}\times(\HALL\bm{\omega}+\bm{b})]
,&\hspace{1em}&
  \partial_t\bm{b}
  =
  \curl[(\bm{v}-\HALL\bm{j})\times\bm{b}]
,
\end{eqnarray}
which implies that the field lines of $\HALL\bm{\omega}+\bm{b}$ ($\bm{b}$) are frozen-in, i.e., advected by $\bm{v}$ ($\bm{v}-\HALL\bm{j}$).
As discussed in \cite{2016arXiv160105477A},
the ``generalized vorticity'' of the Euler, HMHD, and MHD equations have a common mathematical structure that implies the generalized vorticity fields are frozen-in.
Thus, it is expected that the stretching of field lines occurs according to the motion of the ion and electron fluids, which seems to cause dynamo action or turbulent plasma motion.

A conceptual picture is required to understand consistently the coexistence of such non-growing (possibly wavy) and unstable, growing (possibly chaotic or turbulent) motions, but the development of such a picture is left as future work.
\vs

As for the application to turbulence, taking ensemble average and assuming homogeneity and isotropy, we reduced the problem to one of the sectional curvature analysis:
$
  \EnsAve{\big}{
    R(*,\GEV{V}{})\GEV{V}{}
  }(\vec{k})
  =
  \pi^3\ABS{k}^4
  \int d{\p} \gSAC{}{}({\p})\,Q(\ABS{k}{\p})
$,
where ${\p}:=\ABS{p}/\ABS{k}$, $\gSAC{}{}$ 
is the shell-averaged curvature kernel, and $Q=\EnsAve{\big}{\GEC{V}{}\GEC{V}{}}$ 
is the correlation function of the generalized velocity.
Thus, the stability properties does not depend only on the normal-mode sectional curvature, but also on the spectrum of the correlation function.

Sectional curvature analysis conjectures that the \textit{local} mode interactions are relevant to the instability for both the HD and MHD cases.
That is, the interactions of such modes with close characteristic spatial scales evoke the flow instability.
However, the reasons are quite different from each other.
For the HD case, since the sectional curvature is small enough around $\ABS{p}\approx0$, the integral $\int d{\p} \gSAC{}{}({\p})\,Q(\ABS{k}{\p})$ converges when $Q(p)\propto p^{\gamma}$ for $\gamma>-5$ in the infrared side, which includes the famous Kolmogorov 1941 spectrum $Q(p)\propto p^{-11/3}$.
That is, 
although nonlocal interactions are intrinsically destabilizing, 
their influence is small enough.
For the MHD case, on the other hand, the sectional curvature itself is negative only for the local interactions.

\section*{Acknowledgments}

The author is grateful to Prof. S. Yanase for useful comments 
and to Prof. H. Miura for his continuous encouragement.
The author also expresses sincere thanks to the anonymous referees 
for their helpful comments and suggestions for revising manuscript.
This work was performed under the auspices of 
the NIFS Collaboration Research Program (NIFS13KNSS044, NIFS15KNSS065) 
and 
KAKENHI (Grant-in-Aid for Scientific Research(C)) 23540583.
The author would like to thank Enago for the English language review.

\renewcommand{\thesection}{Appendix \arabic{section}}
\setcounter{section}{0}

\renewcommand{\Q}{;\vec{x}}%}%
\renewcommand{\GEM}[2]{\tilde{#1}}%
\renewcommand{\CHM}[2]{\hat{#1}{#2}}%

%\newpage
\section{Galilean invariance of substantial derivative
  \label{Galilean invariance of substantial derivative}}
\renewcommand{\T}[1]{{#1}^{\prime}{}}

Here we consider the Galilean boost wherein the time and space coordinates
are transformed as
\begin{eqnarray}
  \T{t}={t}
,\ \ 
  \T{x}^{i}={x}^{i}-t{U}^{i}
,
\end{eqnarray}
where $\bm{U}=({U}^{i})$ is an assigned constant velocity.
In this appendix,
prime symbols denote the quantities in the frame $\T{K}$,
which moves relative to the reference frame $K$.
Since Galilean boost does not change 
the distance between or the direction of two arbitrary points at the same instant,
arbitrary frozen-in vector fields
(e.g., particle displacement fields,
say $\bm{\xi}=\xi^{i}\dd{x^{i}}{}$)
are invariant under this transformation:
\begin{eqnarray}
  {\xi}^{i}(\vec{x},t)\left(\dd{{x}^{i}}{}\right)_{\vec{x}}
  \longrightarrow
  \T{\xi}^{i}(\T{\vec{x}},\T{t})\left(\dd{\T{x}^{i}}{}\right)_{\T{\vec{x}}}
,
\label{invariance of frozen-in field}
\end{eqnarray}
where each component satisfies
\begin{eqnarray}
  {\xi}^{i}(\vec{x},t)
  =
  \T{\xi}^{i}(\T{\vec{x}},\T{t})
  =
  \T{\xi}^{i}({\vec{x}}-{t}\bm{U},{t})
.
\end{eqnarray}
This equation leads to the following relation between the values of the partial derivatives 
of the components in $K$ and $\T{K}$:
\begin{eqnarray}
  \left(\dd{t}{{\xi}^{i}}\right)_{({\vec{x}},{t})}
  =
  \left(
    \dd{\T{t}}{\T{\xi}^{i}}
  \right)_{(\T{\vec{x}},{t})}
  -{U}^{j}
  \left(
    \dd{\T{x}^{j}}{\T{\xi}^{i}}
  \right)_{(\T{\vec{x}},{t})}
\\
  \left(\dd{{x}^{k}}{{\xi}^{i}}\right)_{({\vec{x}},{t})}
  =
  \left(
    \dd{\T{x}^{k}}{\T{\xi}^{i}}
  \right)_{(\T{\vec{x}},{t})}
.
\end{eqnarray}
Since each component of the position of a fluid particle obeys
$
  \T{X}^{i}(\vec{a},\T{t})={X}^{i}(\vec{a},{t})-t{U}^{i}
,
$
the components of the Eulerian velocity field satisfy
\begin{eqnarray}
\sakujo{
  \dd{\T{t}}{\T{X}^{i}}(\vec{a},\T{t})=\dd{t}{{X}^{i}}(\vec{a},{t})-{U}^{i}
,\ \ 
}
  \T{V}^{i}\big( \T{\vec{X}}(\vec{a},\T{t}), \T{t} \big) + {U}^{i}
  =
  {V}^{i}\big( {\vec{X}}(\vec{a},{t}) , t \big)
.
\end{eqnarray}
Thus,
the time and covariant derivative of frozen-in field ${\bm{\xi}}$ 
that is advected by the fluid flow ${\bm{V}}$
transforms as
\begin{eqnarray}
  \partial_{{t}}{\bm{\xi}}
  & \longrightarrow &
  \partial_{\T{t}}\T{\bm{\xi}}
  -
  (\bm{U}\cdot\T{\nabla})\T{\bm{\xi}}
,
\\
  {\LCC{}{}}_{\bm{V}}{\bm{\xi}}
  & \longrightarrow &
  \T{\LCC{}{}}_{\T{\bm{V}}+\bm{U}}{\T{\bm{\xi}}}
.
\end{eqnarray}
The invariance of the frozen-in vector field 
(\ref{invariance of frozen-in field})
gives rise to
the following transformation of the substantial derivative:
\begin{eqnarray}
  \partial_{{t}}{\bm{\xi}}
  +
  {\LCC{}{}}_{\bm{V}}{\bm{\xi}}
  & \longrightarrow &
  \partial_{\T{t}}\T{\bm{\xi}}
  +
  \T{\LCC{}{}}_{\T{\bm{V}}}{\T{\bm{\xi}}}
.
\end{eqnarray}
Thus, the covariance of the substantial derivative requires
\begin{eqnarray}
  \T{\LCC{}{}}_{\bm{U}}{\T{\bm{\xi}}}
  =
  (\bm{U}\cdot\T{\nabla})\T{\bm{\xi}}
.
\end{eqnarray}
For the HMHD case,
considering the Galilean boost of 
each of the ion and electron fluid motions,
the following conditions are required:
\begin{eqnarray}
&&
  \big(\T{\LCC{}{}}_{\GEV{U}{}}{\T{\GEV{\xi}{}}}\big)_{i}
  =
  (\bm{U}\cdot\T{\nabla})\T{\bm{\xi}}_{i}
,
\hspace{1em}
  \big(\T{\LCC{}{}}_{\GEV{U}{}}{\T{\GEV{\xi}{}}}\big)_{e}
  =
  (\bm{U}\cdot\T{\nabla})\T{\bm{\xi}}_{e}
,
\end{eqnarray}
where 
$\vec{\bm{U}}=({\bm{U}},{\bm{U}})$,
$\vec{\T{\bm{\xi}}}=(\T{\bm{\xi}}_{i},\T{\bm{\xi}}_{e})$,
and
$\T{\bm{\xi}}_{i}$, $\T{\bm{\xi}}_{e}$ are
frozen-in vector fields advected by the motions of 
the ion and electron plasmas, respectively.
The ion component of the covariant derivative by $\bm{U}$
is given by
\newcommand{\myLineBreakFl}[1]{\nonumber\\\fl\hspace{#1em}}
\begin{eqnarray}
\fl
  \Big(\T{\LCC{}{}}_{\GEV{U}{}}{\T{\GEV{\xi}{}}}\Big)_{i}
  =
  C_{1}
  \LieBracket{\big}{\T{\bm{\xi}}_{i}}{\bm{U}}
  +
  C_{2}
  \left( {\ADJ{\dag}{\vec{\T{\bm{\xi}}}}{\vec{\bm{U}}}} \right)_{i}
  -
  C_{3}
  \left( {\ADJ{\dag}{\vec{\bm{U}}}{\vec{\T{\bm{\xi}}}}} \right)_{i}
,
\myLineBreakFl{0}
  =
  C_{1} \T{\nabla}\times(\bm{U}\times\T{\bm{\xi}}_{i})
  +
  C_{2} \DivFree{\big}{
    (\T{\nabla}\times{\bm{U}})\times\T{\bm{\xi}}_{i}
  }
  -
  C_{3} \DivFree{\big}{
    (\T{\nabla}\times{\T{\bm{\xi}}_{i}})\times\bm{U}
  }
,
\myLineBreakFl{0}
  =
  - ( C_{1} + C_{3} )(\bm{U}\cdot\T{\nabla})\T{\bm{\xi}}_{i}
.
\end{eqnarray}
However, by using Equations (\ref{basic Lie bracket}) and (\ref{adjiadje}),
we can obtain the following difference between the ion and electron components of the covariant derivative:
\begin{eqnarray}
\fl
  \Big(\T{\LCC{}{}}_{\GEV{U}{}}\T{\GEV{\xi}{}}\Big)_{i}
  -
  \Big(\T{\LCC{}{}}_{\GEV{U}{}}\T{\GEV{\xi}{}}\Big)_{e}
\myLineBreakFl{0}
  =
  C_{1} (
    \LieBracket{\big}{\T{\bm{\xi}}_{i}}{\bm{U}}
    -
    \LieBracket{\big}{\T{\bm{\xi}}_{e}}{\bm{U}}
  )
  +
  C_{2} \left[
    \left( {\ADJ{\dag}{\vec{\T{\bm{\xi}}}}{\vec{\bm{U}}}} \right)_{i}
    -
    \left( {\ADJ{\dag}{\vec{\T{\bm{\xi}}}}{\vec{\bm{U}}}} \right)_{e}
  \right]
  -
  C_{3} \left[
    \left( {\ADJ{\dag}{\vec{\bm{U}}}{\vec{\T{\bm{\xi}}}}} \right)_{i}
    -
    \left( {\ADJ{\dag}{\vec{\bm{U}}}{\vec{\T{\bm{\xi}}}}} \right)_{e}
  \right]
,
\myLineBreakFl{0}
  =
  C_{1} \Big[
    \T{\nabla}\times(\bm{U}\times\T{\bm{\xi}}_{i})
    -
    \T{\nabla}\times(\bm{U}\times\T{\bm{\xi}}_{e})
  \Big]
  -
  C_{3} \Big[
    \HALL^{-1}(\HALL\T{\nabla}\times)^2(\T{\bm{b}}_{\xi}\times\bm{U})
  \Big]
,
\myLineBreakFl{0}
  =
  C_{1} \Big[
    -
    (\bm{U}\cdot\T{\nabla})\T{\bm{\xi}}_{i}
    +
    (\bm{U}\cdot\T{\nabla})\T{\bm{\xi}}_{e}
  \Big]
  -
  C_{3} 
    \HALL \T{\nabla} \times \big(
      (\bm{U}\cdot\T{\nabla})\T{\bm{b}}_{\xi}
    \big)
,
\myLineBreakFl{0}
  =
  -
  ( C_{1} + C_{3} )
    (\bm{U}\cdot\T{\nabla})(\T{\bm{\xi}}_{i}-\T{\bm{\xi}}_{e})
,
\end{eqnarray}
where 
$
  \T{\bm{b}}_{\xi} :=
  \HALL^{-1}(\curl)^{-1}(\T{\bm{\xi}}_{i}-\T{\bm{\xi}}_{e})
.
$
Thus, 
the Galilean boost covariance of the substantial derivative
consistently yields the condition
$C_{1} + C_{3} = -1$.

\section{Remark on applicable stability problems:
\label{Remark on applicable stability problems}}
The evolution Equation (\ref{jacobi equation matrix})
can treat a wider class of problems than those considered by Arnold,
despite the comment that he made in his textbook:
\begin{quote}
``It should be emphasized that instability of a flow of an ideal fluid is here understood differently than in section $K$:
it is a question of exponential instability of the \textit{motion of the fluid}, not of its velocity field.'' (\cite{arnold1989mathematical}; app. 2, \S L.)
\end{quote}
Since the linear stability problem (\ref{jacobi equation matrix}) 
requires two ${\GEV{V}{}}$-variables as initial conditions,
we can treat the following two kinds of problems as special cases
to the extent that
the existence and uniqueness of solutions of Equations (\ref{eq of mot}) 
and (\ref{perturbation equation}) are guaranteed.
\begin{enumerate}
\item
By setting 
$
  {}^{t}\!\big({\GEV{\xi}{}(0)}\ {\GEV{\eta}{(0)}}\big)
  =
  {}^{t}\!\big({\GEV{0}{}}\ {\GEV{v}{_{0}}}\big)
,
$
which implies that the fluid particle configurations of the ion and electron fluids are not changed and
the $\GEV{V}{}$-variable perturbation ${\GEV{v}{(0)}}={\GEV{v}{_{0}}}$
is initially imposed on the reference solution, we can treat the linear stability problem of
the \textit{ion and electron fluid velocities}.
\item
By setting
$
  {}^{t}\!\big({\GEV{\xi}{(0)}}\ {\GEV{\eta}{(0)}}\big)
  =
  {}^{t}\!\big({\GEV{\xi}{_{0}}}\
    \LCC{\GEV{\xi}{_{0}}}{\GEV{V}{_{0}}}\big)
,
$
we can obtain a perturbation that satisfies ${\GEV{v}{}(0)}={\GEV{0}{}}$,
i.e.,
the initial ion and electron velocities are unchanged.
The perturbation field ${\GEV{\xi}{_{0}}}$ changes
the ``labels'' of the fluid particle and current field lines
so that
we can treat the \textit{relative dispersion of particle pairs}, i.e.,
instability of the \textit{motion of the ion and electron fluids}.
\end{enumerate}
Note that,
since we did not specify a particular material for the HMHD system in the above discussion, 
the latter is applicable in general to geodesically formulated hydrodynamic systems.

\section{Generalized {\Elsasser} variable representation of differential geometrical quantities
\label{Generalized Elsasser variable representation of differential geometrical quantities}}
\renewcommand{\ABS}[1]{|\vec{#1}|}

The GEVs were originally derived by Galtier \cite{galtier2006wave} 
as linear wave modes under the existence of a uniform background magnetic field, $\bm{B}_{0}$.
In particular, when we consider the standard MHD limit, the GEVs helps us to avoid the singularity problems associated with small values of $\HALL$ \cite{araki2015helicity}.
By setting $\HALL\to0$, 
we could obtain the mathematical expressions for the MHD system from the HMHD ones.

The linearized equations for $\bm{u}$ and $\bm{b}$ are 
\begin{eqnarray}
  \partial_{t}(\curl\bm{u})
  =
  \curl[(\curl\bm{b})\times\bm{B}_{0}]
,
\ \ 
  \partial_{t}\bm{b}
  =
  \curl[(\bm{u}-\HALL\curl\bm{b})\times\bm{B}_{0}]
.
\end{eqnarray}
Using a $\GEV{V}{}$-variable, 
we can rewrite these equations as
\begin{eqnarray}
&&
  \dd{t}{}
  \widehat{W}
  \left(\begin{array}{c} {\bm{V}_{i}} \\ {\bm{V}_{e}} \end{array}\right)
  =
  {\bm{B}_{0}}\cdot\nabla
  \left(\begin{array}{c} {\bm{V}_{i }} \\ {\bm{V}_{e}} \end{array}\right)
,
\label{linear wave equations}
\end{eqnarray}
where the operator $\widehat{W}$ is given by
\begin{eqnarray}
  \widehat{W}
  =
  \left(\begin{array}{cc}
    \HALL\nabla\times + (\HALL\nabla\times)^{-1}
  &
    -(\HALL\nabla\times)^{-1}
  \\
    (\HALL\nabla\times)^{-1}
  &
    -(\HALL\nabla\times)^{-1}
  \end{array}\right)
.
\label{particle relabeling operator}
\end{eqnarray}
Note that
this operator is the helicity-based, particle-relabeling operator 
with a specific parameter value \cite{araki2015helicity}.
Since it contains the curl operator and its inverse,
the operator is convenient for decomposing vector fields into
complex helical waves (CHWs),
each of which is the eigenfunction of the curl operator 
on $\mathbb{T}^{3}$ or $E^{3}$
and is given by
\renewcommand{\CHM}[1]{\vec{#1},\HEL_{#1}}%
\begin{eqnarray}
  \CHW{\phi}{(\CHM{k};\vec{x})}
  :=
  2^{-\frac12}
  \Big(\bm{e}_{\theta}(\vec{k})+i\HEL_{k}\bm{e}_{\phi}(\vec{k})\Big)
  e^{2 \pi i \vec{k}\cdot\vec{x}}
,
\end{eqnarray}
where 
$\vec{k}$, 
${\HEL}_{k}=\pm1$, 
$\bm{e}_{\theta}$ and $\bm{e}_{\phi}$
are
the wavenumber, helicity, and 
base vectors of the spherical coordinate system on a wavenumber space
in the $\theta$- and $\phi$-directions,
respectively 
\cite{waleffe1992nature}.
Using CHWs, we can orthogonally decompose 
(\ref{particle relabeling operator})
according to $\vec{k}$ and $\HEL_{k}$
and obtain the eigenvalue 
\footnote{%
The definition of the eigenvalue differs from that in 
\cite{araki2015differential} by the sign of $\POL_{k}$.
Nevertheless, the correspondence of $\POL_{k}=+1$ and $-1$ to
the ion cyclotron and whistler modes, respectively, is unchanged.
}%
\renewcommand{\GEM}[2]{\vec{#1},\HEL_{#1},{#2}\POL_{#1}}%
\renewcommand{\EIGEN}[2]%
%{\EIG\big(\ABS{#1},{\HEL_{#1}},{{#2}\POL_{#1}},\HALL\big)}%
{\EIG\big(\vec{#1},{\HEL_{#1}},{{#2}\POL_{#1}},\HALL\big)}%
\begin{eqnarray}
\label{eigenvalue of W}
  \EIGEN{k}{}
  =
  \HEL_{k}\POL_{k}\left(
    \sqrt{(\pi\HALL\ABS{k})^2+1}
    +
    \POL_{k}
    \pi\HALL\ABS{k}
  \right)
.
\end{eqnarray}
The eigenfunction corresponding to $\EIGEN{k}{}$ is given by
\footnote{%
The definition of the eigenfunction is $\EIGEN{k}{}$ times 
larger than that in \cite{araki2015differential}.
Thus,
the relationship between the expansion coefficients in the present work 
and those in \cite{araki2015differential} is given by
$
  \EIGEN{k}{}\, \GEC{V}{(\GEM{k}{};t)} = \GEC{Z}{(\GEM{k}{};t)}
$.
}%
\begin{eqnarray}
  \GEV{\Phi}{(\GEM{k}{},\HALL;\vec{x})}
  =
  \left(\begin{array}{c}
    \EIGEN{k}{}\,
    \CHW{\phi}{(\CHM{k};\vec{x})}
  \\
    - \EIGEN{k}{-}\,
    \CHW{\phi}{(\CHM{k};\vec{x})}
  \end{array}\right)
.
\label{eigenfunction of L}
\end{eqnarray}
Equations (\ref{linear wave equations}) and (\ref{eigenvalue of W}),
show that the phase velocity of the $(\GEM{k}{})$ 
mode is proportional to the product 
$(\bm{B}_{0}\cdot\vec{k})/\EIGEN{k}{}$.
Thus, the parameter $\POL_{k}$ 
determines the modulus of the phase velocity for each $(\CHM{k})$.
The \textit{low-frequency} modes 
$\GEV{\Phi}{(\vec{k},\HEL_{k},+,\HALL)}$ 
correspond to the \textit{ion cyclotron waves}%
,
while
the \textit{high-frequency} modes 
$\GEV{\Phi}{(\vec{k},\HEL_{k},-,\HALL)}$ 
correspond to the \textit{whistler waves}%
.
Despite the fact that the phase velocity of linear waves tends to zero
in the limit of vanishing $\bm{B}_{0}$ 
or
collapses to $\pm\bm{B}_{0}\cdot\vec{k}$ 
in the $\HALL\to0$ limit,
we call $\GEV{\Phi}{(\vec{k},\HEL_{k},+,\HALL)}$ and
$\GEV{\Phi}{(\vec{k},\HEL_{k},-,\HALL)}$
the ion cyclotron and whistler modes, respectively, for convenience.

Substituting the GEVs, 
we can obtain the values of the components of the Riemannian metric tensor 
(\ref{def riemannian metric})
as follows:
\begin{eqnarray}
\fl
  \Braket{\Big}{
    \,\CC{
      \GEV{\Phi}{(\GEM{k}{},\HALL)}
    }\,
  }{
    \GEV{\Phi}{(\GEM{p}{},\HALL)}
  }
  =
  \big(1+\EIGEN{p}{}^2\big)\,
  \delta_{\HEL_{k},\HEL_{p}}\,
  \delta_{\POL_{k},\POL_{p}}\,
  \delta^3_{\vec{k},\vec{p}}
.
\label{Riemannian metric in Z-coefficient}
\end{eqnarray}
When the $\GEV{V}{}$-variable is expanded in the GEV modes as
\begin{eqnarray}
  \GEV{V}{(\vec{x},t)}
  =
  \sum_{\GEM{k}{}}
    \GEC{V}{(\GEM{k}{};t)}
    \GEV{\Phi}{(\GEM{k}{},\HALL;\vec{x})}
,
\end{eqnarray}
using the relation (\ref{relation vi ve to u w j b}),
we obtain the GEV-coefficient expansion of the ion velocity field $\bm{u}$, 
current density field $\bm{j}$, and magnetic field $\bm{b}$ as
\begin{eqnarray}
  \bm{u}(\vec{x},t) =
  \sum_{\GEM{k}{}}
    \EIGEN{k}{}\, \GEC{V}{(\GEM{k}{};t)}\, \CHW{\phi}{(\CHM{k};\vec{x})}
,
\label{GEV u}
\\
  \bm{j}(\vec{x},t) =
  2 \pi \sum_{\GEM{k}{}}
    \HEL_{k} \ABS{k}\, \GEC{V}{(\GEM{k}{};t)}\,
    \CHW{\phi}{(\CHM{k};\vec{x})}
,
\label{GEV j}
\\
  \bm{b}(\vec{x},t) =
  \sum_{\GEM{k}{}} \GEC{V}{(\GEM{k}{};t)}\, \CHW{\phi}{(\CHM{k};\vec{x})}
,
\label{GEV b}
\end{eqnarray}
where each of the expansion coefficients $\GEC{V}{(\GEM{k}{};t)}$
is given by
\begin{eqnarray}
\fl
  \GEC{V}{(\GEM{k}{};t)}
  :=
  \big(1+\EIGEN{k}{}^2\big)^{-1}\,
  \Braket{\Big}{
    \,\CC{
      \GEV{\Phi}{(\GEM{k}{},\HALL)}
    }\,
  }{
    \GEV{V}{(t)}
  }
.
\end{eqnarray}
Substituting these expressions and using the eigenvalue relations
\begin{eqnarray}
  \EIGEN{k}{}+\EIGEN{k}{-}=2\pi \HALL \HEL_{k} \ABS{k}
,
\\
  \EIGEN{k}{}\EIGEN{k}{-}=-1
,
\end{eqnarray}
we obtain the explicit expression of the combination of the Riemannian metric and the Lie bracket 
(\ref{<a|[b,c]> in uwjb}) 
as follows:
\renewcommand{\LieBracket}[3]{{#1[}#2,#3{#1]}}%
\renewcommand{\myLineBreak}[1]{\nonumber\\\fl\hspace{#1em}}%
\renewcommand{\CHM}[1]{\vec{#1},\HEL_{#1}}%
\renewcommand{\EIGEN}[2]{\EIG(\tilde{#1}_{#2})}%
\renewcommand{\GEM}[2]{\tilde{#1}}%
\renewcommand{\Q}{}%;\vec{x}}%
\begin{eqnarray}
\fl
\renewcommand{\GEM}[2]{\vec{#1},\HEL_{#1},{#2}\POL_{#1}}%
\renewcommand{\EIGEN}[2]%
{\EIG\big(\vec{#1},{\HEL_{#1}},{{#2}\POL_{#1}},\HALL\big)}%
  \Braket{\Big}{
   {\GEV{\Phi}{(\GEM{k}{},\HALL)}}
  }{
    \LieBraket{\Big}
      {\GEV{\Phi}{(\GEM{p}{},\HALL)}}
      {\GEV{\Phi}{(\GEM{q}{},\HALL)}}
  }
\myLineBreak{1}
  =
  \int_{\vec{x}\in M}{\rm{d}}^3\vec{x}\bigg[
    (\nabla\times{\bm{u}_{\GEM{k}{}}})\cdot
    (\bm{u}_{\GEM{p}{}}\times\bm{u}_{\GEM{q}{}})
    +
    \bm{b}_{\GEM{k}{}}
    \cdot
    \Big(
      \bm{u}_{\GEM{p}{}}\times\bm{j}_{\GEM{q}{}}
      +
      \bm{j}_{\GEM{p}{}}\times\bm{u}_{\GEM{q}{}}
      -\HALL
      \bm{j}_{\GEM{p}{}}\times\bm{j}_{\GEM{q}{}}
    \Big)
  \bigg]_{\vec{x}}
\myLineBreak{1}
  =
  \Bigg[
    \frac{\EIGEN{k}{}+\EIGEN{k}{-}}{\HALL}
    \EIGEN{k}{}\EIGEN{p}{}\EIGEN{q}{}
    +
    \Big(
      \EIGEN{p}{}
      \frac{\EIGEN{q}{}+\EIGEN{q}{-}}{\HALL}
      +
      \EIGEN{q}{}
      \frac{\EIGEN{p}{}+\EIGEN{p}{-}}{\HALL}
     \myLineBreak{2}
      -
      \HALL
      \frac{\EIGEN{p}{}+\EIGEN{p}{-}}{\HALL}
      \frac{\EIGEN{q}{}+\EIGEN{q}{-}}{\HALL}
    \Big)
  \Bigg] 
  \renewcommand{\Q}{;\vec{x}}%
  \int_{\vec{x}\in M}
    \CHW{\phi}{(\CHM{k}{\Q})}
    \cdot \Big(
      \CHW{\phi}{(\CHM{p}{\Q})}
      \times
      \CHW{\phi}{(\CHM{q}{\Q})}
    \Big)
  {\rm{d}}^3\vec{x} 
\myLineBreak{1}
  =
  \HALL^{-1}
  \EIGEN{k}{}
  \bigg(
    \EIGEN{k}{} \EIGEN{p}{} \EIGEN{q}{}
    +
    \EIGEN{k}{-} \EIGEN{p}{-} \EIGEN{q}{-}
  \bigg)
  \tripleEuler{\big}{\CHM{k}}{\CHM{p}}{\CHM{q}}
\myLineBreak{1}
\renewcommand{\GEM}[2]{\vec{#1},\HEL_{#1},{#2}\POL_{#1}}%
\renewcommand{\EIGEN}[2]%
{\EIG\big(\vec{#1},{\HEL_{#1}},{{#2}\POL_{#1}},\HALL\big)}%
  =
  \EIGEN{k}{}\ 
  \tripleHMHD{\big}{\GEM{k}{}}{\GEM{p}{}}{\GEM{q}{}}
\label{<k|{p,q}> in Z-space}
,
\end{eqnarray}
where the notations $\EIGEN{k}{}$ and $\EIGEN{k}{-}$
stand for 
$\EIG(\vec{k},\HEL_{k},\POL_{k},\HALL)$ and
$\EIG(\vec{k},\HEL_{k},-\POL_{k},\HALL)$,
\renewcommand{\GEM}[2]{\vec{#1},\HEL_{#1},{#2}\POL_{#1}}%
\renewcommand{\EIGEN}[2]%
{\EIG\big(\vec{#1},{\HEL_{#1}},{{#2}\POL_{#1}},\HALL\big)}%
and the three-mode parenthesis symbols 
$\tripleEuler{\big}{\CHM{k}}{\CHM{p}}{\CHM{q}}$, 
$\tripleHMHD{\big}{\GEM{k}{}}{\GEM{p}{}}{\GEM{q}{}}$ 
are defined by
\begin{eqnarray}
\fl
\label{def(k|p|q)}
  \tripleEuler{\big}{\CHM{k}}{\CHM{p}}{\CHM{q}}
  :=
  \int_{\vec{x}\in M}
    \CHW{\phi}{(\CHM{k};\vec{x})}
    \cdot\Big(
      \CHW{\phi}{(\CHM{p};\vec{x})}
      \times
      \CHW{\phi}{(\CHM{q};\vec{x})}
    \Big)
  {\rm{d}}^3\vec{x}
 \nonumber\\\hspace{1.5em}
  =
  \frac{
    e^{i\Psi\{\vec{k},\vec{p},\vec{q}\}}|\vec{p}\times\vec{q}|
  }{
    2\sqrt{2}\ABS{k}\ABS{p}\ABS{q}
  }
  \big(\HEL_{k}\ABS{k}+\HEL_{p}\ABS{p}+\HEL_{q}\ABS{q}\big)\,
  \delta^3_{\vec{k}+\vec{p}+\vec{q},\vec{0}}
,
\\
\fl
\label{fac str cont HMHD}
  \tripleHMHD{\big}{\GEM{k}{}}{\GEM{p}{}}{\GEM{q}{}}
  :=
  \HALL^{-1}
  \tripleEuler{\big}{\CHM{k}}{\CHM{p}}{\CHM{q}}
 \nonumber\\\hspace{7em}\times
  \bigg(
    \EIGEN{k}{}\, \EIGEN{p}{}\, \EIGEN{q}{}
   \nonumber\\\hspace{8em}
    +
    \EIGEN{k}{-}\, \EIGEN{p}{-}\, \EIGEN{q}{-}
  \bigg)
,
\end{eqnarray}
respectively.
Explicit expression of the phase factor for 
$\{(\vec{k},\vec{p},\vec{q});\vec{k}+\vec{p}+\vec{q}=\vec{0}\}$
is
$$
  e^{i\Psi\{\vec{k},\vec{p},\vec{q}\}}
  =
  \frac{
    \hat{z}\cdot( \bm{e}_n - i\HEL_{p} \bm{e}_b(\vec{p}) )
  }{
    |\hat{z} \times \bm{e}_r(\vec{p})|
  }
  \frac{
    \hat{z}\cdot( \bm{e}_n - i\HEL_{p} \bm{e}_b(\vec{q}) )
  }{
    |\hat{z} \times \bm{e}_r(\vec{q})|
  }
  \frac{
    \hat{z}\cdot( \bm{e}_n - i\HEL_{p} \bm{e}_b(\vec{k}) )
  }{
    |\hat{z} \times \bm{e}_r(\vec{k})|
  }
,
$$
where %$\hat{z}$,
$\bm{e}_r(\vec{k}):=\vec{k}/\ABS{k}$,
$\bm{e}_n:=(\vec{p}\times\vec{q})/|\vec{p}\times\vec{q}|$,
$\bm{e}_b(\vec{k}):=\bm{e}_r(\vec{k})\times\bm{e}_n$. 
Thus,
the GEV representation of the Levi--Civita connection 
(\ref{levi-civita connection in general}) is given by
\begin{eqnarray}
\fl
  \LCC
    {\GEV{\Phi}{(\GEM{p}{},\HALL)}}
    {\GEV{\Phi}{(\GEM{q}{},\HALL)}}
  =
  \sum_{\HEL_{k},\POL_{k}}
  \tripleHMHD{\big}
    {-\vec{p}-\vec{q},\HEL_{k},\POL_{k}}
    {\GEM{p}{}}{\GEM{q}{}}
 \nonumber\\\fl\hspace{3em}\times
  \frac{
    \EIGEN{p}{}
    -
    \EIGEN{q}{}
    -
    \EIG(-\vec{p}-\vec{q},{\HEL_{k}},{{\POL}_{k}},\HALL)
  }{
   2 \left(
    1 +
    \EIG(-\vec{p}-\vec{q},{\HEL_{k}},{{\POL}_{k}},\HALL)^2
   \right)
  }
  \GEV{\Phi}{(\vec{p}+\vec{q},\HEL_{k},\POL_{k},\HALL;\vec{x})}
.
\label{Levi--Civita in HMHD}
\end{eqnarray}

%-----------------------------------------------------------------------
%\newpage
\section{GEV mode representation of the HMHD sectional curvature
and its Hall-term coefficient expansion
\label{GEV mode representation of the HMHD sectional curvature}}
%-----------------------------------------------------------------------

\renewcommand{\GEM}[2]{\vec{#1},\HEL_{#1},{#2}\POL_{#1}}%
\renewcommand{\myLineBreak}[1]{\nonumber\\\fl\hspace{#1em}}
\renewcommand{\ABS}[1]{#1}
Substituting the explicit expression of the eigenvalues 
(\ref{eigenvalue of W})
and the integral of the triple product of the CHM (\ref{def(k|p|q)})
into Eq. (\ref{def R_H})
(i.e., Equations (\ref{fac str cont HMHD}), (\ref{def R_1}), and (\ref{def R_2}))
and calculating with the aid of Mathematica,
we can obtain the following GEV-mode sectional curvature between the modes 
$(\GEM{k}{})$ and $(\GEM{p}{})$:
\renewcommand{\EIGEN}[2]{\EIG(\tilde{#1}_{#2})}%
\renewcommand{\p}{\underline{p}}%{p_*}%
\renewcommand{\q}{\underline{q}}%{q_*}%
\begin{eqnarray}
\fl
  \MSC{H}{}(\ABS{k},\HEL_{k},\POL_{k},\ABS{p},\HEL_{p},\POL_{p},\ABS{q},\HALL)
  =
  \pi^2 \ABS{k}^2
  \frac{
    (1 - {\p} - {\q}) (1 + {\p} - {\q}) (1 - {\p} + {\q}) (1 + {\p} + {\q})
  }{
    16 {\p}^2 {\q}^2 (2 + 2 {\p}^2 - {\q}^2)
    (1+\EIGEN{k}{}^2)
  }
 \myLineBreak{0}
  \times \Bigg\{
    - \big(
      2 - 10 {\p}^2 - 10 {\p}^4 + 2 {\p}^6 + {\q}^2 + 10 {\p}^2 {\q}^2
      + {\p}^4 {\q}^2 + 4 {\q}^4 + 4 {\p}^2 {\q}^4
      - 3 {\q}^6
    \big)
   \myLineBreak{2}
    +
    2 \HEL_{k} \HEL_{p} (1+\EIGEN{k}{}\, \EIGEN{p}{}) {\p}
      \big(2 + 4 {\p}^2 + 2 {\p}^4 - 9 {\q}^2 - 9 {\p}^2 {\q}^2 + 5 {\q}^4\big)
   \myLineBreak{2}
    +
    2 \EIGEN{k}{}\,
      \EIGEN{p}{}\, 
      \big(2 + 2 {\p}^6
        + 2 {\p}^2 
        + 2 {\p}^4 - 5 {\q}^2 - 4 {\p}^2 {\q}^2 - 5 {\p}^4 {\q}^2
       %\myLineBreak{10}
        + 4 {\q}^4 + 4 {\p}^2 {\q}^4 - {\q}^6\big)
   \myLineBreak{0}
    -
    ( 2 \pi \HALL \HEL_{k} \ABS{k} ) \bigg[
      \EIGEN{k}{} \big(
        2 + 2 {\p}^4 + 4 {\p}^6 - 7 {\q}^2 - 11 {\p}^2 {\q}^2 - 
        8 {\p}^4 {\q}^2 + 3 {\q}^4 + 4 {\p}^2 {\q}^4
      \big)
     \myLineBreak{5}
      +
      \EIGEN{p}{} \big(
        2 {\p}^2 + 2 {\p}^4 - 2 {\q}^2 - 5 {\p}^2 {\q}^2 - 
        4 {\p}^4 {\q}^2 + 3 {\q}^4 + 4 {\p}^2 {\q}^4 - {\q}^6
      \big)
    \bigg]
   \myLineBreak{0}
    -
    ( 2 \pi \HALL \HEL_{p} \ABS{p} ) \bigg[
      \EIGEN{k}{} \big(
        2 {\p}^2 + 2 {\p}^4 - 4 {\q}^2 - 5 {\p}^2 {\q}^2
        - 2 {\p}^4 {\q}^2 + 4 {\q}^4 + 3 {\p}^2 {\q}^4 - {\q}^6
      \big)
     \myLineBreak{5}
      +
      \EIGEN{p}{} \big(
        4 + 2 {\p}^2 + 2 {\p}^6 - 8 {\q}^2 - 11 {\p}^2 {\q}^2
        - 7 {\p}^4 {\q}^2 + 4 {\q}^4 + 3 {\p}^2 {\q}^4
      \big)
    \bigg]
   \myLineBreak{0}
    +
    (2 \pi \HALL \HEL_{k} \ABS{k}) (2 \pi \HALL \HEL_{p} \ABS{p}) \bigg[
      \frac{\HEL_{k} \HEL_{p}}{\p}
      \big(2 {\p}^2 + 4 {\p}^4 + 
        2 {\p}^6 - 2 {\q}^2 - 7 {\p}^2 {\q}^2 - 7 {\p}^4 {\q}^2
       \myLineBreak{12}
        - 2 {\p}^6 {\q}^2 - {\q}^4 - {\p}^2 {\q}^4 - {\p}^4 {\q}^4 + {\q}^6 + {\p}^2 {\q}^6
      \big)
     \myLineBreak{7}
      +
      2
      \big(2 {\p}^2 + 2 {\p}^4 
        - 4 {\q}^2 - 7 {\p}^2 {\q}^2 - 4 {\p}^4 {\q}^2 + 4 {\q}^4 + 
        4 {\p}^2 {\q}^4 - {\q}^6\big) 
     \myLineBreak{7}
      +
      \EIGEN{k}{}\,
        \EIGEN{p}{}\, 
        \big((1 - {\p})^2 - {\q}^2\big) \big((1 + {\p})^2 - {\q}^2\big)
        (2 + 2 {\p}^2 - {\q}^2)
    \bigg]
  \Bigg\}
.
\label{GEV curvature general}
\end{eqnarray}
Hereafter ${\p}:=p/k$, ${\q}:=q/k$.
The dependency on the {\polarity} parameters $\POL_{k}$ and $\POL_{p}$
is included in the eigenvalues $\EIGEN{k}{}$ and $\EIGEN{p}{}$.
Integrating Equation (\ref{GEV curvature general}) with respect to $q$, 
we can obtain the following shell-averaged curvature kernel for the HMHD system:
\newcommand{\text}{}%
\newcommand{\LP}{L({\p})}%{L_{P}}%{\ln\left|\frac{1+{\p}}{1-{\p}}\right|}%
\begin{eqnarray}
\fl
  \SAC{H}{}({k},\HEL_{k},\POL_{k},{p},\HEL_{p},\POL_{p},\HALL)
\myLineBreak{0}
  =
  \frac{\pi ^3 k^4 {\p}^3}{1+\EIGEN{k}{}^2} \Bigg\{
    -\frac{{\p}^3}{4}+\frac{7 {\p}}{12}+\frac{7}{12 {\p}}-\frac{1}{4 {\p}^3}
    +\Bigg(\frac{{\p}^4}{8}+{\p}^2-\frac{9}{4}+\frac{1}{{\p}^2}+\frac{1}{8 {\p}^4}\Bigg)
     \LP
   \myLineBreak{2}
    + \HEL_{k} \HEL_{p} ( 1 + \EIGEN{k}{} \EIGEN{p}{} )
      \Bigg[
      -\frac{3}{2}\Big( {\p}^2 + \frac{1}{{\p}^2}\Big) + \frac{11}{3}
      +
      \frac{3}{4}\Big( {\p}^3 - {\p} - \frac{1}{{\p}} + \frac{1}{{\p}^3}
      \Big) \LP
    \Bigg]
   \myLineBreak{2}
    +
    \EIGEN{k}{} \EIGEN{p}{} \Bigg[
      -\frac{{\p}^3}{2}-\frac{7 {\p}}{6}-\frac{7}{6 {\p}}-\frac{1}{2 {\p}^3}
      +
      \Bigg(
        \frac{{\p}^4}{4}+\frac{{\p}^2}{2}-\frac{3}{2}+\frac{1}{2 {\p}^2}+\frac{1}{4 {\p}^4}
      \Bigg)\LP
    \Bigg]
   \myLineBreak{1}
    +
    (\pi\HALL\HEL_{k}k) \Bigg\{
      \EIGEN{p}{} \Bigg[
        \frac{{\p}}{3}+\frac{3}{{\p}}
        +
        \Big(-\frac{3 {\p}^2}{2}+3-\frac{3}{2 {\p}^2}\Big) \LP
      \Bigg]
     \myLineBreak{2}
      +
       \EIGEN{k}{} \Bigg[
        2 {\p}^3-\frac{17 {\p}}{6}-\frac{3}{{\p}}+\frac{1}{2 {\p}^3}
        +
        \Big(-{\p}^4+\frac{7 {\p}^2}{4}-\frac{3}{4}+\frac{1}{4 {\p}^2}-\frac{1}{4 {\p}^4}\Big)
        \LP
      \Bigg]
    \Bigg\}
   \myLineBreak{1}
    +
    (\pi\HALL\HEL_{p}p) \Bigg\{
      \EIGEN{k}{} \Bigg[
        3 {\p}+\frac{1}{3 {\p}}
        +
        \Big(-\frac{3 {\p}^2}{2}+3-\frac{3}{2 {\p}^2}\Big) \LP
      \Bigg]
     \myLineBreak{2}
      +
      \EIGEN{p}{} \Bigg[
        \frac{{\p}^3}{2}-3 {\p}-\frac{17}{6 {\p}}+\frac{2}{{\p}^3}
        +
        \Big(
          -\frac{{\p}^4}{4}+\frac{{\p}^2}{4}-\frac{3}{4}+\frac{7}{4 {\p}^2}-\frac{1}{{\p}^4}
        \Big)\LP
      \Bigg]
    \Bigg\}
   \myLineBreak{1}
    +
    (\pi\HALL\HEL_{k}k) (\pi\HALL\HEL_{p}p) \Bigg\{
      -\frac{2 {\p}}{3}-\frac{2}{3 {\p}}
      +
      \Big(3 {\p}^2-6+\frac{3}{{\p}^2}\Big) \LP
     \myLineBreak{2}
      +
      \EIGEN{k}{} \EIGEN{p}{} \Bigg[
        {\p}^3-\frac{11 {\p}}{3}-\frac{11}{3 {\p}}+\frac{1}{{\p}^3}
        +
        \Big(-\frac{{\p}^4}{2}+2 {\p}^2-3+\frac{2}{{\p}^2}-\frac{1}{2 {\p}^4}\Big)
        \LP
      \Bigg]
     \myLineBreak{2}
      +
      \HEL_{k} \HEL_{p} \bigg[
        \frac{7 {\p}^2}{3}+\frac{22}{3}+\frac{7}{3 {\p}^2}
        +
        \Big(\frac{3 {\p}^3}{2}-\frac{3 {\p}}{2}-\frac{3}{2 {\p}}+\frac{3}{2 {\p}^3}\Big) \LP
      \bigg]
    \Bigg\}
  \Bigg\}
.
\label{GEV curvature kernel general}
\end{eqnarray}
Hereafter $\LP:=\ln|(1+{\p})/(1-{\p})|$.

Because of its dependence on the eigenvalues $\EIGEN{k}{}$ and $\EIGEN{p}{}$,
the GEV-mode sectional curvature is not a homogeneous expression
with respect to the variables $k$, $p$, and $q$.
However,
they can be expanded in powers of $\pi\HALL\ABS{k}$ as follows:
\begin{eqnarray}
\fl
  \MSC{H}{}(\ABS{k},\HEL_{k},\POL_{k},\ABS{p},\HEL_{p},\POL_{p},q,\HALL)
  &=&
  (\pi\ABS{k})^{2}
  \sum_{n=0}^{\infty}
  (\pi\HALL\ABS{k})^{n}
  \gMSC{H}{(n)}(\HEL_{k},\POL_{k},{\p},\HEL_{p},\POL_{p},{\q})
,
\\\fl
  \SAC{H}{}(\ABS{k},\HEL_{k},\POL_{k},\ABS{p},\HEL_{p},\POL_{p},\HALL)
  &=&
  \pi^3\ABS{k}^{4}
  \sum_{n=0}^{\infty}
  (\pi\HALL\ABS{k})^{n}
  \gSAC{H}{(n)}(\HEL_{k},\POL_{k},{\p},\HEL_{p},\POL_{p})
,
\end{eqnarray}
where 
$
  \gMSC{H}{(n)}(\HEL_{k},\POL_{k},{\p},\HEL_{p},\POL_{p},{\q})
$
and
$
  \gSAC{H}{(n)}(\HEL_{k},\POL_{k},{\p},\HEL_{p},\POL_{p})
$
are dimensionless, scale-independent, geometrical factor functions 
determined by the ratio of the wavenumber moduli ${\p}$ and ${\q}$.
The expansion parameter $\pi\HALL\ABS{k}$ measures 
the scale ratio of the observed motion scale to that of the ion skin depth.

%\newpage
%-----------------------------------------------------------------------
\subsection*{MHD sectional curvature ($O(\HALL^0)$ terms in the HMHD sectional curvature)}
%-----------------------------------------------------------------------
%
%
As discussed in \cite{araki2015helicity},
the GEVs work as an orthogonal basis of the HMHD system 
and are the most suitable for avoiding 
problems with singularities that appear 
when the standard magnetohydrodynamic limit
($\HALL\to0$) is considered.
The leading-order terms
$\pi^2\ABS{k}^2\gMSC{H}{(0)}$ and $\pi^3\ABS{k}^4\gSAC{H}{(0)}$
provide the stability information about the MHD dynamics.
The MHD limit of the GEV-mode sectional curvature
is a homogeneous expression of degree 2 
with respect to $k$, $p$, and $q$
and is given as follows:
\begin{eqnarray}
  \lim_{\HALL\to0}
  \MSC{H}{}(k,\HEL_{k},\POL_{k},p,\HEL_{p},\POL_{p},q,\HALL)
  =
  (\pi\ABS{k})^{2} \gMSC{H}{(0)}({\p},{\q},\HEL_{k}\HEL_{p},\POL_{k}\POL_{p})
,
\nonumber
\end{eqnarray}
where the geometric factor $\gMSC{H}{(0)}$ is given by
\begin{eqnarray}
\fl
\gMSC{H}{(0)}({\p},{\q},\HEL_{k}\HEL_{p},\POL_{k}\POL_{p})
 \myLineBreak{0}
  =
    \frac{
      (1 - {\p} - {\q}) (1 + {\p} - {\q}) (1 - {\p} + {\q}) (1 + {\p} + {\q})
    }{
      32 {\p}^2 {\q}^2 (2 + 2 {\p}^2 - {\q}^2)
    }
 \myLineBreak{1}
  \times \Big[
    -
    \Big(
      2 - 10 {\p}^2 - 10 {\p}^4 + 2 {\p}^6 + {\q}^2 + 10 {\p}^2 {\q}^2
      + {\p}^4 {\q}^2 + 4 {\q}^4 + 4 {\p}^2 {\q}^4
      - 3 {\q}^6
    \Big)
   \myLineBreak{2.5}
    +
    2 (\HEL_{k} \HEL_{p} + \POL_{k} \POL_{p}) {\p} \big(
      2 + 4 {\p}^2 + 2 {\p}^4 - 9 {\q}^2 - 9 {\p}^2 {\q}^2 + 5 {\q}^4 \big)
   \myLineBreak{2.5}
    +
    2 \HEL_{k} \HEL_{p} \POL_{k} \POL_{p}
      \big(2 
     %\myLineBreak{2.5}
      + 2 {\p}^2 + 2 {\p}^4 + 2 {\p}^6 - 5 {\q}^2 
      - 4 {\p}^2 {\q}^2 - 5 {\p}^4 {\q}^2 + 4 {\q}^4
      + 4 {\p}^2 {\q}^4 - {\q}^6\big)
  \Big]
.
\end{eqnarray}
The shell-averaged curvature kernel associated with this MHD limit is given by
\begin{eqnarray}
  \lim_{\HALL\to0}
  \SAC{H}{}(k,\HEL_{k},\POL_{k},p,\HEL_{p},\POL_{p},\HALL)
  =
  \pi^3k^4
  \gSAC{H}{(0)}({\p},\HEL_{k}\HEL_{p},\POL_{k}\POL_{p})
,
\nonumber
\end{eqnarray}
where the geometric factor $\gSAC{H}{(0)}$ is given by
\begin{eqnarray}
\fl
\gSAC{H}{(0)}({\p},\HEL_{k}\HEL_{p},\POL_{k}\POL_{p})
\myLineBreak{0}
  =
  \frac{{\p}^3}{2} \Bigg[
    -\frac{{\p}^3}{4}+\frac{7 {\p}}{12}+\frac{7}{12{\p}}-\frac{1}{4{\p}^3}
    +\Bigg(\frac{{\p}^4}{8}+{\p}^2-\frac{9}{4}+\frac{1}{{\p}^2}+\frac{1}{8 {\p}^4}\Bigg)
     \LP
   \myLineBreak{1}%0}%
    + \left( \HEL_{k} \HEL_{p} + \POL_{k} \POL_{p} \right)
      \Bigg(
      -\frac{3 {\p}^2}{2}+\frac{11}{3}-\frac{3}{2 {\p}^2}
      +
      \Big(\frac{3 {\p}^3}{4}-\frac{3 {\p}}{4}-\frac{3}{4 {\p}}+\frac{3}{4 {\p}^3}
      \Big)
      \LP
    \Bigg)
   \myLineBreak{1}
    +
    \HEL_{k} \HEL_{p} \POL_{k} \POL_{p} \Bigg(
      -\frac{{\p}^3}{2}-\frac{7 {\p}}{6}-\frac{7}{6 {\p}}-\frac{1}{2{\p}^3}
      +
      \Bigg(
        \frac{{\p}^4}{4}+\frac{{\p}^2}{2}-\frac{3}{2}+\frac{1}{2 {\p}^2}+\frac{1}{4 {\p}^4}
      \Bigg)
      \LP
    \Bigg)
  \Bigg]
.
\label{gSACH(0)}
\end{eqnarray}
The helicity and {\polarity} parameters appear as the products
$\HEL_{k}\HEL_{p}$ and $\POL_{k}\POL_{p}$,
which implies that, for example, the stability feature of the interactions between the ion cyclotron branch modes is the same as those between the whistler branch modes,
i.e.,
the stability features are invariant under {\polarity} exchange.

%\newpage
%-----------------------------------------------------------------------
\subsection*{$O(\HALL^1)$ terms in the HMHD sectional curvature}
%-----------------------------------------------------------------------
%
%
The lowest order of the Hall term effect can be expressed as follows:
\begin{eqnarray}
  \HALL \lim_{\HALL\to0} \dd{\HALL}{}
    R_{H}(k,\HEL_{k},\POL_{k},p,\HEL_{p},\POL_{p},q,\HALL)
  =
  \HALL \pi^3 k^3
  \widehat{R}_{H}^{(1)}({\p},{\q},\HEL_{k}\HEL_{p},\POL_{k},\POL_{p})
,
\end{eqnarray}
where the geometric factor $\widehat{R}_{H}^{(1)}$ is given by
\begin{eqnarray}
\fl
\widehat{R}_{H}^{(1)}({\p},{\q},\HEL_{k}\HEL_{p},\POL_{k},\POL_{p})
 \myLineBreak{0}
  =
  -
    \frac{
      (1 - {\p} - {\q}) (1 + {\p} - {\q}) (1 - {\p} + {\q}) (1 + {\p} + {\q})
    }{
      32 {\p}^2 {\q}^2 ( 2 + 2 {\p}^2 - {\q}^2 )
    }
 \myLineBreak{1}
  \times \Big[
      2 \HEL_{k} \HEL_{p} \POL_{k} {\p} 
       (6 {\p}^2 + 4 {\p}^4 - 2 {\p}^6 - 12 {\q}^2 - 15 {\p}^2 {\q}^2 + 
        {\p}^4 {\q}^2 + 9 {\q}^4 + 2 {\p}^2 {\q}^4 - {\q}^6)
   \myLineBreak{2}
    + 4 \HEL_{k} \HEL_{p} \POL_{p} 
       (2 {\p}^2 + 2 {\p}^4 - 2 {\q}^2 - 5 {\p}^2 {\q}^2 - 4 {\p}^4 {\q}^2 + 
        3 {\q}^4 + 4 {\p}^2 {\q}^4 - {\q}^6)
   \myLineBreak{2}
    + \POL_{k} 
       (2 + 6 {\p}^2 + 6 {\p}^4 + 2 {\p}^6 - 15 {\q}^2 - 
        14 {\p}^2 {\q}^2 + {\p}^4 {\q}^2 + 2 {\q}^4 - 6 {\p}^2 {\q}^4 + 3 {\q}^6)
   \myLineBreak{2}
    + 2 \POL_{p} {\p}
       (4 + 2 {\p}^2 + 2 {\p}^6 - 8 {\q}^2 - 11 {\p}^2 {\q}^2 - 
        7 {\p}^4 {\q}^2 + 4 {\q}^4 + 3 {\p}^2 {\q}^4)
  \Big]
.
\label{gMSCH(1)}
\end{eqnarray}
The associated shell-averaged sectional curvature is given by
\begin{eqnarray}
  \HALL \lim_{\HALL\to0} \dd{\HALL}{}
    K_{H}(k,\HEL_{k},\POL_{k},p,\HEL_{p},\POL_{p},q,\HALL)
  =
  \HALL \pi^4 \ABS{k}^5
  \widehat{K}_{H}^{(1)}({\p},\HEL_{k}\HEL_{p},\POL_{k},\POL_{p})
,
\nonumber
\end{eqnarray}
where the geometric factor $\widehat{K}_{H}^{(1)}$ is given by
\begin{eqnarray}
\fl
  \widehat{K}_{H}^{(1)}({\p},\HEL_{k}\HEL_{p},\POL_{k},\POL_{p})
\myLineBreak{0}
  =
  \frac{{\p}^3}{2} \Bigg\{
    s_{k}
    \bigg[
      \left(\frac{3 {\p}^3}{4}+\frac{3}{4 {\p}^3}+\frac{{\p}}{4}-\frac{61}{12 {\p}}\right)
      +
      \left(-\frac{3 {\p}^4}{8}-\frac{3}{8 {\p}^4}+\frac{3}{4}\right)
      \LP
    \bigg]
   \myLineBreak{2}
    +
    s_{k} \sigma_{k} \sigma_{p}
    \bigg[
      \left(-\frac{{\p}^4}{2}+\frac{10 {\p}^2}{3}+\frac{1}{{\p}^2}-\frac{9}{2}\right)
      +
      \left(
        \frac{{\p}^5}{4}-\frac{7 {\p}^3}{4}-\frac{1}{2 {\p}^3}+\frac{9 {\p}}{4}
        -\frac{1}{4 {\p}}
      \right)
      \LP
    \bigg]
   \myLineBreak{2}
    +
    s_{p}
    \bigg[
      \left(\frac{{\p}^4}{2}-3 {\p}^2+\frac{2}{{\p}^2}-\frac{17}{6}\right)
      +
      \left(
        -\frac{{\p}^5}{4}+\frac{{\p}^3}{4}-\frac{1}{{\p}^3}-\frac{3 {\p}}{4}+\frac{7}{4 {\p}}
      \right)
      \LP
    \bigg]
   \myLineBreak{2}
    +
    s_{p} \sigma_{k} \sigma_{p}
    \bigg[
      \left(\frac{{\p}}{3}+\frac{3}{{\p}}\right)
      +
      \left(-\frac{3 {\p}^2}{2}-\frac{3}{2 {\p}^2}+3\right)
      \LP
    \bigg]
  \Bigg\}
.
\end{eqnarray}

%-----------------------------------------------------------------------
%\newpage
\section{CHW mode representation of the Euler sectional curvature
\label{CHW mode representation of the Euler sectional curvature}}
%-----------------------------------------------------------------------

For comparison with the HMHD case, 
we examine the HD case, i.e., the case wherein the magnetic field is absent.
As was discussed in \cite{2016arXiv160105477A},
the dissipationless, incompressible HD, MHD, and HMHD systems have a common mathematical structure.
The Riemannian metric and the Lie bracket are given by
\begin{eqnarray}
  \Braket{\big}{
    \bm{u}_{1}
  }{
    \bm{u}_{2}
  }
  :=
  \int
    \bm{u}_{1} \cdot \bm{u}_{2}\,
  {\rm{d}}^3\vec{x}
,\hspace{2em}
  \LieBracket{\big}{\bm{u}_{1}}{\bm{u}_{2}}
  =
  \nabla \times ( \bm{u}_{1} \times \bm{u}_{2} )
,
\label{the HD <*|*> and [*,*]}
\end{eqnarray}
respectively.
The helicity-based, particle-relabeling operator of the HD system is the curl operator, and its eigenfunctions are given by the CHWs.
The values of the Riemannian metric and the Lie bracket, 
which correspond to Equations
(\ref{Riemannian metric in Z-coefficient}) and (\ref{<k|{p,q}> in Z-space})
in the HMHD case, are given by
\begin{eqnarray}
&&
  \Braket{\Big}{
    \CC{
      \CHW{\phi}{(\CHM{k}{})}
    }
  }{
    \CHW{\phi}{(\CHM{p}{})}
  }
  =
  \delta^3_{-\vec{k},\vec{p}}\,
  \delta_{\HEL_{k},\HEL_{p}}
,
\label{Riemannian metric in Euler}
\\&&
  \Braket{\Big}{
    {\CHW{\phi}{(\CHM{k}{})}}
  }{
    \LieBraket{\big}{\CHW{\phi}{(\CHM{p}{})}}{\CHW{\phi}{(\CHM{q}{})}}
  }
  =
  \HEL_{k}\ABS{k}\,
  \tripleEuler{\big}{\CHM{k}{}}{\CHM{p}{}}{\CHM{q}{}}
,
\label{<k|{p,q}> in Euler}
\end{eqnarray}
respectively,
where the three-mode parenthesis symbol
$\tripleEuler{\big}{\CHM{k}}{\CHM{p}}{\CHM{q}}$
is defined by Eq. (\ref{def(k|p|q)}).
The connection that satisfies the same three physical conditions 
discussed in section 
\ref{Linear connection with physically desirable properties}
is given by
\begin{eqnarray}
\fl
  \LCC
    {\CHW{\phi}{(\CHM{p}{})}}
    {\CHW{\phi}{(\CHM{q}{})}}
  &=&
  \frac12
  \sum_{\HEL_{k}}^{\vec{k}+\vec{p}+\vec{q}=\vec{0}}
  \tripleEuler{\big}
    {\vec{k},\HEL_{k}}
    {\CHM{p}{}}{\CHM{q}{}}
 %\nonumber\\\fl\hspace{3em}&&\times
  \left(
    \HEL_{p}\ABS{p}
    -
    \HEL_{q}\ABS{q}
    -
    \HEL_{k}\ABS{k}
  \right)
  \CHW{\phi}{(\vec{p}+\vec{q},\HEL_{k})}
.
\label{Levi--Civita in Euler}
\end{eqnarray}
The geodesic equation corresponding to this connection is
\begin{eqnarray}
  \left(\dd{t}{}+\LCC{\GEV{V}{}}{}\right)\GEV{V}{}
  =
  \dd{t}{\bm{u}}+\DivFree{\Big}{(\bm{u}\cdot\nabla)\bm{u}}
  =
  0
.
\end{eqnarray}
The CHW-mode sectional curvature,
which is the Euler counterpart of Eq. (\ref{GEV curvature general}), becomes
$
  \MSC{E}{}(k,\HEL_k,p,\HEL_p,q)
  =
  (\pi\ABS{k})^{2}
  \gMSC{H}{}({\p},{\q})
,
$
where $\gMSC{H}{}({\p},{\q})$ is the geometric 
(i.e. $|\vec{k}|$-independent) factor given by
\begin{eqnarray}
\fl
  \gMSC{H}{}({\p},{\q})
  =
  - 
  \frac{
    (1 - {\p} - {\q})^2 (1 + {\p} - {\q})^2 (1 - {\p} + {\q})^2 (1 + {\p} +  {\q})^2
  }{
    16 {\p}^2 {\q}^2
  }
.
\label{Euler curvature general}
\end{eqnarray}
The corresponding shell-averaged curvature kernel 
(cf. Equation (\ref{curvature kernel})) is given by
$
  \SAC{E}{}(\ABS{k},\HEL_{k},\ABS{p},\HEL_{p},\HALL)
  =
  \pi^3\ABS{k}^{4}
  \gSAC{E}{}({\p})
,
$
where $\gSAC{E}{}({\p})$ is the geometric factor given by
\begin{eqnarray}
\fl
  \gSAC{E}{}({\p})
  =
  {\p}^3
  \left[
    \frac{{\p}^3}{4}-\frac{11 {\p}}{12}-\frac{11}{12 {\p}}+\frac{1}{4 {\p}^3}
    -
    \left(
      \frac{{\p}^4}{8}-\frac{{\p}^2}{2}+\frac{3}{4}-\frac{1}{2 {\p}^2}+\frac{1}{8 {\p}^4}
    \right)
    \LP
  \right]
.
\label{Euler curvature kernel}
%\nonumber\\
\end{eqnarray}
It is interesting that the obtained CHM-mode sectional curvature does not have a helicity parameter even though one is taken into account in the derivation.

\section{Physical picture of the nonlocal interaction in the MHD case
\label{physical picture of the nonlocal interaction in the MHD case}}
For the mode interaction between the observed perturbation mode ($\bm{u}_k$, $\bm{b}_k$) and the reference flow mode ($\bm{u}_p$, $\bm{b}_p$), which is spatially large compared with the perturbation ($\ABS{p}\ll\ABS{k}$), the evolution equation (\ref{eq of mot u and b}) can be approximated as follows:
$$
\left\{\begin{array}{l}
(\partial_t+\boldsymbol{u}_{p}\cdot\nabla)\boldsymbol{u}_k=(\boldsymbol{b}_{p}\cdot\nabla)\boldsymbol{b}_k
,
\\
(\partial_t+\boldsymbol{u}_{p}\cdot\nabla)\boldsymbol{b}_k=(\boldsymbol{b}_{p}\cdot\nabla)\boldsymbol{u}_k
.
\end{array}\right.
$$
This equation has solutions that are Alfv\'en waves propagating in a frame moving with velocity $\bm{u}_p$.
Hence, the characteristic time scale is given by $\bm{b}_p\cdot\vec{k}$ or $\bm{v}_p\cdot\vec{k}$.
Their values are 
$\bm{u}_p\cdot\vec{k}=\CHC{u}_p\ABS{k}\sin\theta$, 
$\bm{b}_p\cdot\vec{k}=\CHC{b}_p\ABS{k}\sin\theta$, 
respectively,
where $\theta$ is the angle between $\vec{k}$ and $\vec{p}$,
because the velocity and magnetic fields are solenoidal ($\bm{u}_p$, $\bm{b}_p\perp\vec{p}$).
Thus, the square of the characteristic time scale is given by
$|\bm{u}_p\cdot\vec{k}|^2 = |\bm{b}_p\cdot\vec{k}|^2 = \frac12Q(p)k^2\sin^2\theta$.
Integrating this equation over a constant $|\vec{p}|$ shell, we can obtain the net time scale factor arising from the Alfv\'en waves on a constant $|\vec{p}|$ shell as 
$\frac12\int\int Q(p)k^2p^2\sin^3\theta\,d\theta\,d\phi
=\frac{4}{3}\pi \ABS{k}^2 \ABS{p}^2 Q(\ABS{p})
%=\gSAC{H}{(0)}(p) Q(\ABS{p})
$.
As mentioned in the section \ref{O(a0) features of HMHD dynamics: MHD system stability},
the geometric factor of the shell-averaged curvature kernel for the MHD system 
behaved as $\SAC{H}{(0)}(k,p) = \frac{4}{3} \pi^3 k^4 p^{2} + o(p^2)$
for sufficiently small wave numbers $p\ll k$.
Thus, the contribution of small wavenumber component (i.e. the large-scale plasma motion) to the sectional curvature
$
  \SAC{H}{(0)}(p)Q(p) = \frac{4}{3} \pi^3 k^4 p^{2} Q(p)
$
is conjectured to be due to the propagation of the Alfv\'en waves.
The fact that the sectional curvature was positive is consistent with the oscillatory, non-growing motion of the Alfv\'en waves (see section \ref{statistical homogeneity}).

%%%%%%%%%%%%%%%%%%%%%%%%%%%%%%%%%%%%%%%%%%%%%%%%%%%%%%%%%%%%%%%%%%%%%%%%%%%%%%%%
\section*{Reference}
\bibliographystyle{unsrt}
\bibliography{HMHDnotes}

\end{document}